\documentclass{ar2e}[10pts, onecolumn,final]

\usepackage{ARAstroBib,amsmath,amssymb,graphicx}

\usepackage{setspace}
\usepackage{amssymb}
\usepackage{rotating}
\usepackage{array}

\newcommand{\url}{}

\begin{document}
\singlespacing

\input epsf.sty    

\input psfig.sty

\jname{Annu.\ Rev.\ Astron.\ Astrophys.}
\jyear{2014}
\jvol{1}

\title{Numerical Relativity and Astrophysics}

\markboth{Luis Lehner \& Frans Pretorius}{Numerical Relativity and Astrophysics}

\author{Luis Lehner${}^1$ \& Frans Pretorius ${}^2$
\affiliation{${}^1$Perimeter Institute for Theoretical Physics,\\
51 Caroline Street North, Waterloo, Ontario N2L 2Y5, Canada\\
${}^2$Department of Physics, Princeton University,\\
Princeton, New Jersey 08544, USA}}


\begin{abstract}
Throughout the Universe many powerful events are driven by strong gravitational effects
that require general relativity to fully describe them. These include compact binary
mergers, black hole accretion and stellar collapse, where velocities can approach the speed of 
light, and extreme gravitational fields --$\Phi_{\rm Newt}/c^2 \simeq 1$-- mediate the interactions.
Many of these processes trigger emission across a broad range of the electromagnetic
spectrum. Compact binaries further source strong gravitational wave emission
that could directly be detected in the near future. This feat will open up a gravitational wave 
window into our Universe and revolutionize its understanding.
Describing these phenomena requires general relativity, and --where dynamical effects strongly modify gravitational fields--
the full Einstein equations coupled to matter sources.
Numerical relativity is a field within general relativity concerned with studying such
scenarios that cannot be accurately modeled via perturbative or analytical calculations.
In this review, we examine results obtained within this discipline, with
a focus on its impact in astrophysics.
\end{abstract}

\maketitle

\section{Introduction}
Strong gravitational interactions govern many of the most fascinating astrophysical phenomena
and lie behind some of the most spectacular predictions of general relativity, such as black holes and neutron stars.
These objects produce extreme gravitational fields and are believed to be responsible for the most energetic
events in our Universe. Indeed, models for gamma-ray bursts, quasars, AGN, pulsars and a class of ultra-high-energy
cosmic rays all have these still poorly understood compact objects as putative central engines.
Observations across the electromagnetic spectra, soon to be combined with gravitational signals
produced by merging binaries, should provide important insights into their nature. Of course, such understanding
can only be gained by contrasting theoretical models that include all the relevant physics to the 
full front of observations.

It is important to distinguish two sub-classes of strongly gravitating systems. The first is where 
the self-gravitation of any matter/gas/plasma interacting with a compact object or binary
is sufficiently weak such that the gravitational back reaction can be ignored or treated perturbatively. 
Such systems can be analyzed by studying the dynamics of matter on a given fixed
background geometry. Examples include accreting black holes and tidal disruption of
main sequence stars by supermassive black holes.
Widely separated compact binary systems also belong to this sub-class, and suitable post-Newtonian (PN) expansions 
can be adopted to account for the slowly varying gravitational field and its effects.

By contrast, if the interaction is strong and can fundamentally affect the gravitational field of the
system, a fully relativistic, self-gravitating study must be performed.
To this end the Einstein equations, coupled to any relevant matter fields, must
be employed. This task is complex due to the involved nature of Einstein's equations
(a nonlinear, strongly coupled system of equations) in which analytical solutions are only known in highly
specialized scenarios. Consequently, numerical simulations are required and the discipline that concentrates
on the development and application of numerical solutions of Einstein's equations 
is known as {\em numerical relativity} (NR). 

This discipline has, over several decades, steadily progressed to the current epoch in which
studies of relevance to astrophysics can now be performed that address questions both 
of fundamental theoretical interest and that make contact with observations
\footnote{Numerical relativity is also being used to address problems in cosmological contexts.  Applications
that require NR, including bubble collisions~\citep{2012PhRvD..85h3516J,2013arXiv1312.1357W},
the issue of nonlinear structures and voids~\citep{Zhao:2009yp,2013PhRvL.111p1102Y,2014arXiv1404.1435Y},
the evolution near the bounce in cyclic models~\citep{2008PhRvD..78h3537G,2013PhRvD..88h3509X}
and certain aspects of cosmic string dynamics~\citep{Laguna:1989rx}, are still at either a speculative level
or being explored.
In constrast, the paradigm applicable to most of present day observational cosmology can 
effectively be addressed with exact Friedman-Lamaitre-Robison-Walker solutions and perturbations about them, and do not require NR.}.
Of particular interest, spurred by a hope of imminent gravitational wave observation,
are systems capable of producing strong gravitational emission. Detectors include ground-based
interferometers such as LIGO/VIRGO/KAGRA~\citep{Abbott:2007kv,2011CQGra..28k4002A,Somiya:2011np} targeting the
$\simeq 10$Hz-1KHz frequency band, a pulsar timing network (see e.g.~\cite{2013CQGra..30v4010I}) 
sensitive to the 300pHz-100nHz window,
and possible future spaced-based missions (NGO/eLISA, see, e.g.~\cite{AmaroSeoane:2012km}), 
sensitive between $\simeq 10\mu$Hz-0.1Hz. 
Compact binary systems, involving
black holes or neutron stars, are the most natural sources and have thus been the focus of
most recent efforts (see e.g. ~\cite{Andersson:2013mrx} for a recent overview).
In this article we review the key messages obtained by NR relevant to astrophysics.
The discipline is till 
in the midst of rapid development over an increasing breadth 
of applications, promising even more exciting future discoveries
of astrophysical import.

\section{Brief Review of Techniques, Methods and Information Obtainable from Gravitational Waves}
Understanding gravity in highly dynamical/strongly gravitating regimes requires solving Einstein's equations.
This provides the metric tensor, $g_{ab}$, which encodes gravitational effects in geometrical terms. The
covariant character of the equations encode the equivalence principle, hence there is 
no preferred frame of reference to write the particular form of the metric for a given
physical geometry. This further implies that the field equations determining $g_{ab}$ do not lend themselves to a well defined initial value problem unless the spacetime is foliated into a series of surfaces that provide a notion of ``time.'' 
One can then cast Einstein's equations in a form that provides a recipe to evolve the intrinsic metric of each slice 
with time in what has been called ``geometrodynamics.'' There are several options
to carry out this program (see, e.g. the discussion by~\cite{Lehner:2001wq}), though the most common one
is to define these surfaces to be spacelike. This is also most closely related to Newtonian mechanics, and hence provides 
useful intuition in astrophysical scenarios; furthermore,
with some additional assumptions about the coordinates,
the familiar Newtonian potential can easily be extracted from the metric for weakly gravitating systems.  
Current efforts most 
commonly employ one of two particular reformulations of Einstein's equations: the generalized harmonic and the 
BSSN formulations\footnote{For a recent review, see~\cite{Sarbach:2012pr}.}. These equations are hyperbolic
with characteristics given by the speed of light (regardless of the state of the system, as opposed to the
familiar case of hydrodynamics in which perturbations propagate with speeds tied to the state of the fluid).
When coupling in matter sources the equations of relativistic 
hydrodynamics (or magnetohydrodynamics) on a dynamical, curved geometry must also be considered. The relevant
equations can be expressed in a way fully consistent with standard approaches to integrate the Einstein 
equations (for a review on this topic see~\cite{2008LRR....11....7F}).

With the equations defined, they can be discretized for numerical integration.
For the systems considered here, a crucial
observation is that simulations must be carried out in full generality. This means
that time and spatial variations are equally important, and a disparate range of scales need to be resolved (ranging from 
at least the size of each compact object, through the scale where gravitational waves are produced, and to the 
asymptotic region where they are measured). The associated computational cost is quite high, and typical simulations run 
on hundreds to thousands of processors for hours to weeks, even when efficient resolution of the relevant
spatio-temporal scales can be achieved using (for example) adaptive mesh refinement.
It is beyond the scope of this review to describe the techniques employed in detail, though we
briefly mention them and point to some relevant literature for further details. (See also a few textbooks
on the subject written in recent years~\citep{2009LNP...783..171B,2008itnr.book.....A,2010nure.book.....B}).

\begin{itemize}
\item Spatial discretization. As far as the gravitational field itself is concerned, solutions
are generally smooth (except at singularities) provided smooth initial data are defined  
because the equations of motions are {\em linearly degenerate}
(i.e. do not induce shocks from smooth initial data). High-order finite difference approximations (e.g.~\cite{Gustafsson95}) 
or spectral decompositions (e.g.~\cite{Boyd89a,Grandclement:2009LR})
allow for a high degree of accuracy. When matter and the hydrodynamic equations are involved,
finite volume methods and high-resolution shock capturing schemes can be used to determine the future
evolution of the fluid variables (e.g~\cite{Leveque92}).
\item Time integration. The method of lines can be straightforwardly implemented once spatial derivatives
are computed.
\item Constraint enforcement. For systems of interest, several constraints are typically involved.
Those coming from Einstein equations themselves are a nonlinear coupled set of PDEs. In general scenarios,
these constraints are difficult to enforce directly; instead, a strategy of ``constraint damping'' is adopted, whereby
the equations of motion are modified  in a suitable manner via the addition of constraints.
The resulting system is thus not different from the initial one when the constraints are
satified, otherwise the numerical evolution should damp these violations as time progresses.
This desirable behavior can be rigorously shown to hold in perturbations off flat spacetime \cite{Brodbeck:1998az,Gundlach:2005eh} 
and also ``experimentally'' verified in simulations involving black holes and 
neutron stars (e.g.~\cite{Pretorius:2005gq,Pretorius:2006tp,Anderson:2008kz,Chawla:2010sw}). 
This technique --whereby the equations are suitably modified to control constraints-- has
also been extended to other relevant systems of equations. 
For instance, when considering magnetohydrodynamics or electrodynamics, to control the
 no-monopole constraint~\citep{Neilsen:2005rq,Palenzuela:2009hx}.
\item Mesh structure, resolution and adaptivity. As mentioned, several different physical scales need to be
resolved.
For an efficient implementation, techniques like adaptive mesh refinement and 
multiple patches are in common use (e.g.~\cite{Berger:1984zza,Schnetter:2003rb,Lehner:2005vc,Lehner:2005bz,Duez:2008rb,code_paper}).
\item Parallelization. The equations involved are of hyperbolic type and they lend themselves naturally
to a relatively straightforward parallelization. Several computational infrastructures have been
developed for numerical relativity purposes, e.g. BAM, Cactus~\citep{cactus_webpage} and the Einstein Toolkit~\citep{Loffler:2011ay},
Had~\citep{had_webpage}, Whisky, SACRA~\citep{Baiotti:2010ka}.
\end{itemize}

\section{Brief Description of the Dynamics of a Binary System}
Here we review salient properties of the early phase of binary evolution in general relativity
to set the stage for subsequent discussion of the nonlinear regime uncovered by numerical simulations.
For further details the interested reader can consult~\cite{Hughes:2009iq}. \\

An isolated compact binary evolves due to the emission of gravitational waves, and 
consequently a bound system will eventually merge. The end state of compact
binary mergers (i.e. binary black holes,
black hole-neutron star binaries, and all except the least massive binary
neutron stars) will be a single Kerr black hole\footnote{Provided cosmic censorship holds, and there are no indications
yet that it fails for mergers in four-dimensional, asymptotically flat spacetime.}.
At large separations, in which the local velocity of each object in the binary 
is small (relative to the speed of light $c$), a PN expansion (e.g.~\cite{Blanchet:2002LR}),
where objects are taken as point-particles without internal dynamics, suffices to accurately describe the system. 
As the orbit shrinks, the faithfulness of such an expansion decreases as velocities become $O(c)$. If either
compact object is a neutron star tidal effects may be important; these can be modeled within the PN
framework, though again the accuracy of the expansion degrades approaching tidal disruption, which can occur
near merger for stellar mass binaries. During this late stage of inspiral
full numerical solution must be employed to obtain an accurate description of the dynamics of the
geometry and matter. Once a single black hole forms, very shortly afterward (on the order
of a few light-crossing times of the Schwarzschild radius) the spacetime can accurately be
modeled by black hole perturbation theory, and to a good approximation the matter can be
evolved on a stationary black hole background.
In the standard jargon of the field the three different stages just described
are often referred to as the PN inspiral, nonlinear and ``ring-down'' stages. 

The nonlinear phase can
further be subdivided into a late inspiral, plunge and early postmerger epoch. In the first subphase
the binary is still in an orbit, though velocities are high, the orbital frequency quickly 
sweeps upwards, and neutron star tidal dynamics can become relevant (if the
companion is a neutron star or black hole with mass $\lesssim 20 M_\odot$). The second
refers to a rapid increase in the magnitude of the inward radial velocity leading to merger. The plunge
is related to the phenomenon of an innermost stable circular orbit (ISCO) of a black hole, and is thus
most apparent in a high-mass-ratio compact binary (see e.g.~\cite{Buonanno:2006ui,Buonanno:2007sv}).
The last sub-phase begins when either a black hole or a hyper-massive neutron
star forms, and lasts while either object is too ``distorted'' for a straightforward perturbative approach
to be applicable. As mentioned above a black hole settles to a stationary state very rapidly, and hence from
a computational perspective there is little to gain switching to a perturbative treatment 
to measure the ring-down waves. By contrast, in certain ranges of parameter space a hyper-massive 
neutron star can last for several seconds before collapsing to a black hole, which for the
full coupled Einstein-matter equations would be too expensive to evolve at present (a rough estimate
of the cost is O(1000) CPU hours per ms at ``modest'' resolution).

\subsection{Properties of Gravitational Wave Emission}
During the early inspiral stage in which velocities are much smaller than $c$, to leading order the emission
of gravitational waves is proportional to the acceleration of the reduced (trace free) quadrupole
moment tensor $Q_{ij}(t)$ of the system (this is textbook material, though for a couple
of recent review articles see~\cite{Flanagan:2005yc,Buonanno:2007yg}):
\begin{equation}\label{quad}
h_{ij}{}^{TT}(t,\vec{x}) = \frac{ 2 G}{r c^4} \frac{\partial ^2 Q_{kl}(t-r)}{\partial t^2} 
\left[\perp^k{}_i \perp^l{}_j -\frac{1}{2} \perp^{kl} \perp_{ij}\right]
\end{equation}
In the above, $h_{ij}{}^{TT}$ is the perturbation of the spatial components of the
Minkowski metric $\eta_{ij}$ in the transverse traceless (TT) gauge, written in a
Cartesian coordinate system $(t,\vec{x})=(t,x_i)$. In this gauge there are no 
space-time or time-time perturbations of $\eta_{\mu\nu}$, ie $h_{tt}{}^{TT}=h_{tj}{}^{TT}=0$. 
The center of mass of the
source is at the origin, and the above expression assumes the perturbation is
measured at a distance $r=|\vec{x}|$ much greater than the characteristic size
of the source, here $\sim r_p$, the periapse of the orbit\footnote{We use
the periapse here rather than, say, the semi-major axis, because for highly eccentric
systems described later the dominant gravitational wave emission only
occurs around periapse passage. Thus $r_p$ more conveniently characterizes the
relevant scale of gravitational wave emission for all eccentricities.}. 
The projection
tensor $\perp_{ij} = \delta_{ij} - \hat{n}_i \hat{n}_j$, with $\hat{n}_i = x_i/r$, i.e.,
$\hat{n}_i$ is the unit spatial vector from the source to the observer at location $x_i$. 
The above expression is (to leading order) valid in an expanding Universe if the distance $r$ is
replaced by the luminosity distance $D_l$, and time is dilated by a factor $1+z$, where
$z$ is the redshift between the source and the observer.

The projection in (\ref{quad}) encodes the property of general relativity that there are only two
linearly independent propagating degrees of freedom, called the cross and plus
polarizations. Thus the tensor $h_{ij}{}^{TT}$ only has two independent non-zero components,
which are called $h_+$ and $h_{\rm x}$. To illustrate, ignoring back-reaction, a binary on a circular Keplerian
orbit with orbital frequency $\omega=\sqrt{2GM/r_p^3}$ produces a radiation pattern
\begin{eqnarray}\label{wavesPN}
h_+(t,r,\theta,\phi) &=& \frac{4G}{r c^4}\mathcal{M}^{5/3} (2\omega)^{2/3} \cos(2\omega t+\phi)\left[\frac{1+\cos^2\theta}{2}\right],\\
h_{\rm x}(t,r,\theta,\phi) &=& \frac{4G}{r c^4}\mathcal{M}^{5/3} (2\omega)^{2/3} \sin(2\omega t+\phi)\cos(\theta),
\end{eqnarray}
with the so-called chirp mass $\mathcal{M}=\eta^{3/5}M$, the symmetric mass ratio $\eta=m_1 m_2/M^2$, 
$\theta$ is the angle between the observer's line of sight and the axis normal to the plane of the
binary, and $\phi$ is the (arbitrary) initial azimuthal phase. 

The above expressions highlight
several properties about gravitational emission from compact objects relevant for detection.
First, gravitational wave detectors are directly sensitive
to the amplitude of the metric perturbation, and not the energy it carries. The former
decays as $1/r$, whereas the latter decays as $1/r^2$ (and being proportional to the 
square of the third time derivative of $Q_{ij}$), hence an $n$-fold improvement
in the sensitivity of detectors results in an $n^3$ increase in the observable volume
of the Universe. The ``advanced'' upgrades to the first generation of ground-based interferometric
detectors (that will be completed near the end of the decade) are expected to achieve an order-of-magnitude 
increase in sensitivity over initial LIGO, increasing the range over which binary neutron stars could be 
observed to hundreds of Mpc, and binary black holes to over a Gpc~\citep{LIGORate2010}. Note however that
these distances assume matched filtering is used to search for signals that would otherwise be
buried in detector noise. For this to maximize both detection prospects and parameter extraction 
requires template waveforms that are phase-accurate to within a fraction of a cycle over the most sensitive
band of the detectors (which for adLIGO ranges from $\sim10$Hz to $\sim1$Khz).
Over the past two decades this has been the primary goal of the source modeling community;
it is being achieved using high-order perturbative methods for the early inspiral, numerical solution for 
late inspiral and early merger, and perturbations off a single black hole afterward.

Second, the emission is clearly not isotropic. Only plus-polarized waves are radiated along
the equator, and the amplitude is half of that radiated along the pole orthogonal to the orbit. 
Thus the distance to which a source can be observed strongly depends on its relative orientation
to the detector. Importantly however, this radiation is not strongly beamed and so even non-ideal
orientations of the source to the detector can yield detectable signals.

Third, though these expressions only hint at a couple, there are several
degeneracies in the signal that could limit accurate extraction of all relevant parameters from a detection.
Under radiation reaction the orbit shrinks, and a binary will sweep across a range
of frequencies $\omega$, terminating at merger where $\omega_m\approx c^3/GM$. If $\omega_m$ is not
in band (such as for instance with a binary neutron star merger), at leading order there
is essentially complete degeneracy between the chirp mass and the distance to the source.
If an electromagnetic counterpart could be observed and a redshift determined, 
the degeneracy would break. Higher-order effects, in particular if the black holes
spin or the masses are unequal, excite
higher gravitational wave multipoles that can further lift degeneracies. 
This demonstrates the need to understand the full details of the gravitational wave emission,
and if matter is involved possible electromagnetic counterparts. And as is discussed more throughout
this review, such multi-messenger observations could bring us a wealth of information
beyond just measuring binary parameters.

\subsection{Priors on Binary Parameters}\label{sec_priors}

Merger simulations are computationally expensive, taking of order $10^4-10^5$ CPU hours
for a simulation of the last O(10) orbits of a quasi-circular inspiral of a binary black hole
system. This may not sound too extreme, though remember this is just a single
point in an eight-dimensional parameter space---mass ratio, six components of the two spin
vectors, and eccentricity.
The cost goes up with non-vacuum binaries for several reasons. First, in addition to gravity 
the relevant matter equations (relativistic hydrodynamics at least) need to be solved for.
Second, the effective parameter space grows larger. This is 
in part to characterize unknown physics such as 
the equation of state (EOS) of matter at nuclear densities, and in part because of new initial conditions, for example
a neutron star's magnetic field configuration. Third, computational fluid dynamics algorithms
are typically lower order (to be able to deal with shocks and surfaces) than the high-order
finite difference or pseudo-spectral methods used to sove the Einstein equations,
hence higher resolution is required for similar accuracy to a comparable vacuum merger.

The preceding discussion highlights that compact object mergers  
simulations are too demanding to perform a naive, uniform
sampling of parameter space to guide the construction of gravitational wave template banks.
A promising approach to achieve a more optimal sampling
uses the reduced basis method~\citep{Field:2011mf}, though regardless of the method one can
ask what priors can be placed on the range of parameters from either theoretical or
observational considerations? 

A typical neutron star likely has a mass within the
range of $\approx 1$ to $2.5$ $M_{\odot}$, a radius (which for a given mass is
determined by the EOS) in the range $\approx 8$ to $15$ km, 
and they are thought to have low spins (see e.g. ~\cite{Lattimer:2010uk}). 
For black holes,  an obvious theoretical restriction on the spin magnitude is that $|a|\leq 1$. 
Observations of candidate black holes, assuming general relativity
is correct and black holes satisfy the bound, are beginning to provide estimates of 
spins ranging across all possible  magnitudes $|a|\in [0,1]$~\citep{McClintock:2011zq,McClintock:2013vwa}.
(Allowing for the possibility of naked singularities is
not well posed within the framework of classical general relativity, and without
any theoretical/observational guidance perhaps the best one can do with gravitational waves
is to seek for inconsistencies from the predictions of general relativity using something akin to the 
parameterized post-Einsteinian
(ppE) approach~\citep{Yunes:2009ke}.\footnote{For 
tests of gravity with electromagnetic signals see e.g.~\cite{2013arXiv1311.5564B}.}  )
Theoretical models suggest the relative orientation
of spins are not uniform, either due to properties of the progenitor binary for
stellar mass systems, or interactions with surrounding matter or spin-orbit resonant effects
during inspiral~\citep{Bogdanovic:2007hp,Gerosa:2013laa}. 
Nevertheless, neither theory nor observation provides a sufficiently
compelling case to dismiss the full range of spins allowed by general relativity.
For stellar mass black holes, masses are expected to range from a few to possibly hundreds
of solar masses, supermassive black holes lie at least within the range 
$10^6-10^{10}$ M$_\odot$, and evidence
is mounting for intermediate mass black 
holes between this range (see e.g. ~\cite{Greene:2004gy,Godet:2009hn,Shankar:2009ub,Farrell:2010bf,Davis:2011ka,Kamizasa:2012qs,Casares:2013tpa}).
Consequently,
these ranges are sufficiently broad that the mass ratio $q$ is essentially unconstrained, in particular
for the closer-to comparable mass binaries that would require full numerical solution.

One parameter that has been argued {\em can} be constrained, especially for stellar
mass binaries, is the orbital eccentricity.
The reason for this is that the back-reaction of gravitational wave emission on the orbit tends
to reduce eccentricity. To leading order under radiation reaction, the following
is a decent approximation to the relationship between periapse and eccentricity (see
~\cite{Peters:1963ux,Peters:1964zz} for the derivation and full expression)
\begin{equation}
r_p\approx r_{p0}\frac{1+e_0}{1+e}\left(\frac{e}{e_0}\right)^{12/19},
\end{equation}
where $r_{p0},e_0$ are the initial periapse and eccentricity, respectively. For the moderate
initial eccentricities expected when the progenitor of the black hole binary is a 
stellar binary, $e\sim (r_p/r_{p0})^{19/12}$. Such a binary enters the adLIGO
band at $r_p\sim 10^2$km, whereas expected values for $r_{p0}$ are several orders of magnitude
larger (see, e.g.~\cite{Kalogera:2006uj}), hence $e$ will be completely negligible
here. This has focused the majority of work on mergers on the quasi-circular $e=0$
case. However, there are other mechanisms to form binaries, and some could lead
to systems that have high-eccentricity while emitting in the LIGO band. These mechanisms include dynamical 
capture from gravitational wave emission during a close two-body encounter in a dense cluster~\citep{O'Leary:2008xt,lee2010short}, 
a merger induced during a binary-single star interaction in a similar environment~\citep{Samsing:2013kua}, and 
Kozai-resonant enhancement of eccentricity in a hierarchical triple 
system~\citep{Wen:2002km,Kushnir:2013hpa,Seto:2013wwa,Antognini:2013lpa,Antonini:2013tea}. 
Event rates are highly uncertain for both classes of binaries (see~\cite{LIGORate2010} for a review of 
quasi-circular inspiral systems, and~\cite{O'Leary:2008xt,lee2010short,Kocsis_Levin,Tsang:2013mca,East:2012xq} 
for discussions of dynamical capture systems), and though quasi-circular inspirals are likely
dominant, eccentric mergers may not be completely irrelevant as often assumed in the 
gravitational wave community. The formation mechanism for supermassive black hole binaries 
is different (being driven by mergers of the host galaxies of individual black holes), though similarly
there are arguments that in some cases non-negligible eccentricity might remain until merger~\citep{Roedig:2011rn}.

The difficulty with eccentricity is that it is not
``merely'' an additional parameter, but changes the qualitative properties of a merger
in a manner that challenges both source modeling and data analysis strategies. With regard
to modeling, the orbital period increases significantly with $e$ for a given $r_p$, making
numerical simulations of multi-orbit mergers very expensive. Perturbative methods
have not yet been developed to high order for large eccentricity orbits (though see  
~\cite{Bini:2012ji}). Taken together it may be unreasonable to expect templates
accurate enough for data analysis using matched filtering any time soon, and different
(though sub-optimal) strategies may need to be developed, for example power stacking~\citep{East:2012xq,Tai:2014bfa}.
This implies that for practical purposes there are two ``classes'' of binaries, quasi-circular
inspirals, and large eccentricity, small initial pericenter mergers.

Having discussed broad considerations relevant to the three classes of binaries---black hole-black hole,
black hole-neutron star, neutron star-neutron star-- we now discuss
salient features of each class uncovered through numerical simulations.

\section{Binary Black Holes}
Due to the ``no-hair'' property of event horizons in four-dimensional Einstein gravity,
isolated single black holes in our Universe are expected to be described almost exactly by the 
Kerr family of solutions.
This is a two-parameter family, labeled by the total gravitational
mass $M$ and angular momentum $J$.~\footnote{An isolated black hole can also have a conserved
charge, though in astrophysical settings black holes should be neutral to excellent
approximation. ``Exotic'' matter fields could also support additional ``hair'', though we
do not consider such fields here.}
The latter is more conveniently described by a dimensionless spin parameter $a=J/M^2$. As mentioned, an event
horizon is only present if $|a|\leq 1$, otherwise the solution exhibits a naked singularity. If such 
a situation could arise (violating the so-called cosmic censorship conjecture) classical general 
relativity would not be able to describe the exterior solution nor the dynamics of the object 
in our Universe. This would offer a prime opportunity to study quantum gravity, though 
unfortunately to date no theoretical studies of plausible astrophysical processes involving 
dynamical, strong-field gravity, including gravitational collapse 
and compact object mergers, have resulted in a naked singularity.\footnote{Though 
see~\cite{Jacobson:2009kt} for an intriguing suggestion that near extremal black holes could
be ``over spun'', and~\cite{Shapiro:1991zza}, who suggest that collapse of matter with negligible self-pressure
and in a highly prolate configuration could lead to naked singularities.
It is also well known that naked singularities can arise in spherical
collapse of ideal fluids (see e.g.~\cite{Joshi:2013xoa} ), or
critical collapse in a larger class of matter models (see e.g.~\cite{Gundlach:2002sx}).
However, these examples are either non-generic (whether by imposed symmetries
or fine tuning of initial data) or arise in matter that is of arguable relevance
to collapse in astrophysical settings~\citep{1997gr.qc....10068W}.}

Thus, technical details aside, the study of vacuum binary black hole mergers in general
relativity is a well-defined problem characterized by a relatively small set of
parameters : the mass ratio $q$ of the binary, the two initial spin vectors 
$\vec{s}_1,\vec{s}_2$ of each black hole, the initial eccentricity $e_0$
and the size of the orbit (parameterized for example by the initial pericenter distance
$r_{p0}$). There is no intrinsic scale in vacuum Einstein gravity, hence there is
a trivial map from any solution with a given set of these parameters to a desired total
mass $M$ of the binary.
In the remainder of this section we present results from the numerical solution of the Einstein field 
equations for vacuum mergers, discuss some astrophysical consequences, and briefly
comment on issues related to testing general relativity from gravitational wave
observations of vacuum mergers. 
For other review articles discussing similar topics see~\cite{Pretorius:2007nq,2010ARNPS..60...75C,Hannam:2013pra}.

\begin{figure}
\begin{tabular}{ b{2.3in} b{3.0in}}
\includegraphics[width=1.1in,clip]{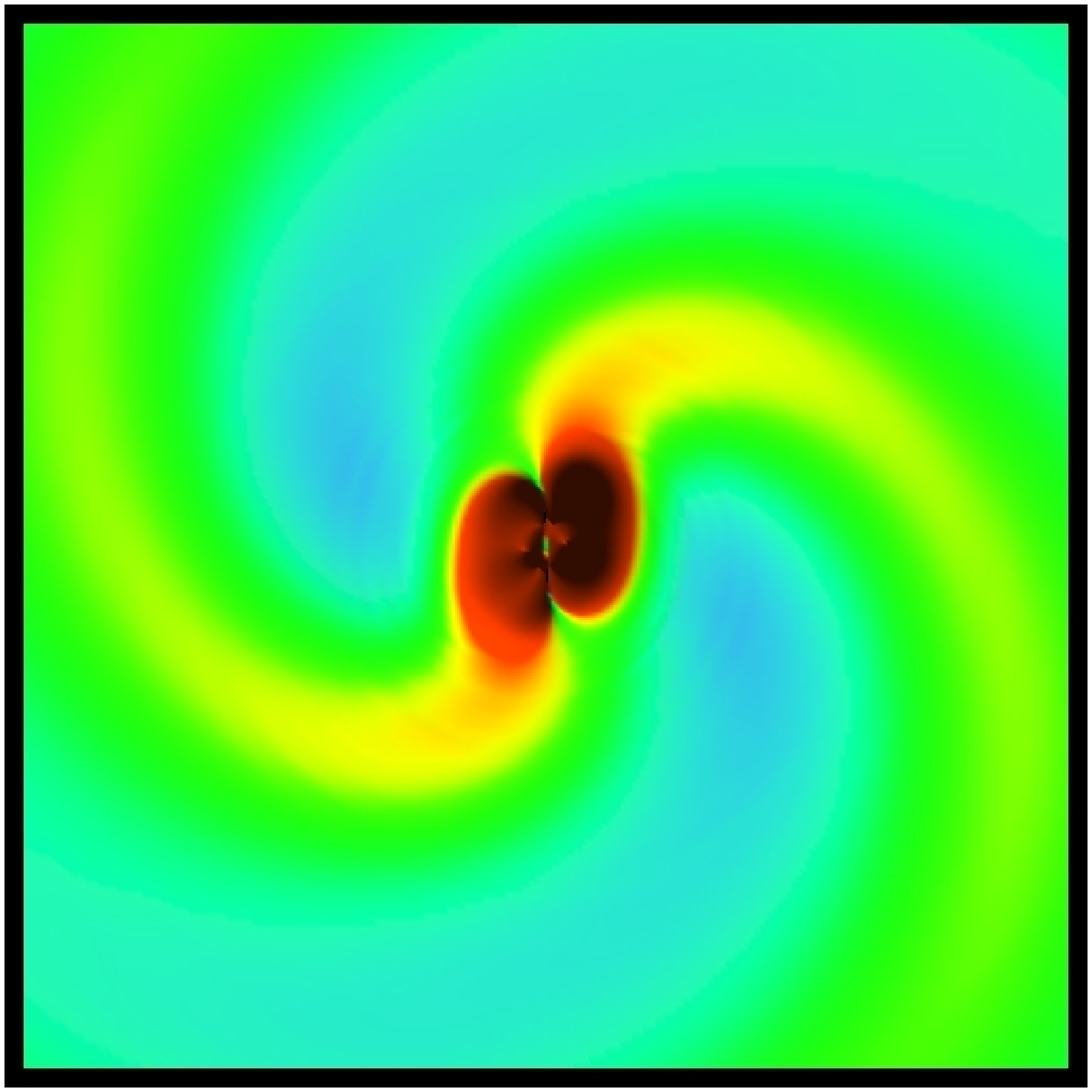} \hspace{-0.225cm}
\includegraphics[width=1.1in,clip]{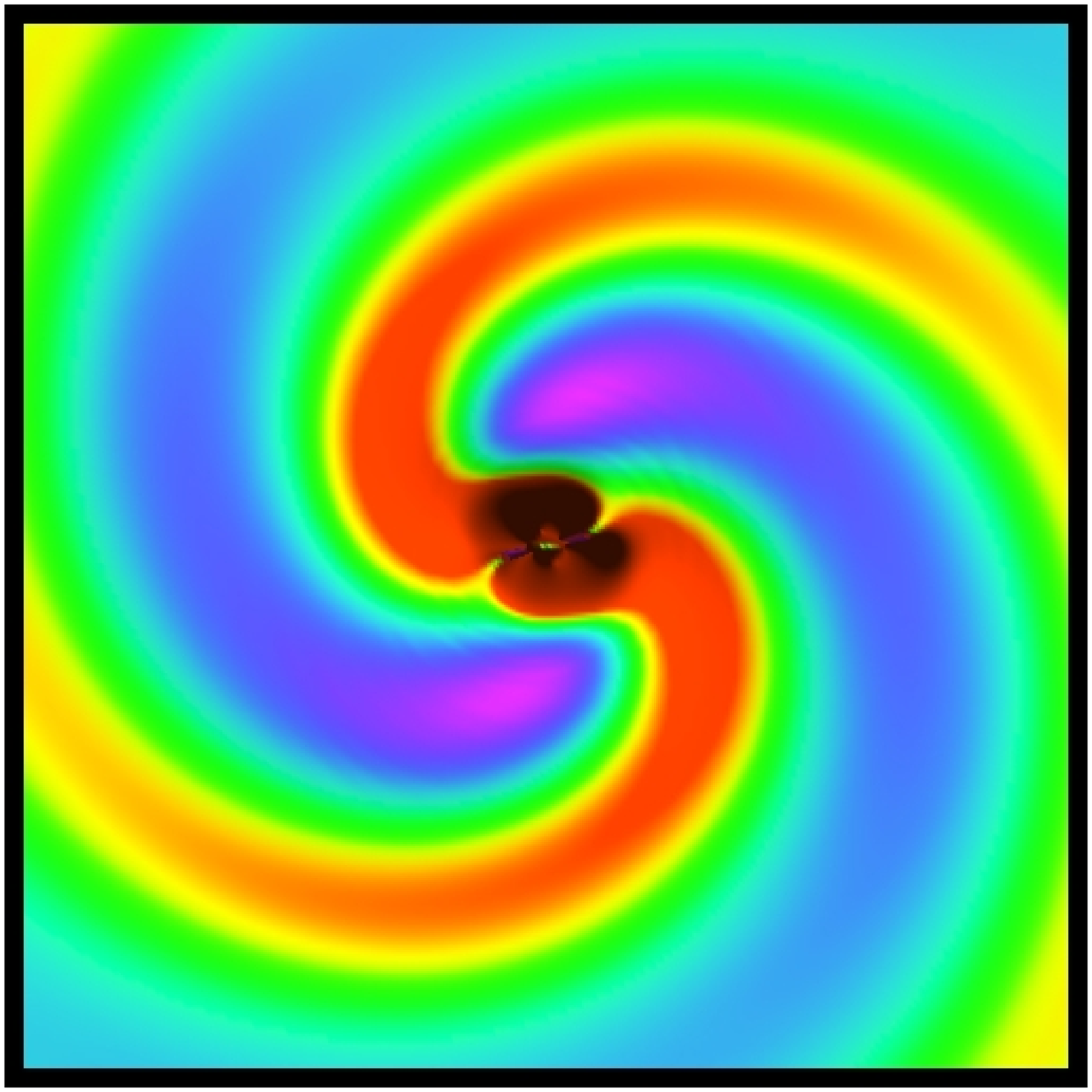} \newline \vspace{0.1cm} \hspace{-.26cm}
\includegraphics[width=1.1in,clip]{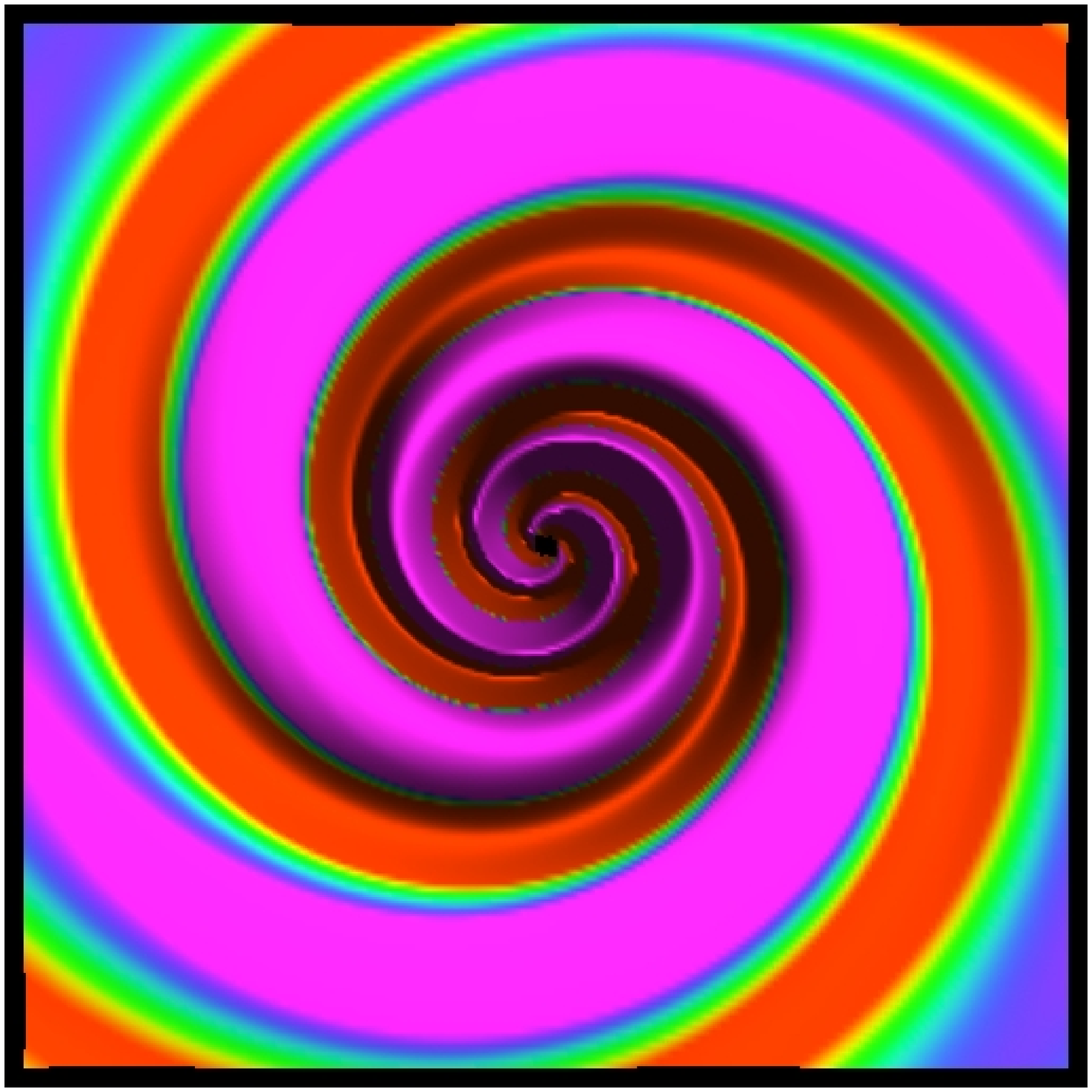} \hspace{-0.22cm} 
\includegraphics[width=1.1in,clip]{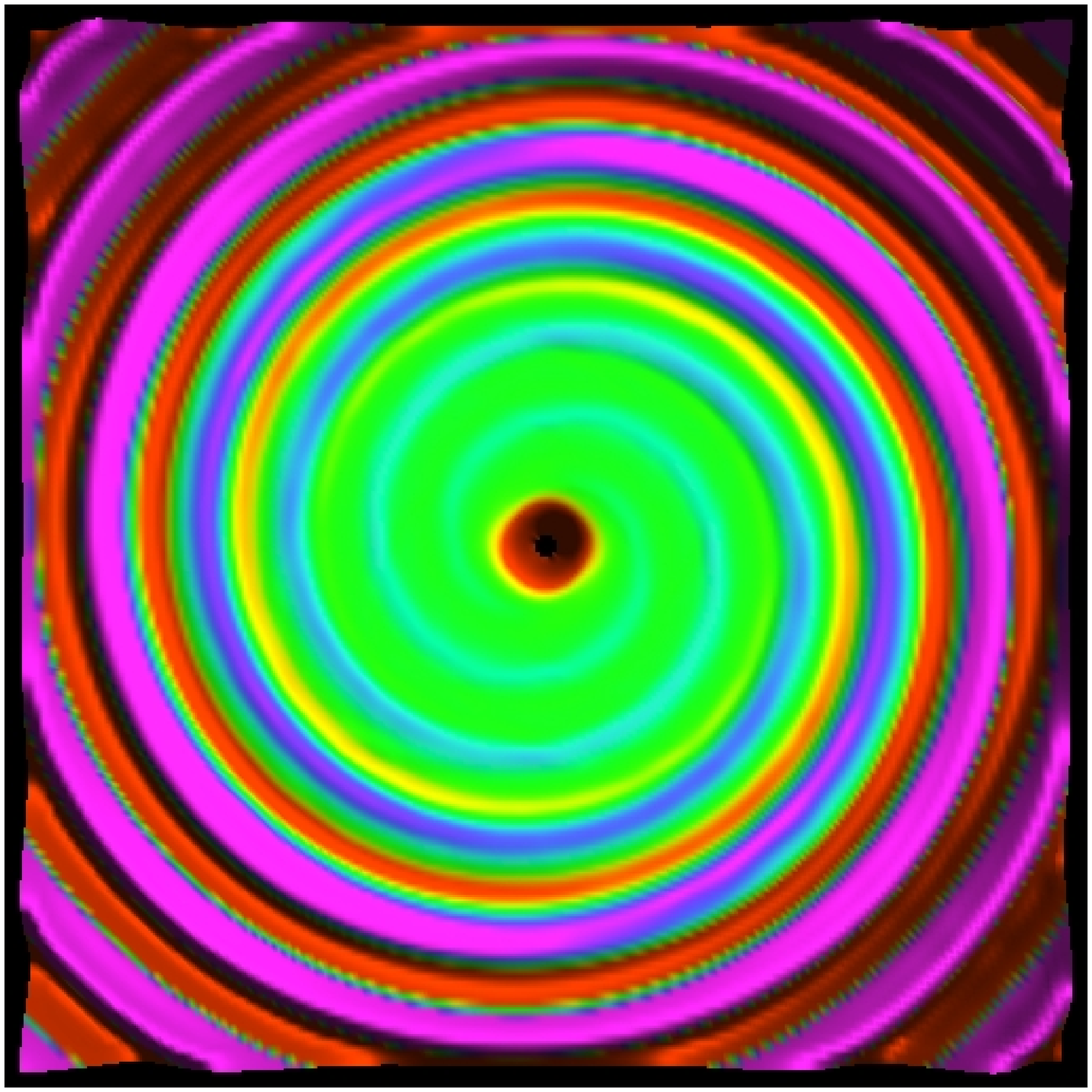} & 
\includegraphics[width=2.70in,clip]{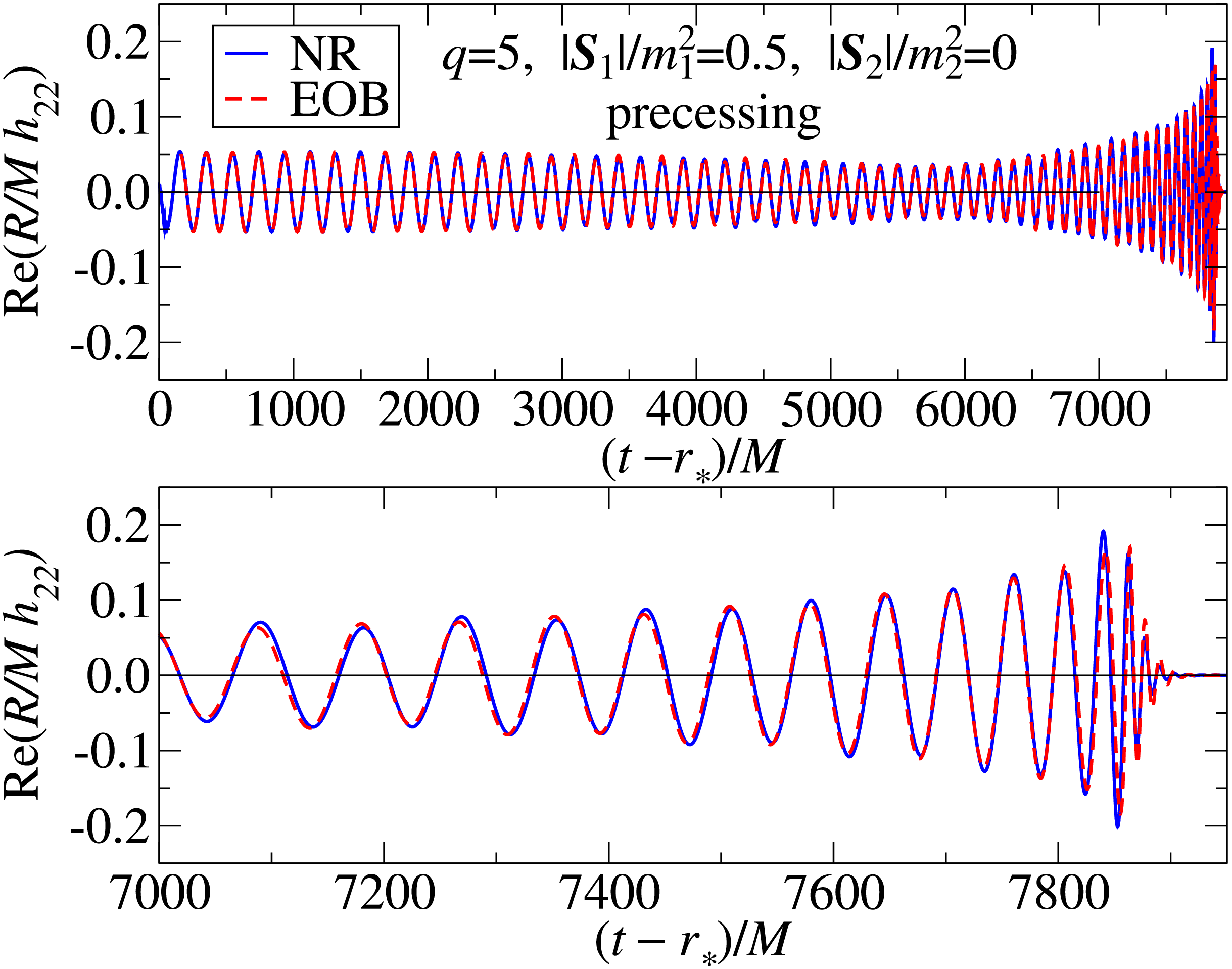} 
\end{tabular}
\caption{
(left)  A depiction of the gravitational waves emitted during
the merger of two equal-mass (approximately) nonspinning black holes, 
from~\cite{Buonanno:2006ui}. Shown is a color-map
of the real component of the Newman-Penrose scalar $\Psi_4$ multiplied
by $r$ along a slice through the orbital plane, which far from
the black holes is proportional to the second time derivative of the
plus polarization (green is $0$, toward violet (red) positive (negative)).
The top left and right panels show the dominantly quadrupolar inspiral
waves $\sim 150M$ and $\sim 75M$ before merger respectively. The
bottom left panel is near the peak of the wave emission at merger, and bottom 
right shows the ringdown waves propagating out $\sim 75M$ after merger.
The size of each box is around $100M^2$.
(right) Gravitational wave emission measured from a numerical relativity simulation of a binary
black hole merger (NR) overlaid with an NR-calibrated effective one body calculation (EOB),
from ~\cite{Taracchini:2013rva}.
The binary has a mass ratio $q=5$, the more massive black hole has
a dimensionless spin of $a=0.5$ with direction of the spin axis initially in the
plane of the orbit, and the second less-massive black hole is non-rotating. That the
amplitude of the wave is not monotonically increasing during inspiral is
a manifestation of the modulation induced by precession of the orbital plane.}
\label{fig:bhbh1}
\end{figure}

\subsection{Results and Applications of Merger Simulations}

A couple of important qualitative questions about the merger process have largely been answered.
The first relates to cosmic censorship : a broad swath of parameter space
has been explored (see for example~\cite{Hinder:2013oqa}), and no naked singularities have been found.
Furthermore, to the level of scrutiny the solutions have been subjected, the late time
behavior is consistent with a spacetime approaching a Kerr solution via quasi-normal
mode decay\footnote{In theory the quasi-normal ringdown should transition to a power-law decay 
at very late times. This has been verified for perturbed single black holes. Binary
simulations have not yet been carried that far beyond merger, though the motivation
for doing so is minimal as the amplitude of these power-law tails is
too small to be observable.}. The second relates to the existence of new ``phases''
of the merger outside the purview of the perturbative treatments governing
the inspiral and ringdown. One line of reasoning
argues that owing to the nonlinearity
of the field equations, and the fact that the late stages of a merger occur in the most
dynamical, strong-field regime of the theory, these new phases would be natural places to expect
novel physics. The opposing argument, which turned out to better describe the simulation results, 
is that the merger is effectively a highly dissipative process that occurs deep within the gravitational
potential well of the combined objects, and very little of the spacetime dynamics that occurs
there will leave an imprint on the waves radiated outwards. Or stated another way,
perturbative methods have been extended to quite high order in $v/c$ in both the conservative
and dissipative dynamics of a binary, and black hole
perturbation theory begins with an exact strong-field solution; these together capture
the ``essential'' nonlinearities of the problem. 
As a consequence of this rather smooth behavior a convenient
approach to constructing templates is the Effective One Body (EOB) method~\citep{Buonanno:1998gg},
where re-summed PN inspiral waveforms are smoothly attached to quasi-normal
ringdown modes via a transition function calibrated by numerical simulations
(for some recent papers
see \cite{2011PhRvL.106x1101A,2013arXiv1307.6232P,2013PhRvD..87h4035D}).
Most of the work in this regard has been conducted on non-precessing orbits (i.e. any net spin
angular momentum is aligned with the orbital angular momentum) or lower-spin black
holes, and it remains to be
seen how well this technique may work for highly spinning black holes in precessing orbits.
See Figure~\ref{fig:bhbh1} for two examples of gravitational wave emission from 
merger simulations, and for one of them a match to EOB calculations.

The science gleaned from numerical simulations of vacuum binary mergers has therefore 
mostly been in details of the process. Important numbers of relevance to astrophysics
include the total energy and angular momentum radiated during merger (and consequently
the final mass and spin of the remnant black hole), and the recoil, or ``kick'' velocity of the final black 
hole to balance net linear momentum radiated. We can not possibly list
these numbers for all cases simulated to date (but we do give some citations to 
relevant literature for further information).
Though to give a sense of the physics and order of magnitude of the numbers we highlight a few
key results and present some of the simpler fitting formulas.\\

\subsubsection{Energy radiated.}
One can think of the energy radiated during a merger coming from two sources : the gravitational
binding energy liberated during inspiral, and energy in the geometry of the 
merger remnant formed during the collision that is emitted 
as the horizon settles down to its stationary Kerr state (on timescales comparable
with the final orbital period).
In the extreme-mass-ratio limit the former dominates, and the total radiated energy
equals the magnitude of the binding energy at the ISCO; for a quasi-circular
inspiral this ranges
from $\sim 3.8-42\%$ of the rest mass of the small black hole depending on the spin
of the large black hole; the lower (upper) limit is a retrograde (prograde)
equatorial orbit about an extremal Kerr black hole (the Schwarzschild case gives $5.7\%$).
As the mass ratio decreases (i.e. the masses become comparable) 
the emitted energy increases, and the amount
coming from the ringdown grows to a comparable fraction approaching the equal mass limit.
Here numerical simulations are required to compute the exact numbers, and
it is more useful to quote
the value as a percentage of the total gravitational (ADM) energy $M$ (the gravitational mass
of the system as measured by an asymptotic observer).
A useful formula interpolating between the analytic extreme-mass-ratio limit (top line below)
and empirical fits to numerical data (bottom line) was derived by~\cite{Barausse:2012qz}
(see ~\cite{2008PhRvD..78h1501T,2010CQGra..27k4006L} for a couple of
alternative formulae):
\begin{eqnarray}
\frac{E_{rad}}{M} \approx & & \eta(1-4\eta) \left[1-\tilde{E}_{ISCO}(\tilde{a})\right] \nonumber \\
&+& 16 \eta^2 \left[p_0 + 4 p_1 \tilde{a}(\tilde{a}+1)\right].
\end{eqnarray}
Here $\tilde{a}=\vec{L}\cdot(\vec{S_1}+\vec{S_2})/M^2$ 
is the projection of the sum of the black hole spin vectors onto the orbital angular momentum prior
to merger, $p_0\approx 0.048, p_1\approx 0.017$,
and $\tilde{E}_{ISCO}(\tilde{a})$ is the energy of the effective
ISCO of the system.
This formula fits existing numerical
simulation results to within better than a percent in most cases
(see~\cite{Barausse:2012qz} for comparisons and more details).

\subsubsection{Final spin.}
There are numerous formulas characterizing the final spin of the merger remnant that have 
been constructed via fits to numerical relativity results
(for e.g. \cite{2008PhRvD..78h1501T,Barausse:2009uz}; see also \cite{2010CQGra..27k4006L} 
for PN-inspired functions, and \cite{2008PhRvL.100o1101B}
for a prescription based on a so-called ``spin expansion'' that uses symmetry arguments to economize the
formulas). Here we give a simple `first-principles' derived expression
from~\cite{2008PhRvD..77b6004B} that captures the 
basic physics, and agrees reasonably well with numerical results.
The following formula is valid for spins
aligned with the orbital angular momentum (see ~\cite{Kesden:2009ds}, and others cited above
for generalizations to precessing binaries):
\begin{equation}\label{fin_a}
a_f M \approx L_{orb}(r_{ISCO},a_f) + m_1 a_1 + m_2 a_2.
\end{equation}
Here $a_f M$ is the spin angular momentum of the remnant (with $M$ approximated by $m_1+m_2$), 
$m_1 a_1$ and $m_2 a_2$ are the
spin angular momenta of the initial black holes, and $L_{orb}(r_{ISCO},a_f)$ is
the orbital angular momentum of a reduced-mass particle equivalent of the
system evaluated at the ISCO of a Kerr black hole using the parameters of the remnant.
The interpretation of this is straight-forward : the system radiates angular momentum
until the plunge to merger, 
after which the majority of the remaining spin plus orbital angular momentum is subsumed
by the final black hole. Some angular momentum is radiated during 
ringdown, but this is taken into account in the above formula
through the use of the effective ISCO of the remnant black hole.
For interest, a quasi-normal mode with frequency $\omega_m$ and azimuthal
multiple number $m$ has the following relationship between the energy
and angular momentum it carries : $J_{rad} \approx (m/\omega_m) E_{rad}$. $m=2$ is the dominant
mode, and for example a Schwarzschild black hole has $\omega_2\approx 0.38/M$~\citep{Berti:2009kk}.
Though it is not possible to clearly differentiate the quasi-normal part of the wave
from the emission that precedes it, a rough estimate is of order
$1-2\%$ of the net energy is emitted in the ringdown for comparable mass mergers.

The formula (\ref{fin_a}) predicts the final
spin to within a few percent in many cases. For example, it gives $a_f/M \approx 0.663$ 
for the merger of equal-mass, non-spinning black holes; comparing
to numerical relativity simulations, an initial estimate was $a_f/M \approx 0.70$~\citep{Pretorius:2005gq},
with the latest high-accuracy simulations refining it to $a_f/M \approx 0.6865$~\citep{Scheel:2008rj}.

\subsubsection{Recoil.}

\begin{figure}
\includegraphics[width=2.3in,clip]{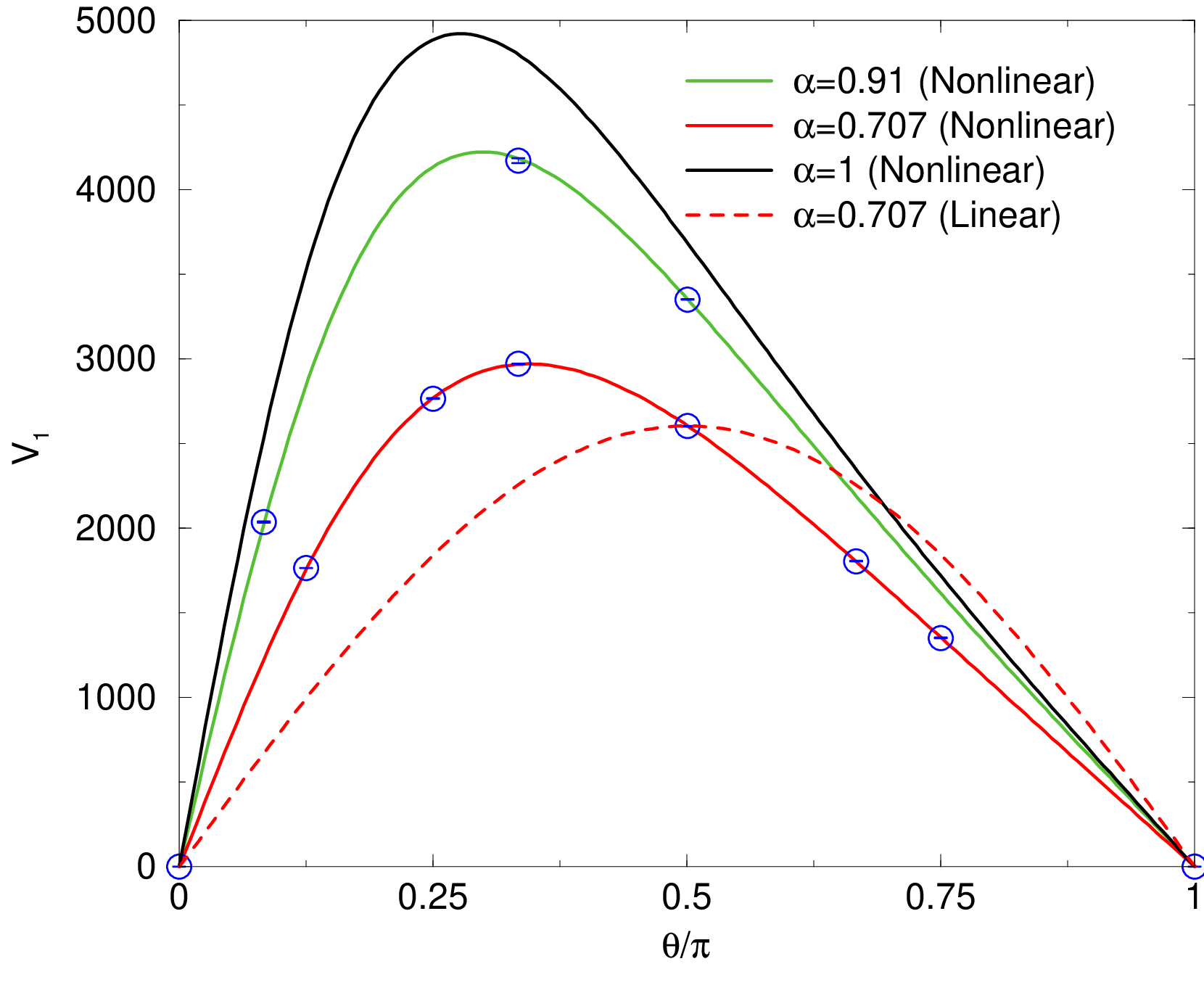} 
\includegraphics[width=3.0in,clip]{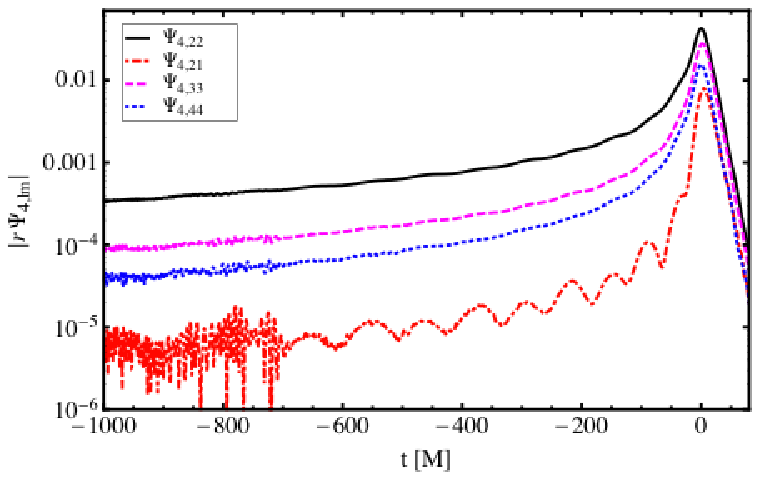} 

\caption{(left) Recoil velocities from equal mass, spinning binary black hole merger
simulations (circles) together with analytical fitting functions.
Each black hole has the same spin magnitude $\alpha$, equal but opposite
components of the spin vector within the orbital plane, and $\theta$ is
the initial angle between each spin vector and the orbital angular momentum. 
The dashed line corresponds
to a fitting formula that depends linearly on the spins, whereas 
solid lines add nonlinear spin contributions.
(from~\cite{Lousto:2011kp}). 
(right) Strength of different modes in the gravitational waves from a binary black
hole merger with mass ratio $M_1/M_2=3$, and spins $a_1=0.75,a_2=0$.
This system exhibits a marked precession that complicates the multipolar
mode structure seen in a fixed observer's frame. However, a transformation to
a co-precessing ``quadrupole aligned'' frame, as shown in this figure, illustrates that the main radiation channel
is still the $l=2,m=2$ mode (from~\cite{Schmidt:2010it}).}
\label{fig:bhbh2}
\end{figure}

A recoil in the remnant, namely a velocity post-merger relative to the initial 
binary center of mass, can arise when there is asymmetric beaming of radiation during the merger. 
Asymmetry comes from unequal masses and black hole spins. The formulas
describing the recoil can be rather involved (see for example~\cite{2011PhRvD..83b4003L}),
so here we just briefly mention some of the salient features and numbers. 
Non-spinning binaries with mass ratio $q$ different from unity give rise to a recoil in the plane 
of the binary, reaching a maximum of $\sim175{\rm km/s}$~\citep{Gonzalez:2006md,Baker:2006vn,2007CQGra..24S..33H,Berti:2007fi}
for $q\approx 1/5$. Spin introduces additional asymmetry in the radiation by causing the orbital plane
to precess and ``bob'', which can induce a recoil both in and orthogonal to the
plane of the binary. The magnitude of the out-of-plane recoil is sinusoidally modulated by
the effectively random initial phase of the binary.
Spin can also allow the onset of a plunge to occur at higher frequency,
and hence give higher gravitational wave luminosity, which further amplifies the recoil.
The bobbing motion (see~\cite{Pretorius:2007nq} for an intuitive description of it)
is associated with the largest recoils, which 
remarkably can reach several thousand km/s for appropriately aligned 
high-magnitude spins~\citep{2007ApJ...659L...5C,2007PhRvL..98w1101G,Lousto:2011kp};
see Figure~\ref{fig:bhbh2} (left panel) for examples.

These largest velocities are well in excess of the escape velocities of even
the most massive galaxies. That observational evidence suggests most galaxies harbour
central supermassive black holes, together with hierarchical structure formation
models of the growth of these galaxies, implies that mergers with very large recoils are rare.
If mergers themselves are common, and black holes can have sizable spins
as implied by current observations~\citep{Reynolds:2013rva}, then the typical recoil 
must be significantly less than the maximum theoretically possible.
One possible explanation for this would be if most mergers take place in gas-rich environments,
as then torques induced by circumbinary material will tend to align the spins
of the black holes with the overall orbital angular momentum, a configuration
that has significantly lower maximum recoil (see e.g.~\cite{Bogdanovic:2007hp,Dotti:2009vz}).
Another, that operates even in vacuum, comes from PN calculations that include
spin-orbit coupling, which shows a tendency for the black hole spins to align (anti-align) with each other 
if the spin of the more massive black hole is initially at least partially aligned (anti-aligned) 
with the orbital angular momentum~\citep{Kesden:2010ji}.

\subsubsection{Tests of General Relativity.}

A further opportunity offered by gravitational wave observations of merging binaries
is to test dynamical, strong-field gravity. With obvious caveats associated with our present
lack of understanding of dark matter and dark energy, general relativity has so
far been shown to provide a consistent description of gravity in all observations
and experiments that are constrained by its predictions (see e.g.~\cite{Will:2006LR}). Lacking here
are tests in the most nonlinear regime of the theory, in particular where black holes
can form. Certainly there is no doubt about the existence of massive, dark, ultra compact objects,
and observations of (for example) X-ray emission from stellar mass candidates or properties
of AGN are consistent with these phenomena being powered by Kerr black holes. However, that
horizon scales cannot be quite resolved at present~\footnote{This may change within a few years
through VLBA observations of our galaxy's central black hole, SGA*~\citep{Broderick:2009ph,Broderick:2011mk} as well
as nearby M87~\citep{Doeleman:2012zc}.}
together with complexities of the matter physics responsible for the emission 
prevents precise determination of local properties of the spacetime.
Binary black hole mergers, in particular stellar mass systems that are expected
to occur largely in vacuum, offer a unique opportunity to study pure, strong-field
gravity. General relativity's ability to predict the entire waveform, which 
is uniquely determined by a small set of numbers ($M,q,e,\vec{s_1},\vec{s_2}$
and detector-source orientation parameters)
and can consist of hundreds or even thousands of cycles in the LIGO band,
can in principle allow for stringent self-consistency tests on high signal-to-noise-ratio (SNR)
events. 

However there are several issues that complicate this promise
to test general relativity in the near future. First,
given the lack of events from the initial LIGO observing runs, it is unlikely
that adLIGO will observe a very loud event. Hence viable tests may
require statistical analysis of a number of low-SNR events, and
little work has yet been done to suggest how this might be carried out (see~\cite{Agathos:2013upa}
for a recent proposal; related work
on constraining the nuclear EOS using multiple mergers events involving
neutron stars is also beginning to be investigated---see the discussion in Sec. 5).
Second, detection and parameter estimation relies on matched filtering with templates.
If the only templates used are those constructed using general relativity, then
all information about possible deviations will be projected out\footnote{
Note also that it is unlikely that general relativity templates will
completely ``miss'' all events even if there are strong-field deviations. This is because
binary pulsar observations confirm the leading order
radiative dynamics of general relativity.}.
If the event has a high SNR, there should be a detectable residual excess power, but again
for the typical SNRs expected for adLIGO this is unlikely\footnote{This is not the case
for possible space-based detectors, like eLISA, as their exquisite SNR could
allow for detecting supermassive binary black holes mergers with masses
in the range $10^4 M_{\odot}<M<10^7 M_{\odot}$ out to redshifts of $z\simeq 20$ with a SNR $\ge 10$.
For a recent review see~\citep{Seoane:2013qna}.}. Compounding the problem,
despite the large number of proposed alternatives or modifications to general
relativity (see for example~\cite{1993tegp.book.....W,Will:2006LR}),
almost none have yet been presented that (i) are consistent with general relativity
in the regimes where it is well tested, (ii) predict observable
deviations in the dynamical strong-field relevant to vacuum mergers, and (iii)
possess a classically well-posed initial value problem to be
amenable to numerical solution in the strong-field. The notable exceptions
are a subset of scalar tensor theories, though these require a time-varying cosmological
scalar field for binary black hole systems~\citep{Horbatsch:2011ye}, or one or more
neutron stars in the merger (see Sec.~\ref{sec_nonvac}).
Thus there is little guidance on what reasonable strong-field deviations
one might expect. 
Proposed solutions to (at least partially) circumvent these problems
include the ppE and related frameworks~\citep{Yunes:2009ke,Agathos:2013upa}, 
modified PN waveforms~\citep{Arun:2006yw}, as well as exploiting properties of the uniqueness of Kerr
and its quasi-normal mode structure~\citep{Collins:2004ex,Berti:2005ys}.

\subsubsection{Eccentric binaries.}\label{sec_ecc_bhbh}
As mentioned above, the majority of the work on binary black hole mergers from the relativity community has
focused on quasi-circular inspiral, except for a handful of recent studies~~\citep{Pretorius:2007jn,2009PhRvL.103m1101H,Gold:2012tk}.
One of the interesting results is that so called ``zoom-whirl'' orbital
dynamics is possible for comparable-mass binaries. In the test particle limit, zoom-whirl orbits are perturbations
of the class of unstable circular geodesics that exist within the ISCO. They exhibit
extreme sensitivity to initial conditions, in which sufficiently fine-tuned data can give an arbitrary number
of near-circular ``whirls'' at periapse for a fixed eccentricity.
Away from the test particle limit gravitational wave emission
adds dissipation to the system; however, what the simulations show is that even in the comparable mass limit the
dissipation is not strong enough to eradicate zoom-whirl dynamics, but merely limits how long it can persist.
Perhaps the most interesting consequences of high-eccentricity mergers could arise when neutron
stars are involved; this is discussed in Sec.~\ref{sec_nonvac}. 

For the vacuum problem, aside from 
providing information on binary formation channels, high-eccentricity events could in principle offer
the most stringent tests of strong-field gravity. The reason is due to the nature of these
orbits compared with quasi-circular inspiral: significantly higher velocities are reached at periapses passage,
a larger fraction of the total power is radiated in this high $v/c$ regime,
and the long time between periapse bursts imply that small deviations in emission could
result in large dephasing of the waveform from one burst to the next.
However, to date no studies have addressed in any quantitative manner how well general relativity
can be constrained using eccentric mergers.

\subsection{Further Physics}\label{sec_further_physics}
With the vacuum merger problem essentially under control, 
the field is now more closely examining the impact a merging black hole binary can have on its astrophysical environment.
The most pertinent scenario is a supermassive binary merger, and questions relate to how the rapidly changing gravitational field,
ensuing gravitational waves, and possible recoil could perturb surrounding gas, plasma, electromagnetic fields, stars, etc.
Here we briefly discuss some of the more interesting and potentially observable phenomena revealed by recent studies.

$\bullet$ {\em Prompt counterparts to supermassive black hole binaries mergers within circumbinary disks}. 
First studies of the interaction of binary black holes with surrounding electromagnetic fields and plasma
were presented by~\cite{Palenzuela:2010nf,Palenzuela:2010xn}. Though not modeled there, the expected
source of these fields and plasma would be a circumbinary disk. More recent work has begun 
to self-consistently model the disk as well~\citep{2012PhRvL.109v1102F,2013arXiv1312.0600G}.
Recall that in the case of a single
black hole, the Blandford-Znajek (BZ) mechanism~\citep{Blandford:1977ds} indicates the 
plasma (coming from an accretion disk) is able to
tap rotational energy from the black hole and power an energetic Poynting flux. Tantalizing observational
evidence linking the strength of radio signals and black hole spin has been presented by~\cite{McClintock:2013vwa}.
In the context of binary black holes, simulations demonstrated that  the spacetime helps stir electromagnetic field
lines and that, akin to the BZ mechanism, the plasma is able to 
tap translational and rotational energy from the system to produce dual jets~\citep{Palenzuela:2010nf,Palenzuela:2010xn}.
These jets would act as spacetime tracers, and their behavior can be modeled reasonably
well by an extension of the BZ formula. That is, prior to merger, the luminosity from the system obeys 
$L \simeq B^2 \sum_{i=1,2}[\Omega_H(i)^2 + \kappa v(i)^2]$ where $\Omega_H(i), v(i)$ and $\kappa$ are the angular rotational velocities 
of the horizons, the black hole velocities and a relative strength parameter respectively\footnote{The dependence on $\Omega_H$ acquires
higher-order corrections close to maximally spinning cases~\cite{2010ApJ...711...50T}.}. 
(The value of $\kappa \simeq 100$, and indicates that black holes
must be moving at $\gtrsim 0.1c$ for a non-trivial contribution unless, of course, they are non-spinning).
Notice that unlike the orignal BZ effect, even if the black holes are non-spinning there could
be a sizable luminosity due to the contribution from the translational kinetic energies 
of the black holes (which can reach $\approx 0.2-0.3 c$ near merger for quasi-circular inspiral).
After merger, a single jet arises with luminosity
$L \simeq B^2 (\Omega_{H_{final}} + \kappa v_{recoil}^2)$ (though the second term is subleading 
unless the final black hole has negligible spin, as $v_{recoil}$ is at most $\approx 0.015 c$). 
This behavior implies interesting possibilities for detection
of gravitational and electromagnetic waves associated with a merger embedded in 
a circumbinary disk (see e.g.~\cite{O'Shaughnessy:2011zz,Moesta:2011bn}).

$\bullet$ {\em Post-merger consequences of binary supermassive black hole mergers}.
The merger event can have several interesting consequences due to the large amounts of energy
radiated and (for appropriate spins and mass ratios) the recoil of the final black hole; we briefly mention a few 
here---for recent reviews of this and other astrophysical consequences 
see~\cite{2012AdAst2012E..14K,2013arXiv1307.3542S}.
With respect to timescales in the disk these effects occur essentially instantaneously.
This near-impulsive perturbation of the gravitational potential
in the outer parts of the accretion disk could lead to the formation of strong shocks;
this, together with subsequent inward migration of the disk, could
producing observable electromagnetic emission on timescales of a month to a year 
afterward~\citep{Milosavljevic:2004cg,Lippai:2008fx}.
The most favored orientations for recoils can produce velocities large enough
to significantly displace the remnant from the galactic core, or 
even eject the black hole from the host galaxy altogether (though as discussed above
this is likely quite rare).
If the system has a circumbinary accretion disk, the recoil would carry the inner part of the
disk with it, and this could be observable in Doppler-displaced emission lines relative to the galactic
rest frame~\citep{Komossa:2008qd}. 
Earlier studies have suggested that prior to merger the accretion rate, and hence the 
luminosity of the nucleus, would
be low as the relatively slow migration of the inner edge of the accretion disk decouples
from the rapidly shrinking orbit of the binary. Post merger then, AGN-like emission could be re-ignited once
the inner edge of the disk reaches the new ISCO of the remnant black hole. This should be displaced from the
galactic center if a large recoil occurred, and could be observable in nearby 
galaxies (see for example~\citep{2007PhRvL..99d1103L}).
However, more recent simulations of circumbinary disks
 using ideal magnetohydrodynamics for the matter
shows that complete decoupling does not occur, and relatively high accretion rates can be maintained
all the way to merger~\citep{2012ApJ...755...51N,2012PhRvL.109v1102F,Bode:2011tq} and
afterwards (e.g~\cite{Shapiro:2009uy}). 
The binary orbit in this case causes a modulation in the luminosity of the system, 
which may be observable. 
A last effect we mention is that a displaced central black hole should also have its loss-cone refilled, increasing the frequency
of close encounters with stars and their subsequent tidal disruption by the black hole, with rates
as high as $0.1/{\rm yr}$; the disruption could produce observable electromagnetic emission~\citep{2011MNRAS.412...75S}.

$\bullet$ {\em Binary black hole mergers and galaxy formation}. 
Ggalaxy formation models have also
been exploited to understanding of the outcome of binary black hole mergers.
There is strong stellar,
gas-dynamical~\citep{1998AJ....115.2285M,1995ARA&A..33..581K,Kormendy:2013dxa} 
and electromagnetic~\citep{2004ApJ...613..682P,2000ApJ...543L...5G}
evidence for the existence of massive black holes at
the centers of galaxies. These central black holes play a fundamental role in
our current paradigm of galaxy formation and evolution; for example, they are required to explain
quasar and AGN emission~\citep{soltan82}, as well as cosmic
downsizing~\citep{AGN_downsizing1,AGN_downsizing2,AGN_downsizing3}.
In the $\Lambda$CDM model galaxies merge into increasingly larger
ones as cosmic time proceeds, and consequently their
massive black holes are expected to merge, initially via processes
such as dynamical friction, with gravitational wave emission only
taking over in the very late stages.
Results from NR simulations have been utilized to follow the evolution
of these black holes through coalescence. More specifically, a number of works studied the
mass and spin evolution of supermassive black holes through cosmic 
time~\citep{berti08,fanidakis11,volonteri_sikora,volonteri13}, in some
cases also accounting for the recoil velocity of the merger remnant~\citep{Lousto:2012su,baraussemodel}.
It has also been suggested that if a space-based detector such as eLISA becomes available,
measurements of the mass ratios of black holes binaries and the precession effects predicted by
PN/NR calculations would help to discriminate between competing models
of galaxy formation~\citep{baraussemodel,Sesana:2010wy}.

\begin{figure}
\includegraphics[width=2.0in,clip]{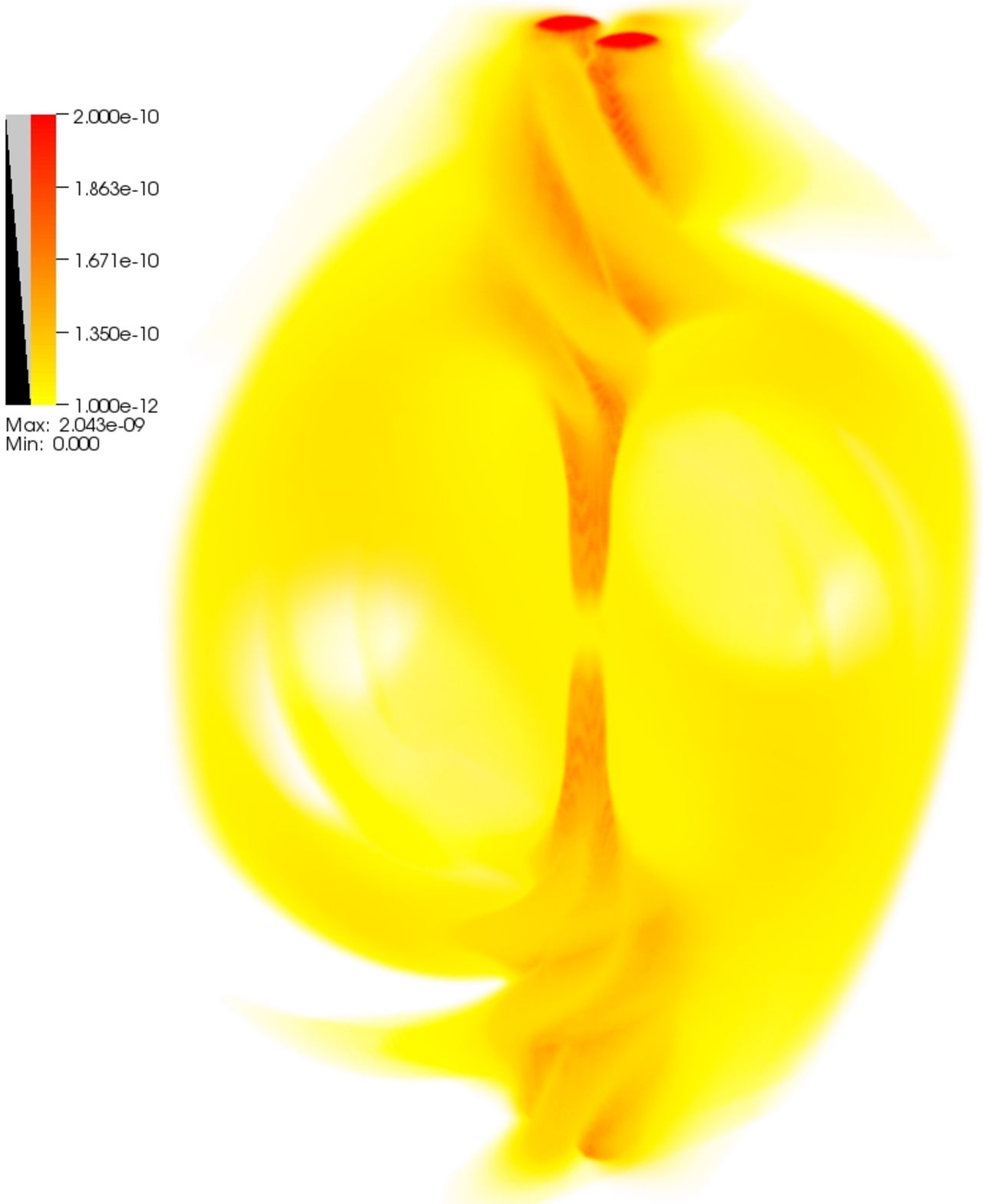} \hspace{0.5in}
\includegraphics[width=2.7in,clip]{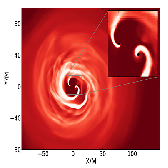} 
\caption{(left) Poynting flux produced by the interaction of an orbiting black hole binary
with a surrounding magnetosphere. The ``braided'' jet structure
is induced by the orbital motion of the black holes (from~\cite{Palenzuela:2010nf}). 
(right) Rest-mass density induced by a supermassive black hole binary 
interacting with a magnetized disk prior to when the binary ``decouples'' from the disk, namely
when the gravitational wave backreaction timescale becomes smaller than the viscous timescale
(from~\cite{2012PhRvL.109v1102F}.}
\label{fig:further_bhbh1}
\end{figure}

\section{Non-vacuum Binaries}\label{sec_nonvac}
As in the binary black hole case, non-vacuum binaries present a well 
defined problem, however they need a larger set of parameters to characterize. 
First, the matter physics introduces a scale, so that unlike the vacuum case
the total mass of the system cannot be factored out.
Thus the set of parameters needed to described the orbit are now the masses $m_1$ and $m_2$ 
of the compact objects, their two initial spin vectors  $\vec{s}_1,\vec{s}_2$ (which however is 
expected to be small for neutron stars), and the initial eccentricity $e_0$ and size of the orbit 
(again which we parameterized by the initial pericenter distance $r_{p0}$).
Second, for neutron star matter one must specify the EOS 
(which for a given mass star determines its radius), and each star's magnetization (strength and dipole direction).

The presence of (magnetized) matter in the system, that is strongly
affected by the rapidly varying geometry during coalescence, can naturally induce electromagnetic 
and neutrino emission in concert with the gravitational waves. A prime example is
short Gamma Ray Bursts (sGRBs), and the evidence is mounting that
non-vacuum binary mergers provide the central engine for these spectacular astrophysical
phenomena (e.g~\cite{Janka99a,Lee:2007js,2012ApJ...746...48M,2012arXiv1204.6242P,2013Natur.500..547T,2013ApJ...774L..23B,Berger:2013jza}).
Thus, in addition to obtaining predictions for the gravitational wave signatures from these events (see
Figure~\ref{fig:nonvac1} for some examples), research using simulations
is also focused on gaining a theoretical understanding of their connections to sGRBs,
and related phenomena.

Widely separated non-vacuum binaries display the same behavior as binary black holes.
Here internal details play a negligible role, as their
effects first appear in a Post-Newtonian expansion at order $(v/c)^{10}$. Closer to merger
tidal forces introduce subtle deviations at first, growing to quite large deformations at
the point of contact in a binary neutron star system, and for a black hole neutron star 
system can even lead to the disruption of the star prior to merger.
For binary neutron stars, if the total mass in the remnant is more than the maximum
mass allowed by the EOS, a black hole will eventually form. 
The intermediate state is called a hypermassive neutron star (HMNS), and is temporarily supported by  
rotation and thermal pressure. An interesting question then, as illustrated in Figure~\ref{fig:nonvac1}, is how
long the HMNS lasts. Once a black hole forms, and following a black hole-neutron
star merger, an accretion disk can form. If there is sufficient mass in the disk, this
could be the beginning of a jet that would eventually produce a sGRB. A host of other
electromagnetic emission is likely as a consequence of these non-vacuum mergers, as is neutrino emission.
In the following sections we discuss these in more detail, highlighting the
information gained from numerical relativity simulations.
For other review articles in this subject area see~\cite{Duez:2009yz,faber_review,Pfeiffer:2012pc}.

\subsection{Binary Neutron Star Mergers}\label{sec_nsns}

Fully general relativistic studies of binary neutron stars have been an active area of research for over a decade.
(For a small sample of recent results in this area see for e.g.~\cite{2008PhRvL.100s1101A,Stephens:2008hu,Rezzolla:2010fd,Sekiguchi:2011zd,Kiuchi:2012mk,2013PhRvD..88d4026H,Read:2013zra,Palenzuela:2013kra,2013arXiv1306.4034K}). 
The initial focus of the research was directed toward understanding broad characteristics of the gravitational wave emission,
and consequently rather simple treatments of the matter were employed (typically an ideal fluid with polytropic equation of state). 
These efforts gave a rather robust understanding of the qualitative dynamics of the system, and prepared a solid foundation
to increase the realism of the matter modeling in the simulations. In recent years the addition of new physical ingredients 
have included more realistic equations of state, magnetic fields and plasmas, and some simplistic
treatments of neutrino and radiation physics. In this section we review the more interesting developments relevant
to astrophysics.

\begin{figure}
\includegraphics[width=2.55in,clip]{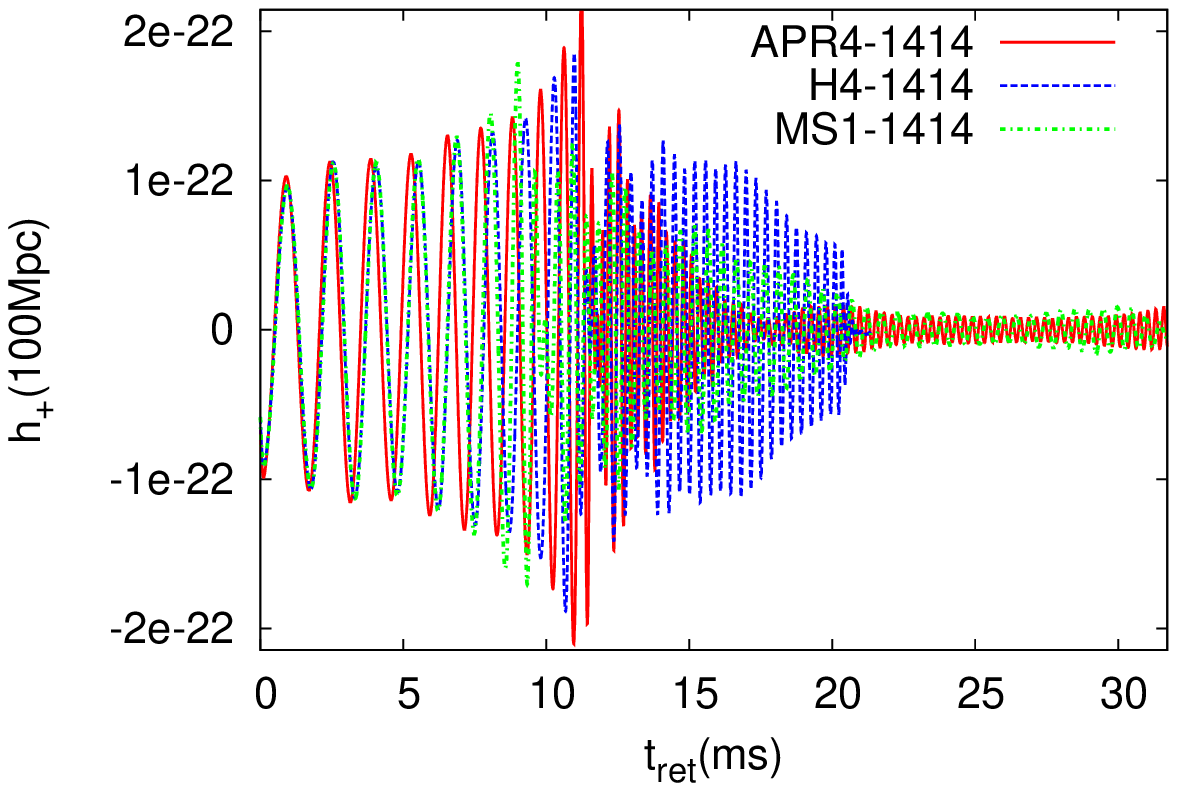} 
\includegraphics[width=2.65in,clip]{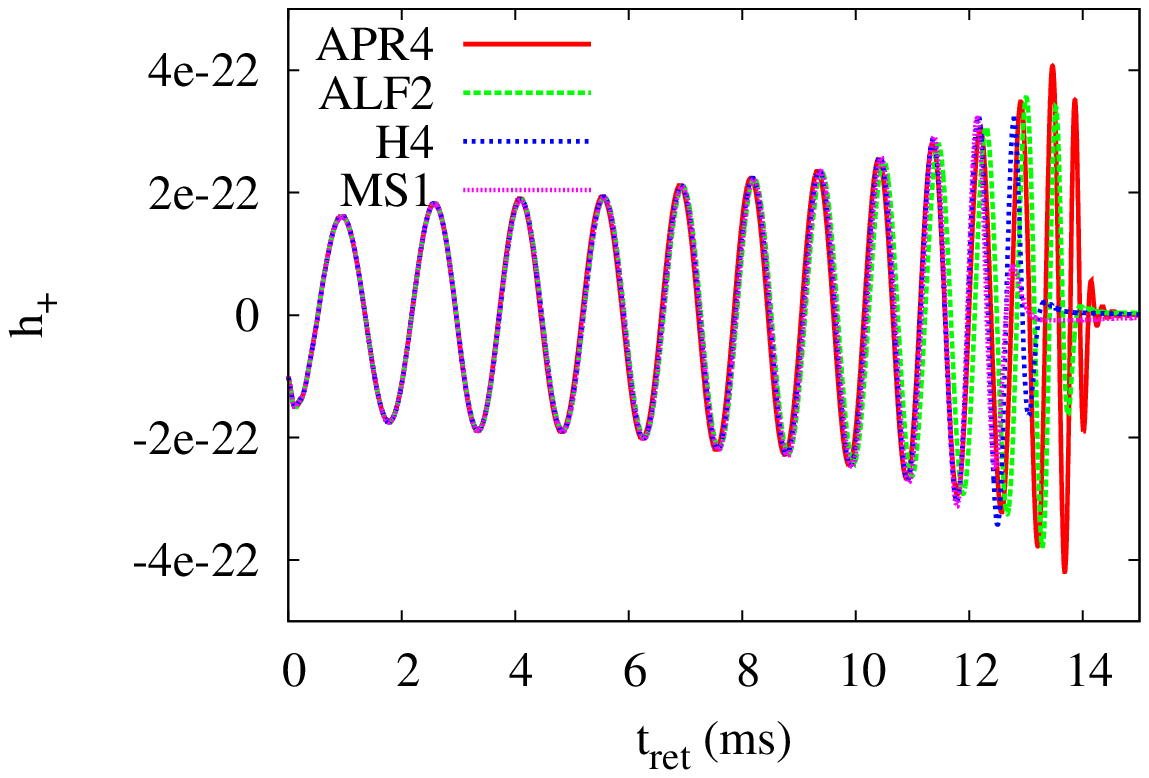}
\caption{Examples of the ``plus'' polarization component of gravitational waves from binary neutron star mergers,
measured $100$ Mpc from the source along the direction of the orbital angular momentum. 
The different curves correspond to different choices of the equations of state (EOS) of the neutron star matter,
labeled APR4,ALF2,H4 and MS1.
For a $1.4M_{\odot}$ neutron star, the APR4,ALF2,H4,MS1 EOS give radii of $11.1,12.4,13.6,14.4$km respectively.
(left) Mergers of an equal mass binary neutron star system (with $m_1=m_2=1.4M_{\odot}$). A hypermassive
neutron star (HMNS) is formed at merger, but how long it survives before collapsing to a black hole
strongly depends on the EOS. The H4 case collapses to a black hole $\approx 10$ms after 
merger; the APR4 and MS1 cases have not yet collapsed $\simeq 35$ms after merger when the
simulations were stopped (the MS1 EOS allows a maximum total mass
of $2.8 M_{\odot}$, so this remnant may be stable). The striking difference in gravitational wave signatures
is self-evident (from~\cite{2013PhRvD..88d4026H}). (right) Emission from black hole-neutron star mergers, with 
$m_{BH} = 4.05 M_{\odot}, m_{NS} = 1.35 M_{\odot}$. Variation with EOS is primarily due to coalescence taking
place earlier for larger radii neutron stars (from~\cite{Kyutoku:2013wxa}). 
}
\label{fig:nonvac1}
\end{figure}

As in the case of binary black holes, several important qualitative questions about the merger process 
have been elucidated. The first relates to behavior post-merger and, for a sufficiently massive
remnant, the onset of collapse to a black hole. A large swath of parameters centered about
the observationally favored initial neutron star masses of $\approx 1.4 M_\odot$, and consistent
with the highest-mass neutron stars observed to date ($\approx 2. M_{\odot}$ 
see, \cite{2010Natur.467.1081D,2013Sci...340..448A} and related discussion 
by~\cite{Lattimer:2010uk}), show
an HMNS forms. 
From a fundamental gravity point of view, an interesting observation is that this intermediate state can have a
highly dense central region, and an effective angular momentum higher than the Kerr bound. 
Yet, obeying the cosmic censorship conjecture, the object does not evolve to a nakedly singular solution, but is able to efficiently 
transport angular momentum outward to eventually allow a black hole to form. The black hole settles down to 
an approximate Kerr solution surrounded by some amount of material in a disk.
A crucially important question then is what the timescale for collapse is. 
This timescale depends sensitively on the initial conditions and several physical 
properties: the individual neutron star masses and eccentricity of the binary (which influences the initial distribution
of mass among the HMNS, bound and ejected material), the EOS,
neutrino and photon cooling, and thermal pressure, as well as diverse mechanisms for angular momentum transport.
The reason this is such an important question is that the timescale is in principle observable, either
directly via the gravitational wave emission (as illustrated in Figure~\ref{fig:nonvac1}, though
note that the frequency of the post-merger waves are sufficiently high that the adLIGO
detectors will not be sensitive to them except for a highly unlikely nearby event), or indirectly
through details of the counterpart electromagnetic/neutrino emission. 

Beyond these broad qualitative issues, theoretical studies have been aimed to analyze in detail
the coalescence process and characteristics of the gravitational wave emission.
As mentioned, early stages of the dynamics are well captured by 
PN treatments. Approaching merger tidal effects do start to influence the evolution
of the orbit, which would be reflected in the gravitational wave emission
and could be detected via delicate data analysis~\citep{Damour:2012yf,Hinderer:2009ca}. 
Another interesting pre-merger consequence of tidal forces during a quasi-circular inspiral is they can 
induce resonant oscillations in the interface modes (i-modes) between the neutron star crust and core
that grow large enough to 
shatter the star's crust, leading to a potentially observable pre-cursor electromagnetic 
outburst~\citep{Tsang:2011ad}. 
For highly eccentric close encounters, the tidal force is impulsive in nature. This can
similarly shatter the crust~\citep{Tsang:2013mca}, and will excite f-mode oscillations
in the star~\citep{Gold2011,bhns_astro_letter}. The f-modes do emit gravitational waves,
though at frequencies that are too high and amplitudes too weak for likely direct detection with adLIGO. 

For low eccentricity encounters 
the stars merge at an orbital frequency that can be estimated by the point at which the stars come into contact, i.e. 
$\Omega_c \simeq [ (m_1 + m_2)(R_1+R_2)^3 ]^{1/2}$. At this stage, the stars are traveling at a considerable fraction 
($\simeq 10-20 \%$) of the speed of light, resulting in a violent collision. In the contact region, shock heating is responsible 
for a considerable amount of mass thrown outwards (some of which becomes unbound) in a rather spheroidal shape. 
Also, strong shearing in this region induces 
Kelvin-Helmholtz instabilities and strong differential rotation develops in the newly formed HMNS.
The temperature of the HMNS can reach values of $\simeq 30-50$Mev and, magnetic fields can grow by several orders of 
magnitude (via winding, tapping kinetic energy and possibly the magneto-rotational
instability (MRI), though for this latter process 
resolutions currently used are still far from that required to adequately resolve it).
Tidal tails form during the earlier stages of the merger and distribute material in the vicinity
of the equatorial plane. As mentioned, because the total mass of the binary likely exceeds the maximum mass 
that a stable, non-rotating and
cold star might achieve, the subsequent behavior of the HMNS divides into two possible cases: prompt or delayed collapse.\\

In the prompt collapse case,  thermal support and differential rotation are unable to overcome the gravitational attraction and
a black hole forms essentially in a free fall time scale. This takes place in binaries
with relatively large total mass $M_{tot} \gtrsim 2.6 - 2.8 M_{\odot}$, though the exact value
depends intimately on the EOS. 

\begin{figure}
\epsfxsize=2.7in
\epsfbox{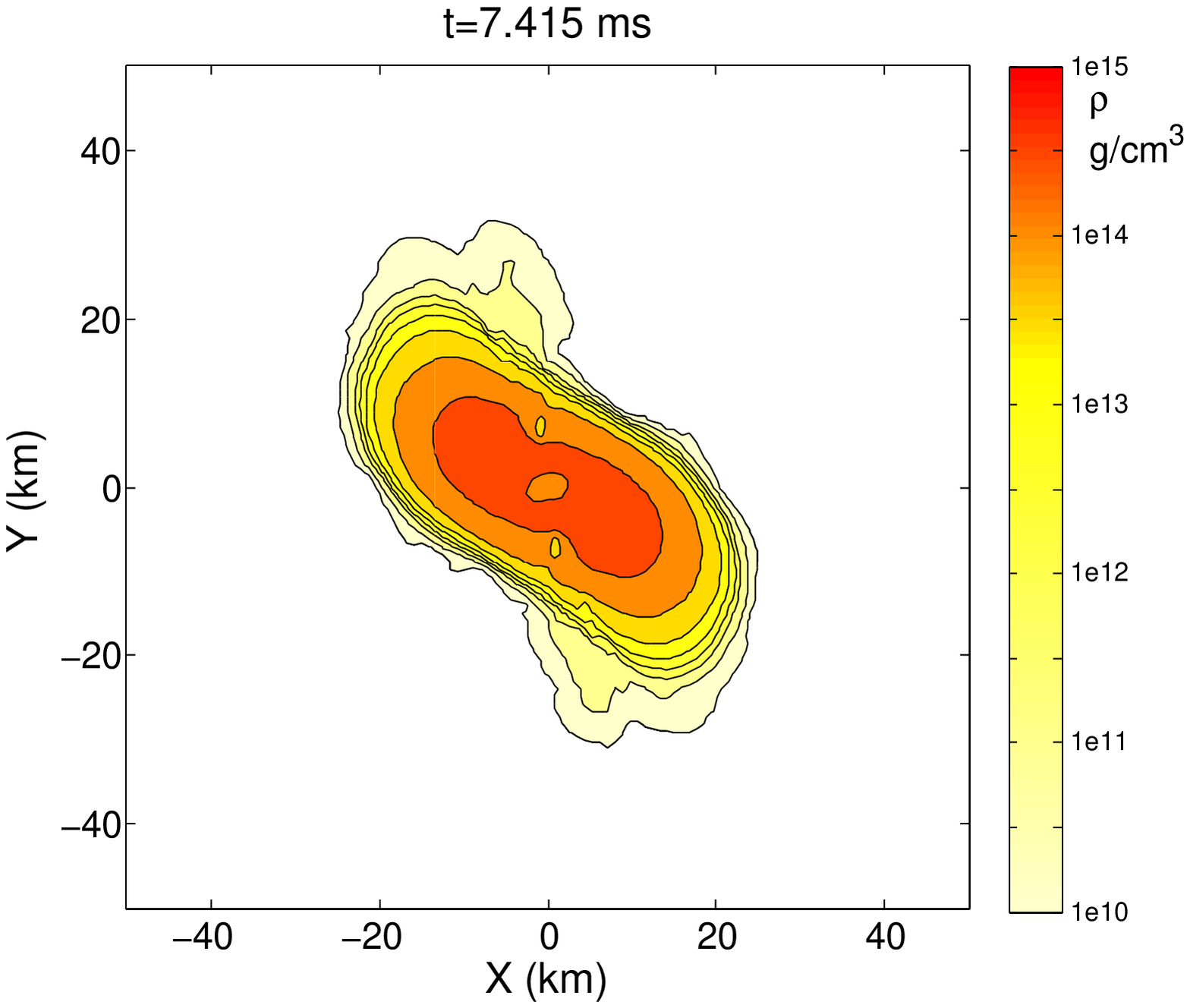}
\epsfxsize=2.7in
\epsfbox{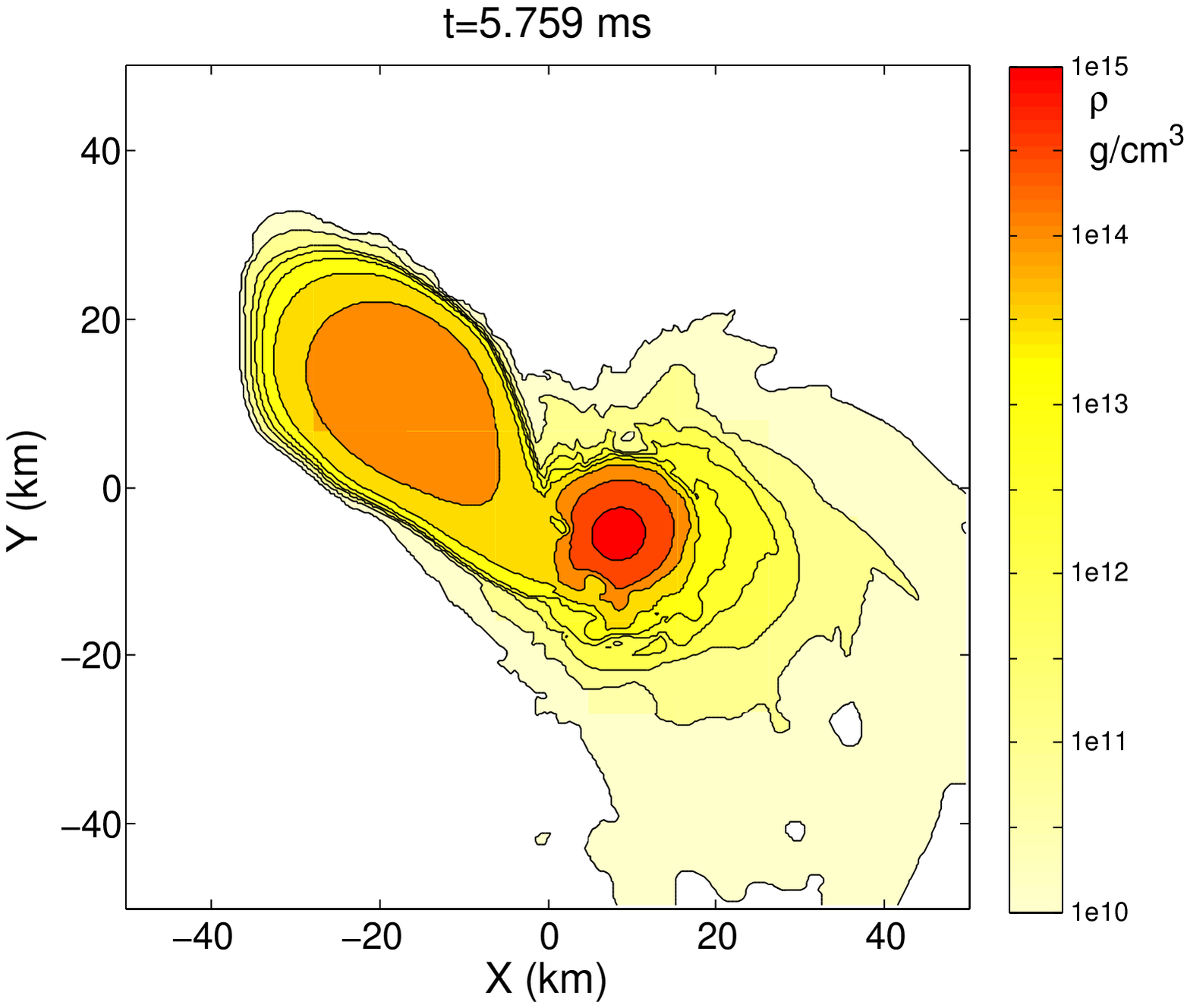}
\caption{Equatorial density profiles $\simeq 3$ms after merger from
an equal (left) and unequal (right) mass binary neutron star system.
The left panel corresponds to a system with $m_1=m_2=1.643 M_{\odot}$ baryonic
mass. The right panel corresponds to $m_1=1.304, m_2=1.805 M_{\odot}$ baryonic masses
(from~\cite{Rezzolla:2010fd}).
}
\label{fig:nonvac2}
\end{figure}

If the collapse does not occur promptly, the post-merger dynamics differs depending on whether the merger involved 
equal masses or not. 
In the former case, the newly formed object resembles a dumbbell composed of two cores (the remnants of the individual stars)
which gradually turns into an ellipsoidal object as a result of angular momentum transport --primarily via hydrodynamics effects-- 
and angular momentum loss via gravitational waves. For example, Figure~\ref{fig:nonvac2} illustrates the equatorial
density of the remnant following an equal mass binary merger, and another that had $m_1/m_2=0.7$. 
The gravitational waves from a post-merger system
has a characteristic frequency in the range $2 \lesssim f \lesssim 4$Khz,
which is proportional (and relatively close) to the Keplerian angular velocity $(M_{HMNS}/R^3_{HMNS})^{1/2}$ (where
$M_{HMNS}, R_{HMNS}$ are the mass and radii of the HMNS, respectively).
If the stars have different masses, the stronger tidal forces induced by the more massive star deforms
the companion, stripping the outer layers and forming an envelope about the newly formed HMNS. This HMNS now displays
two asymmetric cores and behaves as if the more massive core has a satellite that deforms dynamically as time progresses.
Regardless of the mass-ratio, a significant amount of material is estimated to lie beyond the ISCO of the black hole 
that will eventually form, resulting in an accretion disk with mass on the order of $0.01-0.3 M_{\odot}$. Typically
the more massive disks correlate with longer times to black hole formation, a behavior intuitively expected as there
is more time for angular momentum to be transferred outwards to the envelope.

Simulations have also shed light on the processes, and timescales, for such angular momentum transfer. The most important one
is hydrodynamical, which begins to operate efficiently after the merger due to the strong torques induced by asymmetries in the HMNS.
Other significant mechanisms for this transfer are tied to electromagnetic effects: winding and the MRI
can do so by linking the central to outer regions of the HMNS and introducing an effective viscosity in the system.
The angular momentum transport timescale due to winding is of the order of $\tau_{wind} \simeq R_{HMNS}/v_{A}$, with the 
Alfven velocity $v_A \simeq B/\sqrt{\rho}$. A few general relativistic simulations have pointed out that the
strength of $B$ can increase\footnote{In agreement with analogous results obtained in pseudo-Newtonian~\citep{2006Sci...312..719P} or
shearing box studies~\citep{2010A&A...515A..30O}.} from typical premerger values of $10^{10-12}G$, to $10^{15-16}$G via compression, winding and
transfer of hydrodynamical kinetic energy to electromagnetic energy via Kelvin-Helmholtz instabilities~\citep{2008PhRvL.100s1101A,Giacomazzo:2010bx},
which imply timescales $\tau_{wind} \simeq 10-100$ms. We stress however, that present computational resources are still not adequate to give a thorough
analysis of this process. For transport driven by the MRI, simulations are even more challenging, so 
this is still a largely unexplored process within general relativistic simulations of binary neutron star mergers. Nevertheless, estimates
indicate $\tau_{MRI} \sim 100$ms for putative magnetic field strengths of $B \simeq 10^{15}$G. Therefore, either transport mechanism can operate
on timescales $\gtrsim 10 - 100$ms and aid in expediting the collapse. Cooling via neutrino and radiation transport 
reduces thermal-pressure support, so it also helps to shorten the time to collapse. However, the timescale for cooling to
operate in a significant manner is on the order of seconds. Currently, simulations incorporating both electrodynamics and cooling are 
actively being pursued and refined.

Beyond the intricate details of the merger and post-merger behavior, there is strong interest in exploring binary neutron star mergers 
as progenitors of sGRBs and other electromagnetically observable signals. There is already 
tantalizing observational evidence for the connection between non-vacuum compact binaries mergers
and sGRBs (see~\cite{Berger:2013jza} for a recent review), strengthened by 
compelling theoretical models that suggest a merger yielding a rapidly accreting black hole 
could serve as the central energy source through hydrodynamical/plasma or neutrino processes~\citep{Eichler:1989ve,Narayan:1992iy}. 
Other models for the origin of at least a class of sGRBs 
include magnetars produced by binary neutron star mergers, binary white dwarf mergers, or accretion-induced
collapse of a white dwarf~\citep{2006MNRAS.368L...1L,2008MNRAS.385.1455M}, and
the collapse of an accreting neutron star to a black hole~\citep{2005astro.ph.10192M}. 
Simulations of these systems are providing valuable information to
test these models. For instance, once collapse occurs, an initial hyper-accreting stage is observed, followed by a longer
fall back accretion phase with the characteristic
$t^{-5/3}$ power-law dependence expected from analytic calculations~\citep{1988Natur.333..523R}. 
Beyond the burst itself, electromagnetic emission arising from the interaction of 
ejected material with the ambient medium, or through radioactive decay
of r-process elements formed in this material shortly after merger, have been proposed (e.g.~\cite{2012arXiv1204.6242P,2012ApJ...746...48M}).
A candidate for this latter ``kilonova'' event has recently been observed~\citep{2013Natur.500..547T,2013ApJ...774L..23B}.
The time scale for this class of emission can be as long as days or weeks following merger, hence it
is not amenable to ab initio simulations. However, results
from simulations are consistent with properties assumed in these models to give observable signals; in particular,
ejected material of order $\lesssim 10^{-4} M_{eject}/M_{\odot} \lesssim 10^{-2} $ traveling with velocities $\simeq 0.1 - 0.3c$ has
been seen in non-eccentric scenarios (with somewhat larger amounts/higher velocities possible in eccentric mergers).
Finally, a magnetar with magnetic field strength likely in excess
of $10^{15}$G indeed forms during the merger of magnetized neutron stars, 
though its lifetime is typically $\lesssim 100$ms except for the stiffest of EOS and
low-mass binaries~\citep{2008PhRvL.100s1101A,Giacomazzo:2010bx}.\\

\subsubsection{Further Physics}
We conclude this section with a few miscellaneous topics
related to binary neutron star mergers.

\begin{figure}
\includegraphics[width=2.5in,clip]{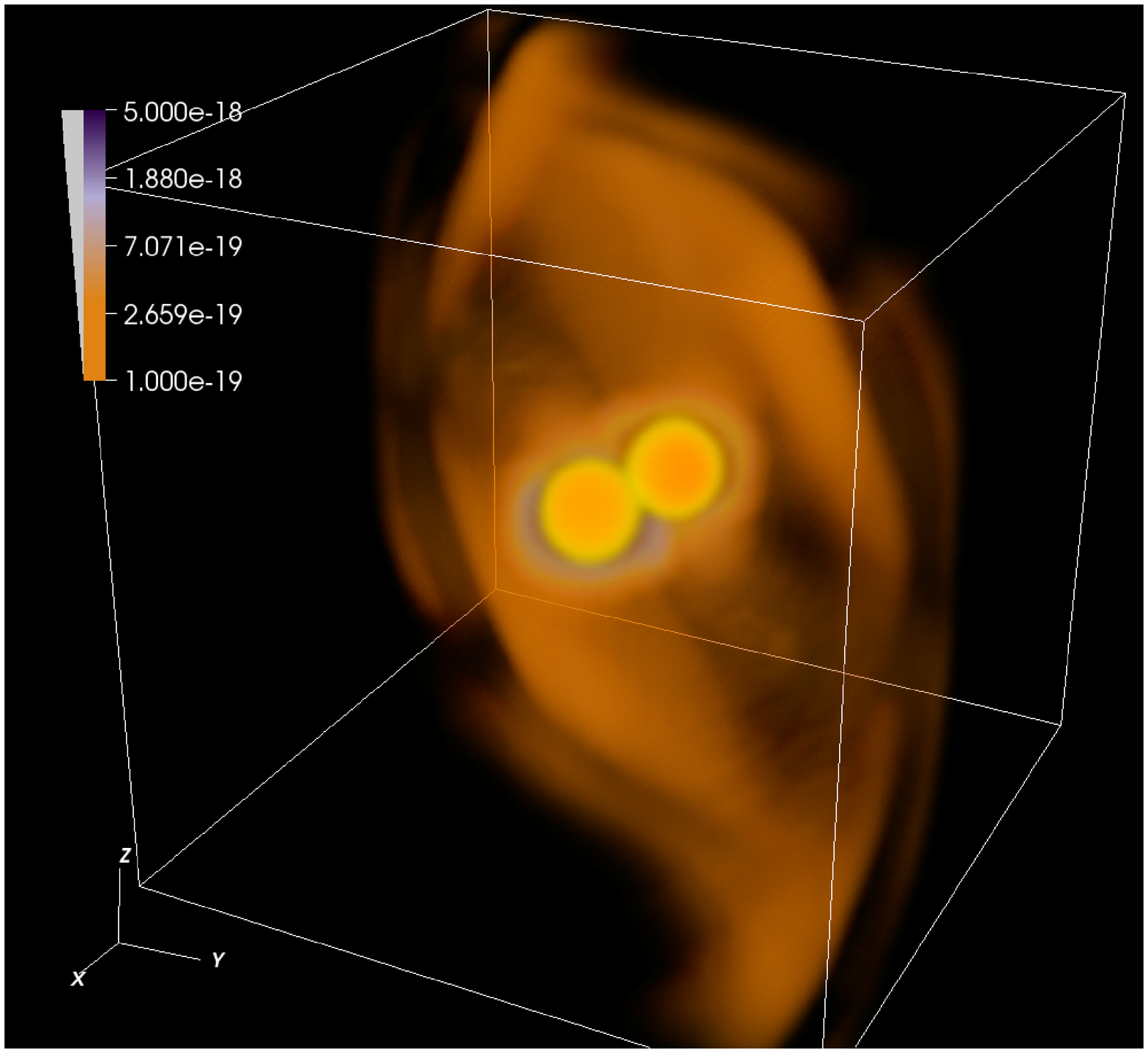} \hspace{0.2in}
\includegraphics[width=2.5in,clip]{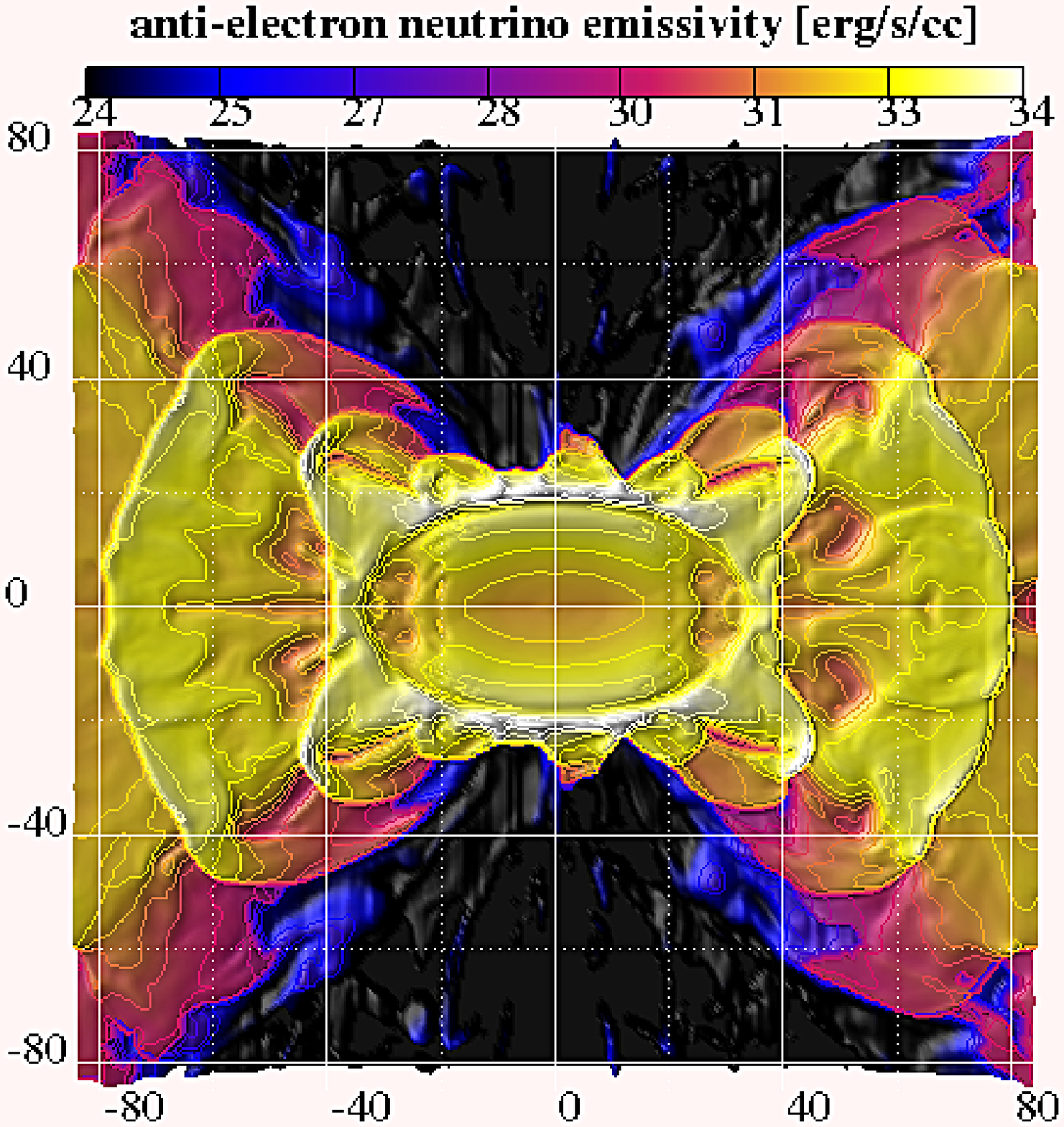}
\caption{(left) Poynting flux produced by the magnetospheric interaction of orbiting, magnetized (with $B=10^{12}$G),
equal-mass ($m_1=m_2=1.4M_{\odot}$) neutron stars $\simeq 1.5$ms before merger (from~\cite{Palenzuela:2013hu}).
 (right) Anti-electron neutrino luminosity in the $x-z$ plane,
$15$ms after an equal mass ($m_1=m_2=1.5M_{\odot}$) binary neutron star merger (from~\cite{Sekiguchi:2011zd}).}
\label{fig:further_bns1}
\end{figure}

\begin{itemize}
\item {\it Magnetosphere interactions}. Neutron stars have among the strongest magnetic fields
in the Universe. As in the case of pulsars, they are surrounded by a magnetosphere
that arises naturally as argued by~\cite{Goldreich:1969sb}. It is thus natural to expect that
a binary interaction can trigger behavior related to that observed in pulsars~\citep{1996A&A...312..937L}, though in this
case with a tight connection to the orbital dynamics. 
This has recently been studied by~\cite{Palenzuela:2013hu,Palenzuela:2013kra},
showing that close to the merger event a strong Poynting flux is emitted ($L\simeq 10^{40-43} B_{11}^2$ erg/s); see
Figure~\ref{fig:further_bns1} (left panel) for an illustration.
As anticipated, many features common to those of pulsars are seen: the existence of gaps in the estimated charge density, 
shear layers, the development of a current sheet and a striped structure in the toroidal magnetic field. 
In the binary case however these features bear tight imprints of the binary's behavior. For instance, 
as the orbit tightens a ramp-up in Poynting luminosity ensues, and the current sheet structure
displays a spiral pattern tied to the orbital evolution of the system. This could provide an important electromagnetic 
counterpart to the gravitational waves. In addition, the 
HMNS --which is likely highly magnetized as a result of the collision-- 
can also trigger a strong Poynting flux as it collapses to a black hole~\citep{Lehner:2011aa}.
The luminosity can be as large as $L\simeq 10^{49} (B/10^{15}G)^2$ erg/s, but shuts off abruptly as the black hole forms.
\item {\it Neutrino emissions}. Incipient works are beginning to incorporate estimates of neutrino effects in the system.
Since, as mentioned, the typical lifetime of the HMNS would likely be limited to $\lesssim 100$ms, whereas the timescale
for neutrino cooling is in the order of seconds, as a first approximation
a full (costly) radiation-transport scheme need not be employed.
Instead, a simplified strategy known as a ``leakage scheme''~\citep{Ruffert:1996by} has become the starting point. The leakage
scheme ignores transport from the diffusion of neutrinos as well as neutrino momentum transfer.
What it does model is the possible equilibration of neutrinos, adopts an opaque, hot stellar matter model
to describe local neutrino sources and sinks, and accounts for charged-current $\beta$ processes,
electron-positron pair-annihilation and plasmon decays. At low optical depth the scheme uses
reaction-rate calculations to estimate the local production and emission of neutrinos. In contrast, at high optical depths it assumes neutrinos
are at their equilibrium abundances, and that neutrino/energy losses occur at the diffusion timescale. In between,
a suitable interpolation is adopted. Early efforts employing this scheme
indicate a binary neutron star merger can produce strong neutrino luminosity of order $\simeq L_{\nu} \simeq 10^{54}$ erg/s~\citep{Sekiguchi:2011zd}.
Figure~\ref{fig:further_bns1} (right panel) illustrates the anti-electron neutrino luminosity shortly after 
an equal-mass ($m_1=m_2=1.5M_{\odot}$) merger.
\item {\it Eccentric binaries}. 
Binaries that emit observable gravitational waves while the orbit has high-eccentricity
show significant qualitative and quantitative differences in properties of the merger
compared with equivalent-mass quasi-circular inspirals.
Because there is more angular momentum in the binary when the two stars
collide, typically more mass is stripped off, some fraction of which is
ejected and the rest forms an accretion disk~\citep{Gold2011,nsns_astro_letter}. This has consequences
for the magnitude of ejecta-powered counterparts, abundance of heavy elements produced through r-processes,
 and the range of initial neutron star masses that can lead to sufficiently
massive disks to power an sGRB. The larger rotational
energy also implies longer lifetimes for HMNS remnants. As mentioned above,
close encounters prior to merger
could induce sufficient strain in each NS to shatter its crust, leading to precursor electromagnetic 
emission~\citep{Tsang:2013mca}. Furthermore, f-modes will be excited in 
each star. This changes the energetics of the orbit and indirectly
affect the subsequent gravitational wave emission. The f-modes will also emit gravitational waves
directly, though because of their relatively low amplitudes and high frequencies (around $1.5$ kHz)
they will not be observable with adLIGO-era detectors.
Regarding the dominant emission from the orbital motion, as with eccentric binary black hole 
mergers, the challenge for detection (issues of rates aside) is to construct waveform models
accurate enough to use in template-based searches, or devise alternative strategies. Also,
even though the integrated energy released is order-of-magnitude comparable to a quasi-circular
inspiral, more of it is radiated at higher frequencies in close periapse, high-eccentricity
mergers. For a binary NS this occurs outside adLIGO's most sensitive range,
making such a system unlikely to be detected beyond $\approx 50$ Mpc even with matched
filtering~\citep{East:2012xq}.
\item {\it Alternative gravity theories}. Binary neutron stars are also good candidates to test alternative theories of
gravity, in particular those that predict deviations depending upon the coupling of matter to geometry.
Scalar-tensor theories posit
the existence of a scalar field, that together with geometry, mediates gravitational phenomena. A sub-class of these
theories allow a phenomena known as {\em scalarization}, whereby a sufficiently compact star spontaneously develops a scalar
charge that modifies its gravitational interaction with other stars, and allows for dipole radiation from the 
system~\citep{Damour:1992we}. Though observations of binary pulsar systems tightly constrain these
theories~\citep{2013Sci...340..448A}, 
recent numerical work has shown that within the allowable region of parameter space
strong departures from general relativity can occur late in the inspiral~\citep{Barausse:2012da}. 
These differences are triggered
close to the merger epoch (yet while the gravitational wave frequencies are still well within the reach of 
near-future detectors), and significantly modify the dynamics, causing an earlier onset of the 
plunge~\citep{Barausse:2012da,Shibata:2013pra,Palenzuela:2013hsa}. 
\end{itemize}

\subsection{Black hole--neutron star mergers}
The remaining binary that is a target for earth-based gravitational wave detectors is composed of
a black hole and a neutron star. Here again,
the regime in which the objects are widely separated is well described by a PN approximation, and the binary's dynamics
proceeds as with the other cases discussed above. However, depending on the relation between two
key radii--the tidal radius ($R_T$) and the radius of the ISCO ($R_{ISCO}$)--markedly
different behavior is expected near merger.
These radii, to leading order, depend on the black hole mass and spin (for $R_{ISCO}$), and the binary mass
ratio, the star's mass and EOS (for $R_T$).  Back of the envelope,
the tidal radius $R_T \propto R_{NS} \left(3 M_{BH}/M_{NS}\right)^{1/3}$; $R_{ISCO}$ 
is $6 M_{BH}$ for a non-spinning black hole, decreasing (increasing)
to $M_{BH}$ ($9 M_{BH}$) for a prograde (retrograde) orbit about a maximally spinning Kerr black hole.
The importance of these two radii stems from the intuition that a plunge precedes tidal disruption if $R_{ISCO} > R_T$,
and the opposite otherwise. This distinction is crucial, as in the former case there would be
little difference in the gravitational wave signal compared with a binary black hole merger having the same masses~\citep{Foucart:2013psa}.
By contrast, if disruption occurs, at its onset gravitational wave emission is sharply
suppressed, not only allowing differentiation from the binary black hole case, but also presenting
clues about the star's EOS as this influences the frequency at which the disruption takes place 
(for a given neutron star mass). It is easy to
convince oneself that the disruption possibility favors high spins/comparable masses, while the plunging behavior favors low spins/high
mass ratios. Note also that there are fewer channels for electromagnetic emission if disruption does not occur; in particular
sGRBs and kilonova require it.

These observations about the nature of black hole/neutron star mergers are clearly born out in simulations.
Early studies began with polytropic equations of state and
non-spinning black holes, and have since steadily progressed to incorporated more realistic equations
of state, now covering a fair range of mass ratios and black hole 
spins~\citep{Foucart:2012nc,Kyutoku:2012fv,Kyutoku:2013wxa}. New physics is also being
modeled, as we discussed above with binary neutron stars (because, of course, the
same code infrastructure can be used for both).
Nevertheless, the same caveats concerning neutron star-neutron star binaries apply to 
black hole-neutron star systems, in that simulations have 
not yet covered the full range of possible parameters, nor are sufficient computational
resources available at present to adequately model all the relevant scales and microphysics.

Regarding the systems in which $R_{ISCO}<R_T$ and disruption occurs, for quasi-circular
mergers numerical simulations
have found as much as $0.3 M_{\odot}$ of material outside the ISCO following merger (with
the largest amounts coming from the low mass ratio/high prograde spin cases).
These results have informed fitting formula predicting the
amount of diskmass~\citep{Foucart:2012nc}, which in turn can be used to estimate the spin 
of the final black hole~\citep{Pannarale:2013jua}.
Usually a larger fraction of stripped material is bound and subsequently
accretes onto the black hole, though 
as much as $\approx 0.05 M_{\odot}$ can be ejected from the system, moving with speeds $\approx 0.2c$. 
Typical maximum temperatures following disruption reach $\simeq 80$Mev. The tail regions are 
substantially cooler $\simeq 10 - 100$Kev,
though shocks can re-heat this material to $\simeq 1-3$Mev.
Interestingly, if the black hole spin and orbital angular momentum direction are misaligned, strong differences
arise. For inclinations $\gtrsim 30^o$ a very low-mass disk seems capable of forming, with most of the material outside $R_{ISCO}$
and following highly eccentric trajectories having large semi-major axes~\citep{Foucart:2012nc,Kyutoku:2012fv,Kyutoku:2013wxa}. 
Based on these trajectories it is estimated this material will return to the black hole to accrete at a rate governed by the familiar 
law $\dot M \propto t^{-5/3}$~\citep{Lee:2007js,Chawla:2010sw,2013ApJ...776...47D}. Interestingly, the behavior of this material
has characteristics consistent with kilonova models. However, the ejecta distribution is mainly around the orbital plane
as opposed to the rather spheroidal one arising in binary neutron star mergers. 
In many cases, the amount of material capable of forming a disk is consistent with estimates for 
triggering short gamma-ray bursts. 

At the other end of the spectrum with low spins and/or high mass ratios in which $R_T < R_{ISCO}$,
the star plunges into the black hole with little or no material left behind. For low-spin black holes this outcome happens
for $m_{BH}/m_{NS} \gtrsim 6$; higher spins (or eccentricity, discussed below) can push this
to somewhat larger mass ratios.
For instance, for $m_{BH} = 10M_{\odot}, m_{NS} = 1.4 M_{\odot}$, significant disruption only takes place
for $a_{BH}/m_{BH} \gtrsim 0.9$~\citep{Foucart:2011mz}. Without disruption electromagnetic
and neutrino counterparts such as sGRBs and kilonova are not expected to occur, though as
we discuss below there may still be electromagnetic emission if the neutron star has
a strong enough magnetic field.

\subsubsection{Further physics.}
Naturally, as in the binary neutron star case, a plethora of phenomena can be triggered by the
system's dynamics, and diverse works are proceeding to examine these scenarios, several
of which we describe here.

\begin{figure}

\centerline{
\includegraphics[width=3.0in,clip]{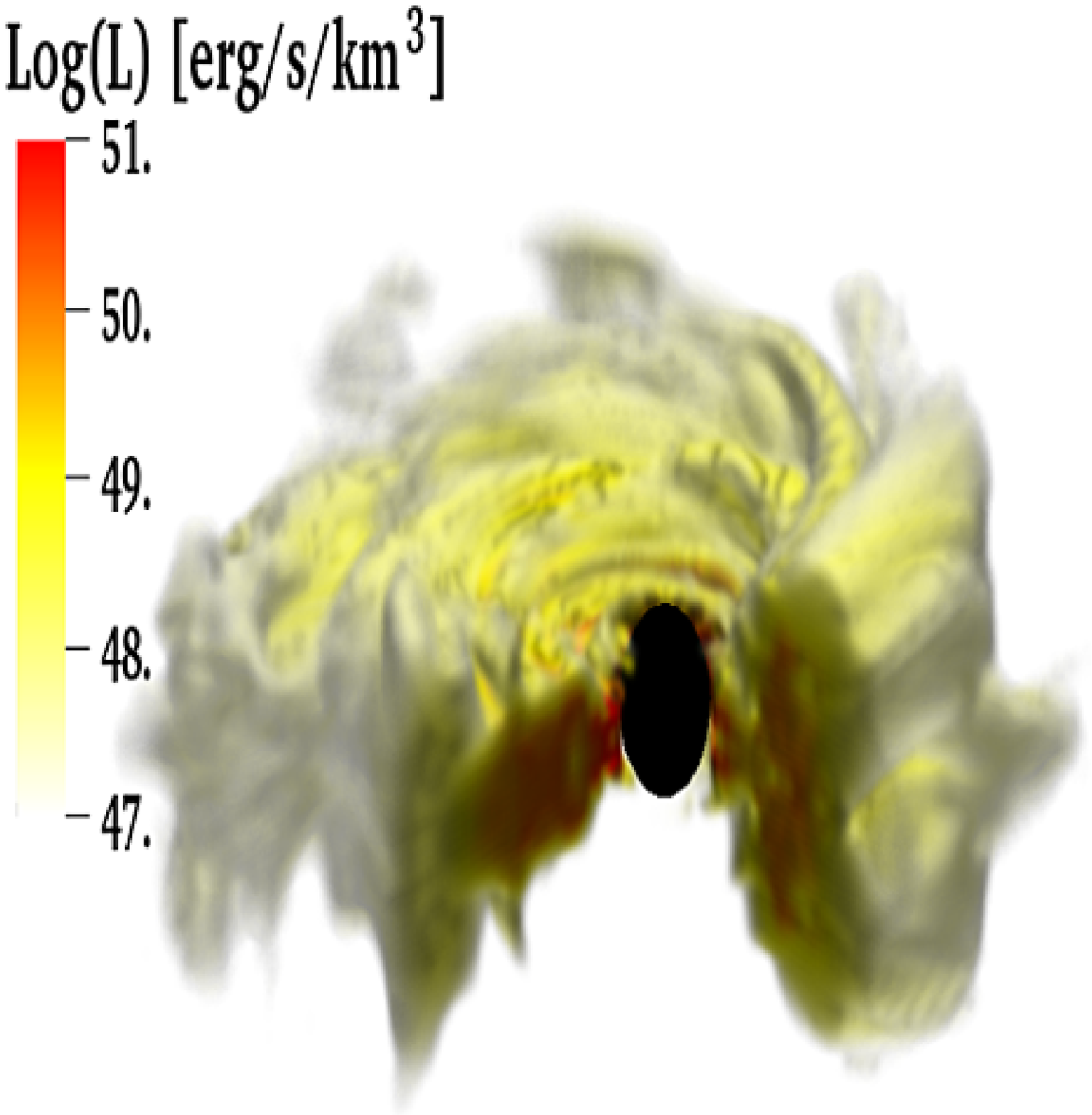}
}
\caption{Luminosity from all neutrino species at $\simeq 12.5$ms after the merger of
a black hole ($M_{BH} = 7 M_{\odot}$) with a neutron star
($M_{NS}=1.4 M_{\odot}$, described by the ``LS220'' equation of state).
The emission region coincides roughly with the disk; namely
densities $\rho>3\times 10^{9}$g/cm$^3$ are approximately within the white region, $\rho>2\, 10^{10}$ g/c$^3$  
the red/orange regions, and the maximum density in the disk is $\simeq 6,\ 10^{11}$ g/cm$^3$. (Figure from
F. Foucart for the SXS Collaboration.)}
\label{fig:bhnsneutrino}
\end{figure}

\begin{itemize}
\item {\it Magnetized stars}. A few studies have explored the behavior of the
system when the neutron star is magnetized. Although electromagnetic effects are too subleading
to alter the orbit and gravitational wave emission, the binary's dynamics can affect properties
of the electromagnetic field after merger.
In particular, the resulting field topology
in the newly formed accretion disk is relevant to assessing whether a jet can be 
launched from the system. Notice that in the absence of spin-orbit-induced precession near the onset of disruption, the initial
poloidal field gets twisted to a mainly toroidal configuration, implying further
processes, such as a dynamo/MRI~\citep{Balbus:1991ay}, would need to take over to reinstate a
poloidal configuration for an
efficient jet mechanism to operate~\citep{McKinney:2008ev}. The orbit
will precess if the black hole is spinning and the spin axis is misaligned with the orbital angular momentum; then
the resulting magnetic field topology after disruption has both poloidal and toroidal components. This might aid 
in giving rise to favorable configurations for jet launching (see the discussion by~\cite{Foucart:2012nc}). 
Full simulations accurately resolving all of 
these effects for both precessing and non-precessing configurations
have yet to be performed, owing to their heavy computational requirements.
\item {\it Magnetosphere interactions}. As mentioned when discussing black holes interacting with
a magnetosphere, the latter is able to tap kinetic energy --rotational or translational-- from
the black hole if there is a relative motion between them. Such a scenario naturally arises
during the inspiral of a magnetized neutron star with a black hole (the binary will not be tidally locked, and
so there will be relative motion of the black hole through the magnetic field lines sourced by the neutron star). Basic estimates using a
simple ``unipolar induction model'' indicate the possibility of a strong Poynting luminosity
produced by the system~\citep{Hansen:2000am,McWilliams:2011zi}. First simulations in this direction
have recently been completed, obtaining consistent values with $L \simeq 10^{41} B_{12}^2$~\citep{Paschalidis:2013jsa}.
Though lacking a detailed account of how this Poynting flux could be converted to observable
photons, this offers the possibility of an electromagnetic counterpart preceding the merger.
\item {\it Neutrino emissions}. As in the case of binary neutron stars, simulations are just beginning to
incorporate neutrino effects, again using the leakage scheme.
Figure~\ref{fig:bhnsneutrino} shows an example of the neutrino luminosity from one
such ongoing investigation, a follow-up study to~\citep{2013ApJ...776...47D}. In this
follow-up study
they found that the merger of a $1.4M_{\odot}$ star with a black hole having mass $7M_{\odot}$ and spin 
parameter $a/M \equiv J/M^2=0.9$ (so significant disruption takes place) yields a peak
neutrino luminosity on the order of $\simeq 10^{54}$ erg/s shortly after disruption, decreasing by an order
of magnitude after $50$ms.
\item {\it Eccentric binaries}. In eccentric black hole-neutron star encounters,
similar quantitative and qualitative differences arise compared with quasi-circular inspiral
as discussed above for the other systems
(possibility of zoom-whirl orbital dynamics, neutron star crust cracking and/or excitation
of f-modes during close encounters, typically larger amounts of
ejecta and accretion disk mass, etc.). In addition, because the effective ISCO 
for eccentric orbits of particles orbiting a non-extremal black hole
is closer to the  event horizon
(e.g. for a Schwarzschild black hole the geodesic ISCO moves in from $r=6M$
to $r=4M$ going from $e=0$ to $1$), the limit for the onset of observable tidal disruption moves
to slightly higher mass ratios\footnote{Note by ``observable'' tidal disruption
we mean disruption that can influence the gravitational waveform and 
the matter ejected or left to accrete. In theory the location of the event horizon
sets the ultimate location for this, though for black hole/neutron star interactions the effective
ISCO appears to be a better proxy. The reason is that once a matter parcel crosses the ISCO,
barring the rise of strong non-gravitational forces, it will reach the event horizon
in of order the light-crossing time of the black hole. Though to date simulations
have not included all the relevant matter microphysics, it is unlikely that
effects triggered by tidal disruption, e.g. shock heating, shearing of magnetic
fields, etc., could grow large enough on such a short time-scale to prevent
immediate accretion of the matter.}.
Furthermore, in systems where tidal disruption
begins outside the ISCO there is the possibility of multiple partial disruptions
and accretion episodes prior to the final disruption/merger~\citep{bhns_astro_letter,bhns_astro_paper}.
\end{itemize}

\section{Gravitational Collapse to a Neutron Star or Black Hole}
\begin{figure}
\centerline{
\includegraphics[width=4.5in,clip]{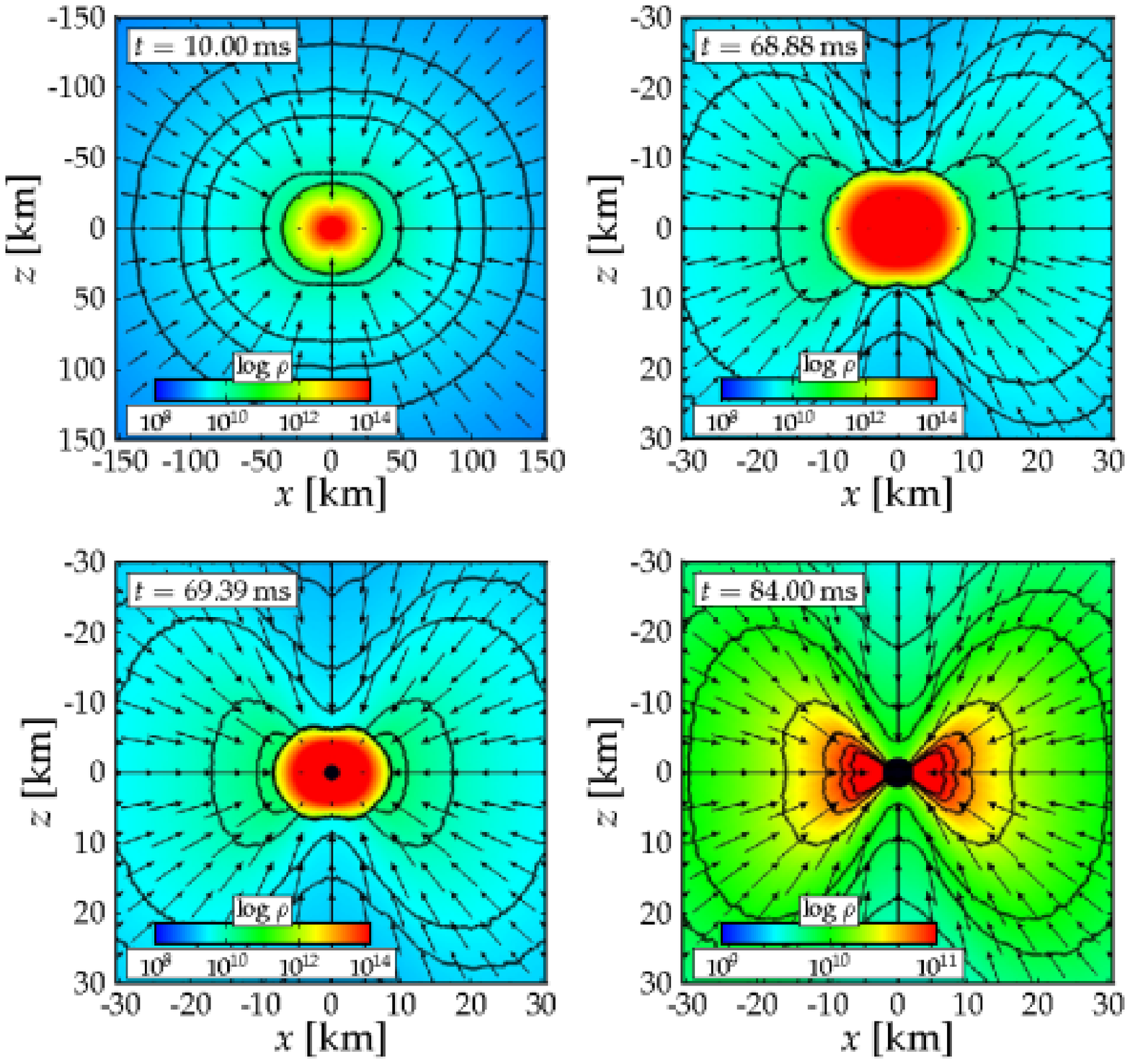}
}
\caption{Density colormaps of the meridional plane of a collapsing
$75 M_{\odot}$ star, superposed with velocity vectors at various
times after bounce (and note the different scale of the top left panel from the rest).
The collapse first forms a proto-neutron star (top panels) which
later collapses to a BH (bottom panels). (From~\cite{2011PhRvL.106p1103O}).
}
\label{fig:core-collase}
\end{figure}

Considerable efforts have been undertaken to study gravitational collapse to a neutron star or a black hole,
in particular within the context of core-collapse supernovae.
Here, stars with masses in the range $10 M_{\odot} \lesssim M \lesssim 100 M_{\odot}$ at zero-age main sequence
form cores that can exceed the Chandrasekhar mass and become gravitationally unstable. 
This leads to collapse that compresses the inner core to nuclear densities, at which point
the full consequences of general relativity must be accounted for. Depending upon the mass of the core, it
can ``bounce'' or collapse to a black hole. 
Figure~\ref{fig:core-collase} displays representative snapshots of the behavior of a collapsing $75 M_{\odot}$ star at different
times. The collapse forms a proto-neutron star that later collapses to a black hole.
In the case of a bounce, an outward propagating shock wave is launched that collides with still infalling material 
and stalls. Observations of core-collapse supernovae imply some mechanism is capable of reviving the shock, 
which is then able to plow through  the stellar envelope and blow up the star. This process is extremely energetic, 
releasing energies on the order of $10^{53}$erg,  the majority of which is emitted in 
neutrinos (for a recent review see~\cite{2009CQGra..26f3001O}). For several decades now, the primary motivation driving
theoretical and numerical studies has been to understand what process (or combination of processes)
mediates such revival, and how.
Several suspects have been identified: heating by neutrinos, (multidimensional) hydrodynamical instabilities, magnetic
fields and nuclear burning (see e.g.~\cite{2007PhR...442...38J,2007PhR...442...23B}). With the very disparate
time- and space-scales involved, a multitude of
physically relevant effects to consider, and the intrinsic cost to accurately model them
(e.g. radiation transport is a seven-dimensional problem) progress
has been slow. Moreover, electromagnetic observations
do not provide much guidance to constrain possible mechanisms as they can not peer deep into the central
engine. By contrast, observations of gravitational waves and neutrinos have the potential to do so, provided
the exploding star is sufficiently nearby. 
Thus, in addition to exploring mechanisms capable of reviving the stalled
shock, simulations have also concentrated on predicting specific gravitational wave and neutrino signatures.
Modeling gravity using full general relativity has only been undertaken recently~\citep{Ott:2012mr}, though prior
to this some of the more relevant relativistic affects were incorporated
(e.g.~\cite{Dimmelmeier:2002bk,Obergaulinger:2006qr,2010ApJS..189..104M,Wongwathanarat:2012zp}).
Although the full resolution of the problem is still likely years ahead, 
interesting insights into fundamental questions and observational prospects have been garnered.
For example, simulations have shown that in rotating core-collapse scenarios, gravitational waves
can be produced and their characteristics are strongly dependent on properties of the collapse :
the precollapse central angular velocity, the development of non-axisymmetric rotational instabilities, postbounce
convective overturn, the standing accretion shock instability, protoneutron star pulsations, etc. 
If a black hole forms, gravitational wave
emission is mainly determined by the quasi-normal modes of the newly formed black hole.
The typical frequencies of gravitational radiation can lie in the range $\simeq 100-1500$Hz, and so are potential sources 
for advanced Earth-based gravitational wave detectors (though the amplitudes are sufficiently
small that it would need to be a Galactic event). As mentioned, the characteristics of these waveforms
depend on the details of the collapse, and, hence, could allow 
us to distinguish the mechanism inducing the explosion. 
Neutrino signals have also been calculated,
revealing possible correlations between oscillations of gravitational waves
and variations in neutrino luminosities. However, current estimates suggest neutrino detections
would be difficult for events taking place at kpc distances~\citep{Ott:2012mr}.

\section{Further Frontiers}
Beyond comparable mass-ratio and comparable radii binaries, NR simulations are starting to 
explore binaries involving higher mass-ratios or less dense stars: 
black hole-white dwarf, neutron star-white dwarf,  intermediate-mass
black holes and main sequence stars, 
black hole binaries involving intermediate and stellar masses~\citep{2012ApJ...749..117H}, etc. 
Here, more rapid progress is hampered
by the computational cost, as it is considerably higher to simulate the larger range of spatial
and temporal scales over which the interesting dynamics takes place.
Several approaches have been suggested to address, at least in part, this difficulty.
These include the use of implicit-explicit methods to tackle large-mass-ratio
binary black holes~\citep{Lau:2011we}, a suitable re-scaling of physical parameters to 
model neutron star-white dwarf binaries~\citep{Paschalidis:2011ez}, a ``background subtraction'' technique to study extreme-mass-ratio systems in which the solution of the dominant gravitational body is known~\citep{East:2013iwa}, and a reformulation of the problem
in terms of a Post-Newtonian approximation incorporating both black hole and matter effects to allow
straight-forward modification of existing ``Newtonian-based'' astrophysical codes~\citep{Barausse:2013ysa}. These are illustrative
examples indicating how the field is progressing beyond traditional boundaries. To date, however, as
far as astrophysical applications are concerned, the predominant focus of NR has been the compact binary problem
(and more recently including the study of core-collapse supernovae~\citep{Ott:2012mr}). Complementary efforts have been directed towards
understanding fundamental questions about strongly gravitating settings, some of which have clear astrophysical implications. 
One example is the question of whether gravitational collapse always leads to a black hole which is 
described by the Kerr solution, or to a naked singularity. Although most of the cases studied so far have indeed shown the black hole result to be the case, especially
in astrophysically relevant contexts, counter examples have been constructed. In $d=4$ spacetime
dimensions these include collapsing matter configurations finely tuned to the threshold of black hole 
formation (in so-called Type II critical collapse~\citep{Choptuik:1992jv}, see 
e.g.~\citep{Gundlach:2002sx} for a review and~\citep{Joshi:2013xoa} for spherical collapse 
of ideal fluids.). 
Due to the fine-tuning required to reach Planck-scale curvatures visible outside an event horizon, it is unlikely
critical collapse occurs naturally in the Universe (though see~\cite{Niemeyer:1997mt} for 
arguments suggesting it would be relevant if certain primordial black hole formation scenarios occurred).
By contrast, this is not the case in higher dimensions in which
 simulations of a class of black holes (black strings) have shown violations of cosmic
censorship can arise generically~\citep{Lehner:2010pn}. This not only highlights that Einstein gravity 
still holds secrets that could be revealed by theoretical studies, but also that surprises of astrophysical significance 
might be in store if our Universe were in fact higher dimensional.

\section{Final Comments}
In this review we have described the status of numerical relativity applied in astrophysical contexts. 
We have focused our presentation on events in which strong gravitational interactions require 
the full Einstein field equations to unravel all details of the phenomena. 
Due to page limitations, we have had to choose a representative subset of
all relevant activities; nevertheless we hope it is clear that the field has ``come of age.''
Yet there is still much to learn, and continued efforts will refine
numerical relativity's predictions and application in astrophysics.

\section{Acknowledgements}
We want to thank our collaborators on work mentioned here: M. Anderson, E. Barausse, W. East, E. Hirschman, 
S. Liebling, S. McWilliams, P. Motl, D. Neilsen, C. Palenzuela, F. Ramazanoglou, B. Stephens and N. Yunes. 
We also acknowledge helpful discussions with A. Buonanno, K. Belczynski, E. Berger,  D. Brown, C. Fryer, T. Janka,
W. Lee, B. Metzger, R. Narayan, E. Poisson, E. Quataert, E. Ramirez-Ruiz, S. Rosswog, A. Spitkovsky and J. Stone.
For figures: C. Ott, B. Giacommazzo, F. Foucart, C. Lousto, A. Buonanno, H. Pfeiffer, S. Shapiro, R. Gold, 
M. Hannam, K. Kyutoku, K. Hotokezaka, M. Shibata.
We also thank the participants of the ``Chirps, Mergers and Explosions: The Final Moments of
Coalescing Compact Binaries'' workshop at the Kavli Institute for Theoretical Physics for
stimulating discussions. This work was supported in part by an NSERC Discovery Grant and CIFAR (LL),
NSF grants PHY-1065710, PHY1305682 and the Simons Foundation(FP).
Research at Perimeter Institute
is supported by the Government of Canada through Industry Canada and by the Province of Ontario through the
Ministry of Research and Innovation.

\bibliography{SFG_NR}\label{refs}

\begin{thebibliography}{100}

\bibitem{LIGORate2010}
J.~Abadie et~al.
\newblock {TOPICAL REVIEW: Predictions for the rates of compact binary
  coalescences observable by ground-based gravitational-wave detectors}.
\newblock {\em Classical and Quantum Gravity}, 27(17):173001, Sept. 2010,
  1003.2480.

\bibitem{Abbott:2007kv}
B.~P. Abbott et~al.
\newblock {LIGO: The Laser Interferometer Gravitational-Wave Observatory}.
\newblock {\em Rep. Prog. Phys.}, 72:076901, 2009, {arXiv:0711.3041 [gr-qc]}.

\bibitem{2011CQGra..28k4002A}
T.~Accadia et~al.
\newblock {Status of the Virgo project}.
\newblock {\em Classical and Quantum Gravity}, 28(11):114002, June 2011.

\bibitem{Agathos:2013upa}
M.~Agathos, W.~Del~Pozzo, T.~G.~F. Li, C.~V.~D. Broeck, J.~Veitch, et~al.
\newblock {TIGER: A data analysis pipeline for testing the strong-field
  dynamics of general relativity with gravitational wave signals from
  coalescing compact binaries}.
\newblock 2013, 1311.0420.

\bibitem{2011PhRvL.106x1101A}
P.~{Ajith} et~al.
\newblock {Inspiral-Merger-Ringdown Waveforms for Black-Hole Binaries with
  Nonprecessing Spins}.
\newblock {\em Physical Review Letters}, 106(24):241101, June 2011, 0909.2867.

\bibitem{2008itnr.book.....A}
M.~{Alcubierre}.
\newblock {\em {Introduction to 3+1 Numerical Relativity}}.
\newblock Oxford University Press, 2008.

\bibitem{AmaroSeoane:2012km}
P.~Amaro-Seoane, S.~Aoudia, S.~Babak, P.~Binetruy, E.~Berti, et~al.
\newblock {eLISA: Astrophysics and cosmology in the millihertz regime}.
\newblock 2012, 1201.3621.

\bibitem{2008PhRvL.100s1101A}
M.~{Anderson} et~al.
\newblock {Magnetized Neutron-Star Mergers and Gravitational-Wave Signals}.
\newblock {\em Physical Review Letters}, 100(19):191101, May 2008, 0801.4387.

\bibitem{Anderson:2008kz}
M.~Anderson, E.~W. Hirschmann, L.~Lehner, S.~L. Liebling, P.~M. Motl,
  D.~Neilsen, C.~Palenzuela, and J.~E. Tohline.
\newblock Simulating binary neutron stars: Dynamics and gravitational waves.
\newblock {\em Phys. Rev. D}, 77(2):024006, 2008, {arXiv:0708.2720 [gr-qc]}.

\bibitem{Andersson:2013mrx}
N.~Andersson, J.~Baker, K.~Belczynski, S.~Bernuzzi, E.~Berti, et~al.
\newblock {The Transient Gravitational-Wave Sky}.
\newblock {\em Class.Quant.Grav.}, 30:193002, 2013, 1305.0816.

\bibitem{Antognini:2013lpa}
J.~M. Antognini, B.~J. Shappee, T.~A. Thompson, and P.~Amaro-Seoane.
\newblock {Rapid Eccentricity Oscillations and the Mergers of Compact Objects
  in Hierarchical Triples}.
\newblock 2013, 1308.5682.

\bibitem{2013Sci...340..448A}
J.~{Antoniadis}, P.~C.~C. {Freire}, N.~{Wex}, T.~M. {Tauris}, R.~S. {Lynch},
  M.~H. {van Kerkwijk}, M.~{Kramer}, C.~{Bassa}, V.~S. {Dhillon}, T.~{Driebe},
  J.~W.~T. {Hessels}, V.~M. {Kaspi}, V.~I. {Kondratiev}, N.~{Langer}, T.~R.
  {Marsh}, M.~A. {McLaughlin}, T.~T. {Pennucci}, S.~M. {Ransom}, I.~H.
  {Stairs}, J.~{van Leeuwen}, J.~P.~W. {Verbiest}, and D.~G. {Whelan}.
\newblock {A Massive Pulsar in a Compact Relativistic Binary}.
\newblock {\em Science}, 340:448, Apr. 2013, 1304.6875.

\bibitem{Antonini:2013tea}
F.~Antonini, N.~Murray, and S.~Mikkola.
\newblock {Black hole triple dynamics: implications for gravitational wave
  detections}.
\newblock 2013, 1308.3674.

\bibitem{Arun:2006yw}
K.~G. Arun, B.~R. Iyer, M.~S.~S. Qusailah, and B.~S. Sathyaprakash.
\newblock {Testing post-Newtonian theory with gravitational wave observations}.
\newblock {\em Class. Quantum Grav.}, 23:L37--L43, 2006, {arXiv:gr-qc/0604018}.

\bibitem{Baiotti:2010ka}
L.~Baiotti, M.~Shibata, and T.~Yamamoto.
\newblock {Binary neutron-star mergers with Whisky and SACRA: First
  quantitative comparison of results from independent general-relativistic
  hydrodynamics codes}.
\newblock {\em Phys.Rev.}, D82:064015, 2010, 1007.1754.

\bibitem{Baker:2006vn}
J.~G. Baker et~al.
\newblock Getting a kick out of numerical relativity.
\newblock {\em Astrophys. J.}, 653:L93--L96, 2006, {arXiv:astro-ph/0603204}.

\bibitem{Balbus:1991ay}
S.~A. Balbus and J.~F. Hawley.
\newblock A powerful local shear instability in weakly magnetized disks. 1.
  linear analysis. 2. nonlinear evolution.
\newblock {\em Astrophys. J.}, 376:214--233, 1991.

\bibitem{baraussemodel}
E.~{Barausse}.
\newblock {The evolution of massive black holes and their spins in their
  galactic hosts}.
\newblock {\em Mon. Not. R. Astron. Soc.}, 423:2533--2557, July 2012,
  1201.5888.

\bibitem{Barausse:2012da}
E.~Barausse et~al.
\newblock {Neutron-star mergers in scalar-tensor theories of gravity}.
\newblock {\em Phys.Rev.}, D87:081506, 2013, 1212.5053.

\bibitem{Barausse:2013ysa}
E.~Barausse and L.~Lehner.
\newblock {A Post-Newtonian approach to black hole-fluid systems}.
\newblock {\em Phys.Rev.}, D88:024029, 2013, 1306.5564.

\bibitem{Barausse:2012qz}
E.~Barausse, V.~Morozova, and L.~Rezzolla.
\newblock On the mass radiated by coalescing black-hole binaries.
\newblock {\em Astrophys.J.}, 758:63, 2012, 1206.3803.

\bibitem{Barausse:2009uz}
E.~Barausse and L.~Rezzolla.
\newblock Predicting the direction of the final spin from the coalescence of
  two black holes.
\newblock {\em Astrophys. J.}, 704:L40--L44, 2009, {arXiv:0904.2577 [gr-qc]}.

\bibitem{2010nure.book.....B}
T.~W. {Baumgarte} and S.~L. {Shapiro}.
\newblock {\em {Numerical Relativity: Solving Einstein's Equations on the
  Computer}}.
\newblock June 2010.

\bibitem{Berger:2013jza}
E.~Berger.
\newblock {Short-Duration Gamma-Ray Bursts}.
\newblock 2013, 1311.2603.

\bibitem{2013ApJ...774L..23B}
E.~{Berger}, W.~{Fong}, and R.~{Chornock}.
\newblock {An r-process Kilonova Associated with the Short-hard GRB~130603B}.
\newblock {\em \apjl}, 774:L23, Sept. 2013, 1306.3960.

\bibitem{Berger:1984zza}
M.~J. Berger and J.~Oliger.
\newblock Adaptive mesh refinement for hyperbolic partial differential
  equations.
\newblock {\em J. Comp. Phys.}, 53:484, 1984.

\bibitem{Berti:2009kk}
E.~Berti, V.~Cardoso, and A.~O. Starinets.
\newblock Quasinormal modes of black holes and black branes.
\newblock {\em Class. Quantum Grav.}, 26:163001, 2009, {arXiv:0905.2975
  [gr-qc]}.

\bibitem{Berti:2005ys}
E.~Berti, V.~Cardoso, and C.~M. Will.
\newblock {Gravitational-wave spectroscopy of massive black holes with the
  space interferometer LISA}.
\newblock {\em Phys. Rev. D}, 73:064030, 2006, {arXiv:gr-qc/0512160}.

\bibitem{Berti:2007fi}
E.~Berti et~al.
\newblock {Inspiral, merger, and ringdown of unequal mass black hole binaries:
  A multipolar analysis}.
\newblock {\em Phys. Rev. D}, 76:064034, 2007, {arXiv:gr-qc/0703053}.

\bibitem{berti08}
E.~{Berti} and M.~{Volonteri}.
\newblock {Cosmological Black Hole Spin Evolution by Mergers and Accretion}.
\newblock {\em Astrophys. J.}, 684:822--828, Sept. 2008, 0802.0025.

\bibitem{Bini:2012ji}
D.~Bini and T.~Damour.
\newblock {Gravitational radiation reaction along general orbits in the
  effective one-body formalism}.
\newblock {\em Phys.Rev.}, D86:124012, 2012, 1210.2834.

\bibitem{Blanchet:2002LR}
L.~Blanchet.
\newblock {Gravitational Radiation from Post-Newtonian Sources and Inspiralling
  Compact Binaries}.
\newblock {\em Living Rev. Relativity}, 5(3), 2002, {arXiv:gr-qc/0202016}.
\newblock \url{http://www.livingreviews.org/lrr-2002-3}.

\bibitem{Blandford:1977ds}
R.~D. Blandford and R.~L. Znajek.
\newblock {Electromagnetic extraction of energy from Kerr black holes}.
\newblock {\em Mon. Not. R. Astron. Soc.}, 179:433--45, 1977.

\bibitem{Bode:2011tq}
T.~Bode, T.~Bogdanovic, R.~Haas, J.~Healy, P.~Laguna, and D.~M. Shoemaker.
\newblock {Mergers of Supermassive Black Holes in Astrophysical Environments}.
\newblock {\em Astrophys. J.}, 744:45, 2012, {arXiv:1101.4684 [gr-qc]}.

\bibitem{Bogdanovic:2007hp}
T.~Bogdanovic, C.~S. Reynolds, and M.~C. Miller.
\newblock Alignment of the spins of supermassive black holes prior to
  coalescence.
\newblock {\em Astrophys. J.}, 661:L147--L150, 2007, {arXiv:astro-ph/0703054}.

\bibitem{2009LNP...783..171B}
C.~{Bona}, C.~{Bona-Casas}, and C.~{Palenzuela-Luque}.
\newblock {Matter Spacetimes}.
\newblock In C.~{Bona}, C.~{Palenzuela-Luque}, and C.~{Bona-Casas}, editors,
  {\em Elements of Numerical Relativity and Relativistic Hydrodynamics}, volume
  783 of {\em Lecture Notes in Physics, Berlin Springer Verlag}, page 171,
  2009.

\bibitem{AGN_downsizing3}
R.~G. {Bower}, A.~J. {Benson}, R.~{Malbon}, J.~C. {Helly}, C.~S. {Frenk}, C.~M.
  {Baugh}, S.~{Cole}, and C.~G. {Lacey}.
\newblock {Breaking the hierarchy of galaxy formation}.
\newblock {\em Mon. Not. R. Astron. Soc.}, 370:645--655, Aug. 2006,
  astro-ph/0511338.

\bibitem{Boyd89a}
J.~P. Boyd.
\newblock {\em {C}hebyshev and {F}ourier {S}pectral {M}ethods}.
\newblock Springer-Verlag, New York, 1989.

\bibitem{2008PhRvL.100o1101B}
L.~{Boyle}, M.~{Kesden}, and S.~{Nissanke}.
\newblock {Binary Black-Hole Merger: Symmetry and the Spin Expansion}.
\newblock {\em Physical Review Letters}, 100(15):151101, Apr. 2008, 0709.0299.

\bibitem{Brodbeck:1998az}
O.~Brodbeck et~al.
\newblock {Einstein's equations with asymptotically stable constraint
  propagation}.
\newblock {\em J. Math. Phys.}, 40:909--923, 1999, {arXiv:gr-qc/9809023}.

\bibitem{Broderick:2011mk}
A.~E. Broderick, V.~L. Fish, S.~S. Doeleman, and A.~Loeb.
\newblock {Constraining the Structure of Sagittarius A*'s Accretion Flow with
  Millimeter-VLBI Closure Phases}.
\newblock {\em Astrophys.J.}, 738:38, 2011, 1106.2550.

\bibitem{2013arXiv1311.5564B}
A.~E. {Broderick}, T.~{Johannsen}, A.~{Loeb}, and D.~{Psaltis}.
\newblock {Testing the No-Hair Theorem with Event Horizon Telescope
  Observations of Sagittarius A*}.
\newblock {\em ArXiv e-prints}, Nov. 2013, 1311.5564.

\bibitem{Broderick:2009ph}
A.~E. Broderick, A.~Loeb, and R.~Narayan.
\newblock {The Event Horizon of Sagittarius A*}.
\newblock {\em Astrophys.J.}, 701:1357--1366, 2009, 0903.1105.

\bibitem{Buonanno:2007yg}
A.~Buonanno.
\newblock {Gravitational waves}.
\newblock 2007, 0709.4682.

\bibitem{Buonanno:2006ui}
A.~Buonanno, G.~B. Cook, and F.~Pretorius.
\newblock Inspiral, merger and ring-down of equal-mass black-hole binaries.
\newblock {\em Phys. Rev. D}, 75:124018, 2007, {arXiv:gr-qc/0610122}.

\bibitem{Buonanno:1998gg}
A.~Buonanno and T.~Damour.
\newblock Effective one-body approach to general relativistic two-body
  dynamics.
\newblock {\em Phys. Rev. D}, 59:084006, 1999, {arXiv:gr-qc/9811091}.

\bibitem{Buonanno:2007sv}
A.~Buonanno, L.~E. Kidder, and L.~Lehner.
\newblock Estimating the final spin of a binary black hole coalescence.
\newblock {\em Phys. Rev. D}, 77:026004, 2008, {arXiv:0709.3839 [astro-ph]}.

\bibitem{2008PhRvD..77b6004B}
A.~{Buonanno}, L.~E. {Kidder}, and L.~{Lehner}.
\newblock {Estimating the final spin of a binary black hole coalescence}.
\newblock {\em \prd}, 77(2):026004, Jan. 2008, 0709.3839.

\bibitem{2007PhR...442...23B}
A.~{Burrows} et~al.
\newblock {Multi-dimensional explorations in supernova theory}.
\newblock {\em \physrep}, 442:23--37, 2007, astro-ph/0612460.

\bibitem{2007ApJ...659L...5C}
M.~{Campanelli} et~al.
\newblock {Large Merger Recoils and Spin Flips from Generic Black Hole
  Binaries}.
\newblock {\em \apjl}, 659:L5--L8, Apr. 2007, arXiv:gr-qc/0701164.

\bibitem{Casares:2013tpa}
J.~Casares and P.~Jonker.
\newblock {Mass Measurements of Stellar and Intermediate Mass Black-Holes}.
\newblock 2013, 1311.5118.

\bibitem{2010ARNPS..60...75C}
J.~{Centrella} et~al.
\newblock {The Final Merger of Black-Hole Binaries}.
\newblock {\em Annual Review of Nuclear and Particle Science}, 60:75--100, Nov.
  2010, 1010.2165.

\bibitem{Chawla:2010sw}
S.~Chawla et~al.
\newblock {Mergers of Magnetized Neutron Stars with Spinning Black Holes:
  Disruption, Accretion, and Fallback}.
\newblock {\em Phys. Rev. Lett.}, 105:111101, 2010, {arXiv:1006.2839 [gr-qc]}.

\bibitem{Choptuik:1992jv}
M.~W. Choptuik.
\newblock Universality and scaling in gravitational collapse of a massless
  scalar field.
\newblock {\em Phys. Rev. Lett.}, 70:9--12, 1993.

\bibitem{Collins:2004ex}
N.~A. Collins and S.~A. Hughes.
\newblock {Towards a formalism for mapping the space-times of massive compact
  objects: Bumpy black holes and their orbits}.
\newblock {\em Phys.Rev.}, D69:124022, 2004, gr-qc/0402063.

\bibitem{AGN_downsizing2}
D.~J. {Croton}, V.~{Springel}, S.~D.~M. {White}, G.~{De Lucia}, C.~S. {Frenk},
  L.~{Gao}, A.~{Jenkins}, G.~{Kauffmann}, J.~F. {Navarro}, and N.~{Yoshida}.
\newblock {The many lives of active galactic nuclei: cooling flows, black holes
  and the luminosities and colours of galaxies}.
\newblock {\em \mnras}, 365:11--28, Jan. 2006, astro-ph/0508046.

\bibitem{Damour:1992we}
T.~Damour and G.~Esposito-Farese.
\newblock {Tensor multiscalar theories of gravitation}.
\newblock {\em Class.Quant.Grav.}, 9:2093--2176, 1992.

\bibitem{2013PhRvD..87h4035D}
T.~{Damour}, A.~{Nagar}, and S.~{Bernuzzi}.
\newblock {Improved effective-one-body description of coalescing nonspinning
  black-hole binaries and its numerical-relativity completion}.
\newblock {\em \prd}, 87(8):084035, Apr. 2013, 1212.4357.

\bibitem{Damour:2012yf}
T.~Damour, A.~Nagar, and L.~Villain.
\newblock {Measurability of the tidal polarizability of neutron stars in
  late-inspiral gravitational-wave signals}.
\newblock {\em Phys.Rev.}, D85:123007, 2012, 1203.4352.

\bibitem{Davis:2011ka}
S.~W. Davis, R.~Narayan, Y.~Zhu, D.~Barret, S.~A. Farrell, et~al.
\newblock {The Cool Accretion Disk in ESO 243-49 HLX-1: Further Evidence of an
  Intermediate Mass Black Hole}.
\newblock {\em Astrophys.J.}, 734:111, 2011, 1104.2614.

\bibitem{2013ApJ...776...47D}
M.~B. {Deaton}, M.~D. {Duez}, F.~{Foucart}, E.~{O'Connor}, C.~D. {Ott}, L.~E.
  {Kidder}, C.~D. {Muhlberger}, M.~A. {Scheel}, and B.~{Szilagyi}.
\newblock {Black Hole-Neutron Star Mergers with a Hot Nuclear Equation of
  State: Outflow and Neutrino-cooled Disk for a Low-mass, High-spin Case}.
\newblock {\em \apj}, 776:47, Oct. 2013, 1304.3384.

\bibitem{2010Natur.467.1081D}
P.~B. {Demorest}, T.~{Pennucci}, S.~M. {Ransom}, M.~S.~E. {Roberts}, and
  J.~W.~T. {Hessels}.
\newblock {A two-solar-mass neutron star measured using Shapiro delay}.
\newblock {\em \nat}, 467:1081--1083, Oct. 2010, 1010.5788.

\bibitem{Dimmelmeier:2002bk}
H.~Dimmelmeier, J.~A. Font, and E.~M\"{u}ller.
\newblock Relativistic simulations of rotational core collapse. {I}. {Methods},
  initial models, and code tests.
\newblock {\em Astron. Astrophys.}, 388:917--935, 2002,
  {arXiv:astro-ph/0204288}.

\bibitem{Doeleman:2012zc}
S.~S. Doeleman, V.~L. Fish, D.~E. Schenck, C.~Beaudoin, R.~Blundell, et~al.
\newblock {Jet Launching Structure Resolved Near the Supermassive Black Hole in
  M87}.
\newblock {\em Science}, 338:355, 2012, 1210.6132.

\bibitem{Dotti:2009vz}
M.~Dotti, M.~Volonteri, A.~Perego, M.~Colpi, M.~Ruszkowski, and F.~Haardt.
\newblock {Dual black holes in merger remnants – II. Spin evolution and
  gravitational recoil}.
\newblock {\em Mon. Not. R. Astron. Soc.}, 402:682--690, 2010, {arXiv:0910.5729
  [astro-ph.HE]}.

\bibitem{Duez:2009yz}
M.~D. Duez.
\newblock {Numerical relativity confronts compact neutron star binaries: a
  review and status report}.
\newblock {\em Class.Quant.Grav.}, 27:114002, 2010, 0912.3529.

\bibitem{Duez:2008rb}
M.~D. Duez, F.~Foucart, L.~E. Kidder, H.~P. Pfeiffer, M.~A. Scheel, et~al.
\newblock {Evolving black hole-neutron star binaries in general relativity
  using pseudospectral and finite difference methods}.
\newblock {\em Phys.Rev.}, D78:104015, 2008, 0809.0002.

\bibitem{East:2012xq}
W.~E. East et~al.
\newblock {Observing complete gravitational wave signals from dynamical capture
  binaries}.
\newblock {\em Phys.Rev.}, D87(4):043004, 2013, 1212.0837.

\bibitem{nsns_astro_letter}
W.~E. East and F.~Pretorius.
\newblock {Dynamical Capture Binary Neutron Star Mergers}.
\newblock {\em Astrophys.J.}, 760:L4, 2012, 1208.5279.

\bibitem{East:2013iwa}
W.~E. East and F.~Pretorius.
\newblock {Simulating extreme-mass-ratio systems in full general relativity}.
\newblock {\em Phys.Rev.}, D87:101502, 2013, 1303.1540.

\bibitem{bhns_astro_paper}
W.~E. East, F.~Pretorius, and B.~C. Stephens.
\newblock Eccentric black hole-neutron star mergers: Effects of black hole spin
  and equation of state.
\newblock {\em Phys. Rev. D}, 85:124009, Jun 2012.

\bibitem{code_paper}
W.~E. East, F.~Pretorius, and B.~C. Stephens.
\newblock {Hydrodynamics in full general relativity with conservative AMR}.
\newblock {\em Phys.Rev.}, D85:124010, 2012, 1112.3094.

\bibitem{Eichler:1989ve}
D.~Eichler, M.~Livio, T.~Piran, and D.~N. Schramm.
\newblock Nucleosynthesis, neutrino bursts and $\gamma$-rays from coalescing
  neutron stars.
\newblock {\em Nature}, 340:126--128, 1989.

\bibitem{faber_review}
J.~A. {Faber} and F.~A. {Rasio}.
\newblock {Binary Neutron Star Mergers}.
\newblock {\em ArXiv e-prints}, Apr. 2012, 1204.3858.

\bibitem{fanidakis11}
N.~{Fanidakis}, C.~M. {Baugh}, A.~J. {Benson}, R.~G. {Bower}, S.~{Cole},
  C.~{Done}, and C.~S. {Frenk}.
\newblock {Grand unification of AGN activity in the {$\Lambda$}CDM cosmology}.
\newblock {\em Mon. Not. R. Astron. Soc.}, 410:53--74, Jan. 2011, 0911.1128.

\bibitem{Farrell:2010bf}
S.~Farrell, N.~Webb, D.~Barret, O.~Godet, and J.~Rodrigues.
\newblock {An Intermediate-mass Black Hole of Over 500 Solar Masses in the
  Galaxy ESO 243-49}.
\newblock {\em Nature}, 460:73--75, 2009, 1001.0567.

\bibitem{2012PhRvL.109v1102F}
B.~D. {Farris} et~al.
\newblock {Binary Black-Hole Mergers in Magnetized Disks: Simulations in Full
  General Relativity}.
\newblock {\em Physical Review Letters}, 109(22):221102, Nov. 2012, 1207.3354.

\bibitem{Field:2011mf}
S.~E. Field, C.~R. Galley, F.~Herrmann, E.~Oschner, and M.~Tiglio.
\newblock {Reduced Basis Catalogs for Gravitational Wave Templates}.
\newblock {\em Phys. Rev. Lett.}, 106:221102, 2011, {arXiv:1101.3765 [gr-qc]}.

\bibitem{Flanagan:2005yc}
E.~E. Flanagan and S.~A. Hughes.
\newblock The basics of gravitational wave theory.
\newblock {\em New J. Phys.}, 7:204, 2005, {arXiv:gr-qc/0501041}.

\bibitem{2008LRR....11....7F}
J.~A. {Font}.
\newblock {Numerical Hydrodynamics and Magnetohydrodynamics in General
  Relativity}.
\newblock {\em Living Reviews in Relativity}, 11:7, Sept. 2008.

\bibitem{Foucart:2012nc}
F.~Foucart.
\newblock {Black Hole-Neutron Star Mergers: Disk Mass Predictions}.
\newblock {\em Phys.Rev.}, D86:124007, 2012, 1207.6304.

\bibitem{Foucart:2013psa}
F.~Foucart, L.~Buchman, M.~D. Duez, M.~Grudich, L.~E. Kidder, et~al.
\newblock {First direct comparison of non-disrupting neutron star-black hole
  and binary black hole merger simulations}.
\newblock {\em Phys.Rev.}, D88:064017, 2013, 1307.7685.

\bibitem{Foucart:2011mz}
F.~Foucart, M.~D. Duez, L.~E. Kidder, M.~A. Scheel, B.~Szilagyi, et~al.
\newblock {Black hole-neutron star mergers for 10 solar mass black holes}.
\newblock {\em Phys.Rev.}, D85:044015, 2012, 1111.1677.

\bibitem{2008PhRvD..78h3537G}
D.~{Garfinkle}, W.~C. {Lim}, F.~{Pretorius}, and P.~J. {Steinhardt}.
\newblock {Evolution to a smooth universe in an ekpyrotic contracting phase
  with w>1}.
\newblock {\em \prd}, 78(8):083537, Oct. 2008, 0808.0542.

\bibitem{2000ApJ...543L...5G}
K.~{Gebhardt}, J.~{Kormendy}, L.~C. {Ho}, R.~{Bender}, G.~{Bower},
  A.~{Dressler}, S.~M. {Faber}, A.~V. {Filippenko}, R.~{Green}, C.~{Grillmair},
  T.~R. {Lauer}, J.~{Magorrian}, J.~{Pinkney}, D.~{Richstone}, and
  S.~{Tremaine}.
\newblock {Black Hole Mass Estimates from Reverberation Mapping and from
  Spatially Resolved Kinematics}.
\newblock {\em Astrophys. J. Lett.}, 543:L5--L8, Nov. 2000, astro-ph/0007123.

\bibitem{Gerosa:2013laa}
D.~Gerosa, M.~Kesden, E.~Berti, R.~O'Shaughnessy, and U.~Sperhake.
\newblock {Resonant-plane locking and spin alignment in stellar-mass black-hole
  binaries: a diagnostic of compact-binary formation}.
\newblock {\em Phys.Rev.}, D87:104028, 2013, 1302.4442.

\bibitem{Giacomazzo:2010bx}
B.~Giacomazzo, L.~Rezzolla, and L.~Baiotti.
\newblock {Accurate evolutions of inspiralling and magnetized neutron stars:
  Equal-mass binaries}.
\newblock {\em Phys. Rev. D}, 83:044014, 2011, {arXiv:1009.2468 [gr-qc]}.

\bibitem{Godet:2009hn}
O.~Godet, D.~Barret, N.~Webb, S.~Farrell, and N.~Gehrels.
\newblock {First evidence for spectral state transitions in the ESO243-49 hyper
  luminous X-ray source HLX-1}.
\newblock {\em Astrophys.J.}, 705:L109--L112, 2009, 0909.4458.

\bibitem{Gold:2012tk}
R.~Gold and B.~Bruegmann.
\newblock {Eccentric black hole mergers and zoom-whirl behavior from elliptic
  inspirals to hyperbolic encounters}.
\newblock {\em Phys.Rev.}, D88:064051, 2013, 1209.4085.

\bibitem{Gold2011}
R.~Gold et~al.
\newblock {Eccentric binary neutron star mergers}.
\newblock {\em Phys.Rev.}, D86:121501, 2012, 1109.5128.

\bibitem{2013arXiv1312.0600G}
R.~{Gold}, V.~{Paschalidis}, Z.~B. {Etienne}, S.~L. {Shapiro}, and H.~P.
  {Pfeiffer}.
\newblock {Accretion disks around binary black holes of unequal mass: GRMHD
  simulations near decoupling}.
\newblock {\em ArXiv e-prints}, Dec. 2013, 1312.0600.

\bibitem{Goldreich:1969sb}
P.~Goldreich and W.~H. Julian.
\newblock Pulsar electrodynamics.
\newblock {\em Astrophys. J.}, 157:869--880, 1969.

\bibitem{Gonzalez:2006md}
J.~A. Gonz\'{a}lez et~al.
\newblock Maximum kick from nonspinning black-hole binary inspiral.
\newblock {\em Phys. Rev. Lett.}, 98:091101, 2007, {arXiv:gr-qc/0610154}.

\bibitem{2007PhRvL..98w1101G}
J.~A. {Gonz{\'a}lez} et~al.
\newblock {Supermassive Recoil Velocities for Binary Black-Hole Mergers with
  Antialigned Spins}.
\newblock {\em Physical Review Letters}, 98(23):231101, 2007,
  arXiv:gr-qc/0702052.

\bibitem{Grandclement:2009LR}
P.~Grandclement and J.~Novak.
\newblock Spectral methods for numerical relativity.
\newblock {\em Living Rev. Relativity}, 12(1), 2009, {arXiv:0706.2286 [gr-qc]}.

\bibitem{Greene:2004gy}
J.~E. Greene and L.~C. Ho.
\newblock {Active galactic nuclei with candidate intermediate - mass black
  holes}.
\newblock {\em Astrophys.J.}, 610:722--736, 2004, astro-ph/0404110.

\bibitem{Gundlach:2002sx}
C.~Gundlach.
\newblock {Critical phenomena in gravitational collapse}.
\newblock {\em Phys.Rept.}, 376:339--405, 2003, gr-qc/0210101.

\bibitem{Gundlach:2005eh}
C.~Gundlach et~al.
\newblock Constraint damping in the {Z4} formulation and harmonic gauge.
\newblock {\em Class. Quantum Grav.}, 22:3767--3774, 2005,
  {arXiv:gr-qc/0504114}.

\bibitem{Gustafsson95}
B.~Gustafsson, H.-O. Kreiss, and J.~Oliger.
\newblock {\em Time dependent problems and difference methods}.
\newblock Wiley, New York, 1995.

\bibitem{2012ApJ...749..117H}
R.~{Haas}, R.~V. {Shcherbakov}, T.~{Bode}, and P.~{Laguna}.
\newblock {Tidal Disruptions of White Dwarfs from Ultra-close Encounters with
  Intermediate-mass Spinning Black Holes}.
\newblock {\em \apj}, 749:117, Apr. 2012, 1201.4389.

\bibitem{Hannam:2013pra}
M.~Hannam.
\newblock {Modelling gravitational waves from precessing black-hole binaries:
  Progress, challenges and prospects}.
\newblock 2013, 1312.3641.

\bibitem{Hansen:2000am}
B.~M. Hansen and M.~Lyutikov.
\newblock {Radio and x-ray signatures of merging neutron stars}.
\newblock {\em Mon.Not.Roy.Astron.Soc.}, 322:695, 2001, astro-ph/0003218.

\bibitem{2009PhRvL.103m1101H}
J.~{Healy}, J.~{Levin}, and D.~{Shoemaker}.
\newblock {Zoom-Whirl Orbits in Black Hole Binaries}.
\newblock {\em Physical Review Letters}, 103(13):131101, Sept. 2009, 0907.0671.

\bibitem{2007CQGra..24S..33H}
F.~{Herrmann} et~al.
\newblock {Unequal mass binary black hole plunges and gravitational recoil}.
\newblock {\em Classical and Quantum Gravity}, 24:33, June 2007,
  arXiv:gr-qc/0601026.

\bibitem{Hinder:2013oqa}
I.~Hinder et~al.
\newblock {Error-analysis and comparison to analytical models of numerical
  waveforms produced by the NRAR Collaboration}.
\newblock {\em Class.Quant.Grav.}, 31:025012, 2013, 1307.5307.

\bibitem{Hinderer:2009ca}
T.~Hinderer et~al.
\newblock {Tidal deformability of neutron stars with realistic equations of
  state and their gravitational wave signatures in binary inspiral}.
\newblock {\em Phys. Rev. D}, 81:123016, 2010, {arXiv:0911.3535 [astro-ph.HE]}.

\bibitem{cactus_webpage}
{\sc CACTUS}.~home page.
\newblock \url{http://www.cactuscode.org}, 2013.

\bibitem{had_webpage}
{\sc HAD}.~home page.
\newblock \url{http://had.liu.edu}, 2013.

\bibitem{Horbatsch:2011ye}
M.~Horbatsch and C.~Burgess.
\newblock {Cosmic Black-Hole Hair Growth and Quasar OJ287}.
\newblock {\em JCAP}, 1205:010, 2012, 1111.4009.

\bibitem{2013PhRvD..88d4026H}
K.~{Hotokezaka} et~al.
\newblock {Remnant massive neutron stars of binary neutron star mergers:
  Evolution process and gravitational waveform}.
\newblock {\em \prd}, 88(4):044026, 2013, 1307.5888.

\bibitem{Hughes:2009iq}
S.~Hughes.
\newblock {Gravitational waves from merging compact binaries}.
\newblock {\em Ann.Rev.Astron.Astrophys.}, 47:107--157, 2009, 0903.4877.

\bibitem{2013CQGra..30v4010I}
R.~N.~M.~t. {IPTA}.
\newblock {The International Pulsar Timing Array}.
\newblock {\em Classical and Quantum Gravity}, 30(22):224010, Nov. 2013.

\bibitem{Jacobson:2009kt}
T.~Jacobson and T.~P. Sotiriou.
\newblock {Over-spinning a black hole with a test body}.
\newblock {\em Phys.Rev.Lett.}, 103:141101, 2009, 0907.4146.

\bibitem{Janka99a}
H.-T. Janka, T.~Eberl, M.~Ruffert, and C.~L. Fryer.
\newblock Black hole-neutron star mergers as central engines of gamma-ray
  bursts.
\newblock {\em Astrophys. J.}, 527:L39--L42, 1999, {arXiv:astro-ph/9908290}.

\bibitem{2007PhR...442...38J}
H.-T. {Janka} et~al.
\newblock {Theory of core-collapse supernovae}.
\newblock {\em \physrep}, 442:38--74, Apr. 2007, astro-ph/0612072.

\bibitem{2012PhRvD..85h3516J}
M.~C. {Johnson}, H.~V. {Peiris}, and L.~{Lehner}.
\newblock {Determining the outcome of cosmic bubble collisions in full general
  relativity}.
\newblock {\em \prd}, 85(8):083516, Apr. 2012, 1112.4487.

\bibitem{Joshi:2013xoa}
P.~S. Joshi.
\newblock {Spacetime Singularities}.
\newblock 2013, 1311.0449.

\bibitem{Kalogera:2006uj}
V.~Kalogera, K.~Belczynski, C.~Kim, R.~W. O'Shaughnessy, and B.~Willems.
\newblock {Formation of Double Compact Objects}.
\newblock {\em Phys.Rept.}, 442:75--108, 2007, astro-ph/0612144.

\bibitem{Kamizasa:2012qs}
N.~Kamizasa, Y.~Terashima, and H.~Awaki.
\newblock {A New Sample of Candidate Intermediate-Mass Black Holes Selected by
  X-ray Variability}.
\newblock 2012, 1205.2772.

\bibitem{2013arXiv1306.4034K}
J.~D. {Kaplan}, C.~D. {Ott}, E.~P. {O'Connor}, K.~{Kiuchi}, L.~{Roberts}, and
  M.~{Duez}.
\newblock {The Influence of Thermal Pressure on Hypermassive Neutron Star
  Merger Remnants}.
\newblock {\em ArXiv e-prints}, June 2013, 1306.4034.

\bibitem{Kesden:2009ds}
M.~Kesden, G.~Lockhart, and E.~S. Phinney.
\newblock {Maximum black-hole spin from quasicircular binary mergers}.
\newblock {\em Phys. Rev. D}, 82:124045, 2010, {arXiv:1005.0627 [gr-qc]}.

\bibitem{Kesden:2010ji}
M.~Kesden, U.~Sperhake, and E.~Berti.
\newblock Relativistic suppression of black hole recoils.
\newblock {\em Astrophys. J.}, 715:1006--1011, 2010, {arXiv:1003.4993
  [astro-ph.CO]}.

\bibitem{Kiuchi:2012mk}
K.~Kiuchi, Y.~Sekiguchi, K.~Kyutoku, and M.~Shibata.
\newblock {Gravitational waves, neutrino emissions, and effects of hyperons in
  binary neutron star mergers}.
\newblock {\em Class.Quant.Grav.}, 29:124003, 2012, 1206.0509.

\bibitem{Kocsis_Levin}
B.~Kocsis and J.~Levin.
\newblock Repeated bursts from relativistic scattering of compact objects in
  galactic nuclei.
\newblock {\em Phys. Rev. D}, 85:123005, Jun 2012.

\bibitem{2012AdAst2012E..14K}
S.~{Komossa}.
\newblock {Recoiling Black Holes: Electromagnetic Signatures, Candidates, and
  Astrophysical Implications}.
\newblock {\em Advances in Astronomy}, 2012, 2012, 1202.1977.

\bibitem{Komossa:2008qd}
S.~Komossa, H.~Zhou, and H.~Lu.
\newblock A recoiling supermassive black hole in the quasar {SDSS}
  {J092712.65+294344.0}.
\newblock {\em Astrophys. J.}, 678:L81--L84, 2008, {arXiv:0804.4584
  [astro-ph]}.

\bibitem{Kormendy:2013dxa}
J.~Kormendy and L.~C. Ho.
\newblock {Coevolution (Or Not) of Supermassive Black Holes and Host Galaxies}.
\newblock {\em Ann.Rev.Astron.Astrophys.}, 51:511--653, 2013, 1304.7762.

\bibitem{1995ARA&A..33..581K}
J.~{Kormendy} and D.~{Richstone}.
\newblock {Inward Bound---The Search For Supermassive Black Holes In Galactic
  Nuclei}.
\newblock {\em Annu. Rev. Astron. Astrophys.}, 33:581, 1995.

\bibitem{Kushnir:2013hpa}
D.~Kushnir, B.~Katz, S.~Dong, E.~Livne, and R.~Fernández.
\newblock {Head-on Collisions of White Dwarfs in Triple Systems Could Explain
  Type Ia Supernovae}.
\newblock {\em Astrophys.J.}, 778:L37, 2013, 1303.1180.

\bibitem{Kyutoku:2012fv}
K.~Kyutoku, K.~Ioka, and M.~Shibata.
\newblock {Ultra-Relativistic Counterparts to Binary Neutron Star Mergers in
  Every Direction, X-ray-to-Radio Bands and Second-to-Day Timescales}.
\newblock 2012, 1209.5747.

\bibitem{Kyutoku:2013wxa}
K.~Kyutoku, K.~Ioka, and M.~Shibata.
\newblock {Anisotropic mass ejection from black hole-neutron star binaries:
  Diversity of electromagnetic counterparts}.
\newblock {\em Phys.Rev.}, D88:041503, 2013, 1305.6309.

\bibitem{Laguna:1989rx}
P.~Laguna and D.~Garfinkle.
\newblock {Space-time of Supermassive U(1) Gauge Cosmic Strings}.
\newblock {\em Phys.Rev.}, D40:1011--1016, 1989.

\bibitem{Lattimer:2010uk}
J.~M. Lattimer and M.~Prakash.
\newblock {What a Two Solar Mass Neutron Star Really Means}.
\newblock 2010, 1012.3208.

\bibitem{Lau:2011we}
S.~R. Lau, G.~Lovelace, and H.~P. Pfeiffer.
\newblock {Implicit-explicit (IMEX) evolution of single black holes}.
\newblock {\em Phys.Rev.}, D84:084023, 2011, 1105.3922.

\bibitem{Lee:2007js}
W.~H. Lee and E.~Ramirez-Ruiz.
\newblock {The Progenitors of Short Gamma-Ray Bursts}.
\newblock {\em New J.Phys.}, 9:17, 2007, astro-ph/0701874.

\bibitem{lee2010short}
W.~H. Lee, E.~Ramirez-Ruiz, and G.~Van~de Ven.
\newblock {Short gamma-ray bursts from dynamically assembled compact binaries
  in globular clusters: Pathways, rates, hydrodynamics, and cosmological
  setting}.
\newblock {\em The Astrophysical Journal}, 720(1):953, 2010.

\bibitem{Lehner:2001wq}
L.~Lehner.
\newblock Numerical relativity: A review.
\newblock {\em Class. Quantum Grav.}, 18:R25--R86, 2001, {arXiv:gr-qc/0106072}.

\bibitem{Lehner:2011aa}
L.~Lehner et~al.
\newblock {Intense Electromagnetic Outbursts from Collapsing Hypermassive
  Neutron Stars}.
\newblock {\em Phys.Rev.}, D86:104035, 2012, 1112.2622.

\bibitem{Lehner:2005vc}
L.~Lehner, S.~L. Liebling, and O.~A. Reula.
\newblock Amr, stability and higher accuracy.
\newblock {\em Class. Quantum Grav.}, 23:S421--S446, 2006,
  {arXiv:gr-qc/0510111}.

\bibitem{Lehner:2010pn}
L.~Lehner and F.~Pretorius.
\newblock {Black Strings, Low Viscosity Fluids, and Violation of Cosmic
  Censorship}.
\newblock {\em Phys.Rev.Lett.}, 105:101102, 2010, 1006.5960.

\bibitem{Lehner:2005bz}
L.~Lehner, O.~Reula, and M.~Tiglio.
\newblock {Multi-block simulations in general relativity: High order
  discretizations, numerical stability, and applications}.
\newblock {\em Class.Quant.Grav.}, 22:5283--5322, 2005, gr-qc/0507004.

\bibitem{2006MNRAS.368L...1L}
A.~J. {Levan}, G.~A. {Wynn}, R.~{Chapman}, M.~B. {Davies}, A.~R. {King}, R.~S.
  {Priddey}, and N.~R. {Tanvir}.
\newblock {Short gamma-ray bursts in old populations: magnetars from white
  dwarf-white dwarf mergers}.
\newblock {\em \mnras}, 368:L1--L5, May 2006, astro-ph/0601332.

\bibitem{Leveque92}
R.~J. LeVeque.
\newblock {\em Numerical Methods for Conservation Laws}.
\newblock Birkhauser Verlag, Basel, 1992.

\bibitem{Lippai:2008fx}
Z.~Lippai, Z.~Frei, and Z.~Haiman.
\newblock Prompt shocks in the gas disk around a recoiling supermassive black
  hole binary.
\newblock {\em Astrophys. J.}, 676:L5--L8, 2008, {arXiv:0801.0739 [astro-ph]}.

\bibitem{1996A&A...312..937L}
V.~M. {Lipunov} and I.~E. {Panchenko}.
\newblock {Pulsars revived by gravitational waves.}
\newblock {\em Astron. Astrophys.}, 312:937--940, Aug. 1996, astro-ph/9608155.

\bibitem{2007PhRvL..99d1103L}
A.~{Loeb}.
\newblock {Observable Signatures of a Black Hole Ejected by
  Gravitational-Radiation Recoil in a Galaxy Merger}.
\newblock {\em Physical Review Letters}, 99(4):041103, July 2007,
  arXiv:astro-ph/0703722.

\bibitem{Loffler:2011ay}
F.~L\"{o}ffler, J.~Faber, E.~Bentivegna, T.~Bode, P.~Diener, R.~Haas,
  I.~Hinder, B.~C. Mundim, C.~D. Ott, E.~Schnetter, G.~Allen, M.~Campanelli,
  and P.~Laguna.
\newblock {The Einstein Toolkit: A Community Computational Infrastructure for
  Relativistic Astrophysics}.
\newblock {\em Class. Quantum Grav.}, 29:115001, 2011, {arXiv:1111.3344
  [gr-qc]}.

\bibitem{2010CQGra..27k4006L}
C.~O. {Lousto}, M.~{Campanelli}, Y.~{Zlochower}, and H.~{Nakano}.
\newblock {Remnant masses, spins and recoils from the merger of generic black
  hole binaries}.
\newblock {\em Classical and Quantum Gravity}, 27(11):114006, June 2010,
  0904.3541.

\bibitem{Lousto:2011kp}
C.~O. Lousto and Y.~Zlochower.
\newblock {Hangup Kicks: Still Larger Recoils by Partial Spin-Orbit Alignment
  of Black-Hole Binaries}.
\newblock {\em Phys. Rev. Lett.}, 107:231102, 2011, {arXiv:1108.2009 [gr-qc]}.

\bibitem{2011PhRvD..83b4003L}
C.~O. {Lousto} and Y.~{Zlochower}.
\newblock {Modeling maximum astrophysical gravitational recoil velocities}.
\newblock {\em \prd}, 83(2):024003, Jan. 2011, 1011.0593.

\bibitem{Lousto:2012su}
C.~O. Lousto, Y.~Zlochower, M.~Dotti, and M.~Volonteri.
\newblock {Gravitational recoil from accretion-aligned black-hole binaries}.
\newblock {\em Phys. Rev. D}, 85:084015, 2012, {arXiv:1201.1923 [gr-qc]}.

\bibitem{2005astro.ph.10192M}
A.~I. {MacFadyen}, E.~{Ramirez-Ruiz}, and W.~{Zhang}.
\newblock {X-ray flares following short gamma-ray bursts from shock heating of
  binary stellar companions}.
\newblock {\em ArXiv Astrophysics e-prints}, Oct. 2005, astro-ph/0510192.

\bibitem{1998AJ....115.2285M}
J.~{Magorrian}, S.~{Tremaine}, D.~{Richstone}, R.~{Bender}, G.~{Bower},
  A.~{Dressler}, S.~M. {Faber}, K.~{Gebhardt}, R.~{Green}, C.~{Grillmair},
  J.~{Kormendy}, and T.~{Lauer}.
\newblock {The Demography of Massive Dark Objects in Galaxy Centers}.
\newblock {\em Astron. J.}, 115:2285--2305, June 1998, astro-ph/9708072.

\bibitem{McClintock:2011zq}
J.~E. McClintock, R.~Narayan, S.~W. Davis, L.~Gou, A.~Kulkarni, J.~A. Orosz,
  R.~F. Penna, R.~A. Remillard, and J.~F. Steiner.
\newblock {Measuring the Spins of Accreting Black Holes}.
\newblock {\em Class. Quantum Grav.}, 28:114009, 2011, {arXiv:1101.0811
  [astro-ph.HE]}.

\bibitem{McClintock:2013vwa}
J.~E. McClintock, R.~Narayan, and J.~F. Steiner.
\newblock {Black Hole Spin via Continuum Fitting and the Role of Spin in
  Powering Transient Jets}.
\newblock 2013, 1303.1583.

\bibitem{McKinney:2008ev}
J.~C. {McKinney} and R.~D. {Blandford}.
\newblock Stability of relativistic jets from rotating, accreting black holes
  via fully three-dimensional magnetohydrodynamic simulations.
\newblock {\em Mon. Not. R. Astron. Soc.}, 394:L126--L130, 2009,
  arXiv:0812.1060.

\bibitem{McWilliams:2011zi}
S.~T. McWilliams and J.~J. Levin.
\newblock {Electromagnetic extraction of energy from black hole-neutron star
  binaries}.
\newblock {\em Astrophys. J.}, 742:90, 2011.

\bibitem{2012ApJ...746...48M}
B.~D. {Metzger} and E.~{Berger}.
\newblock {What is the Most Promising Electromagnetic Counterpart of a Neutron
  Star Binary Merger?}
\newblock {\em \apj}, 746:48, Feb. 2012, 1108.6056.

\bibitem{2008MNRAS.385.1455M}
B.~D. {Metzger}, E.~{Quataert}, and T.~A. {Thompson}.
\newblock {Short-duration gamma-ray bursts with extended emission from
  protomagnetar spin-down}.
\newblock {\em \mnras}, 385:1455--1460, Apr. 2008, 0712.1233.

\bibitem{Milosavljevic:2004cg}
M.~Milosavljevi\'{c} and E.~S. Phinney.
\newblock The afterglow of massive black hole coalescence.
\newblock {\em Astrophys. J.}, 622:L93--L96, 2005, {arXiv:astro-ph/0410343}.

\bibitem{Moesta:2011bn}
P.~M\"{o}sta, D.~Alic, L.~Rezzolla, O.~Zanotti, and C.~Palenzuela.
\newblock {On the detectability of dual jets from binary black holes}.
\newblock {\em Astrophys. J.}, 749:L32, 2012, {arXiv:1109.1177 [gr-qc]}.

\bibitem{2010ApJS..189..104M}
B.~{M{\"u}ller}, H.-T. {Janka}, and H.~{Dimmelmeier}.
\newblock {A New Multi-dimensional General Relativistic Neutrino Hydrodynamic
  Code for Core-collapse Supernovae. I. Method and Code Tests in Spherical
  Symmetry}.
\newblock {\em Astrophys. J. Supp. Ser.}, 189:104--133, July 2010, 1001.4841.

\bibitem{Narayan:1992iy}
R.~Narayan, B.~Paczynski, and T.~Piran.
\newblock Gamma-ray bursts as the death throes of massive binary stars.
\newblock {\em Astrophys. J.}, 395:L83, 1992, {arXiv:astro-ph/9204001}.

\bibitem{Neilsen:2005rq}
D.~Neilsen, E.~W. Hirschmann, and R.~S. Millward.
\newblock {Relativistic MHD and excision: formulation and initial tests}.
\newblock {\em Class. Quantum Grav.}, 23:S505, 2006, {arXiv:gr-qc/0512147}.

\bibitem{Niemeyer:1997mt}
J.~C. Niemeyer and K.~Jedamzik.
\newblock {Near-critical gravitational collapse and the initial mass function
  of primordial black holes}.
\newblock {\em Phys.Rev.Lett.}, 80:5481--5484, 1998, astro-ph/9709072.

\bibitem{2012ApJ...755...51N}
S.~C. {Noble} et~al.
\newblock {Circumbinary Magnetohydrodynamic Accretion into Inspiraling Binary
  Black Holes}.
\newblock {\em \apj}, 755:51, 2012, 1204.1073.

\bibitem{2010A&A...515A..30O}
M.~{Obergaulinger}, M.~A. {Aloy}, and E.~{M{\"u}ller}.
\newblock {Local simulations of the magnetized Kelvin-Helmholtz instability in
  neutron-star mergers}.
\newblock {\em Astron. Astrophys.}, 515:A30, June 2010, 1003.6031.

\bibitem{Obergaulinger:2006qr}
M.~Obergaulinger et~al.
\newblock {Axisymmetric simulations of magnetorotational core collapse:
  approximate inclusion of general relativistic effects}.
\newblock {\em Astron.Astrophys.}, 457:209--222, 2006, astro-ph/0602187.

\bibitem{O'Leary:2008xt}
R.~M. O'Leary, B.~Kocsis, and A.~Loeb.
\newblock Gravitational waves from scattering of stellar-mass black holes in
  galactic nuclei.
\newblock {\em Mon. Not. R. Astron. Soc.}, 395:2127--2146, 2009,
  {arXiv:0807.2638 [astro-ph]}.

\bibitem{O'Shaughnessy:2011zz}
R.~O'Shaughnessy, D.~Kaplan, A.~Sesana, and A.~Kamble.
\newblock {Blindly detecting orbital modulations of jets from merging
  supermassive black holes}.
\newblock {\em Astrophys.J.}, 743:136, 2011, 1109.1050.

\bibitem{2009CQGra..26f3001O}
C.~D. {Ott}.
\newblock {The gravitational-wave signature of core-collapse supernovae}.
\newblock {\em Classical and Quantum Gravity}, 26(6):063001, 2009, 0809.0695.

\bibitem{2011PhRvL.106p1103O}
C.~D. {Ott} et~al.
\newblock {Dynamics and Gravitational Wave Signature of Collapsar Formation}.
\newblock {\em Physical Review Letters}, 106(16):161103, 2011, 1012.1853.

\bibitem{Ott:2012mr}
C.~D. Ott et~al.
\newblock {General-Relativistic Simulations of Three-Dimensional Core-Collapse
  Supernovae}.
\newblock {\em Astrophys.J.}, 768:115, 2013, 1210.6674.

\bibitem{Palenzuela:2013hsa}
C.~Palenzuela, E.~Barausse, M.~Ponce, and L.~Lehner.
\newblock {Dynamical scalarization of neutron stars in scalar-tensor gravity
  theories}.
\newblock 2013, 1310.4481.

\bibitem{Palenzuela:2013hu}
C.~Palenzuela et~al.
\newblock {Gravitational and electromagnetic outputs from binary neutron star
  mergers}.
\newblock {\em Phys.Rev.Lett.}, 111:061105, 2013, 1301.7074.

\bibitem{Palenzuela:2010xn}
C.~Palenzuela, T.~Garrett, L.~Lehner, , and S.~L. Liebling.
\newblock {Magnetospheres of black hole systems in force-free plasma}.
\newblock {\em Phys. Rev. D}, 82:044045, 2010, {arXiv:1007.1198 [gr-qc]}.

\bibitem{Palenzuela:2010nf}
C.~Palenzuela, L.~Lehner, and S.~L. Leibling.
\newblock Dual jets from binary black holes.
\newblock {\em Science}, 329:927--930, 2010, {arXiv:1005.1067 [astro-ph.HE]}.

\bibitem{Palenzuela:2013kra}
C.~Palenzuela, L.~Lehner, S.~L. Liebling, M.~Ponce, M.~Anderson, et~al.
\newblock {Linking electromagnetic and gravitational radiation in coalescing
  binary neutron stars}.
\newblock {\em Phys.Rev.}, D88:043011, 2013, 1307.7372.

\bibitem{Palenzuela:2009hx}
C.~Palenzuela, L.~Lehner, and S.~Yoshida.
\newblock {Understanding possible electromagnetic counterparts to loud
  gravitational wave events: Binary black hole effects on electromagnetic
  fields}.
\newblock {\em Phys. Rev. D}, 81:084007, 2010, {arXiv:0911.3889 [gr-qc]}.

\bibitem{2013arXiv1307.6232P}
Y.~{Pan} et~al.
\newblock {Inspiral-merger-ringdown waveforms of spinning, precessing
  black-hole binaries in the effective-one-body formalism}.
\newblock 2013, 1307.6232.
\newblock [arXiv:1307.6232].

\bibitem{Pannarale:2013jua}
F.~Pannarale.
\newblock {Black hole remnant of black hole-neutron star coalescing binaries
  with arbitrary black hole spin}.
\newblock 2013, 1311.5931.

\bibitem{Paschalidis:2013jsa}
V.~Paschalidis, Z.~B. Etienne, and S.~L. Shapiro.
\newblock {General relativistic simulations of binary black hole-neutron stars:
  Precursor electromagnetic signals}.
\newblock {\em Phys.Rev.}, D88:021504, 2013, 1304.1805.

\bibitem{Paschalidis:2011ez}
V.~Paschalidis, Y.~T. Liu, Z.~Etienne, and S.~L. Shapiro.
\newblock {The merger of binary white dwarf--neutron stars: Simulations in full
  general relativity}.
\newblock {\em Phys.Rev.}, D84:104032, 2011, 1109.5177.

\bibitem{Peters:1964zz}
P.~C. Peters.
\newblock Gravitational radiation and the motion of two point masses.
\newblock {\em Phys. Rev.}, 136:B1224--B1232, 1964.

\bibitem{Peters:1963ux}
P.~C. Peters and J.~Mathews.
\newblock Gravitational radiation from point masses in a {Keplerian} orbit.
\newblock {\em Phys. Rev.}, 131:435--440, 1963.

\bibitem{2004ApJ...613..682P}
B.~M. {Peterson}, L.~{Ferrarese}, K.~M. {Gilbert}, S.~{Kaspi}, M.~A. {Malkan},
  D.~{Maoz}, D.~{Merritt}, H.~{Netzer}, C.~A. {Onken}, R.~W. {Pogge},
  M.~{Vestergaard}, and A.~{Wandel}.
\newblock {Central Masses and Broad-Line Region Sizes of Active Galactic
  Nuclei. II. A Homogeneous Analysis of a Large Reverberation-Mapping
  Database}.
\newblock {\em Astrophys. J.}, 613:682--699, Oct. 2004, astro-ph/0407299.

\bibitem{Pfeiffer:2012pc}
H.~P. Pfeiffer.
\newblock {Numerical simulations of compact object binaries}.
\newblock {\em Class.Quant.Grav.}, 29:124004, 2012, 1203.5166.

\bibitem{2012arXiv1204.6242P}
T.~{Piran}, E.~{Nakar}, and S.~{Rosswog}.
\newblock {The electromagnetic signals of compact binary mergers}.
\newblock {\em Mon. Not. R. Astron. Soc.}, 430:2121--2136, Apr. 2013,
  1204.6242.

\bibitem{Pretorius:2005gq}
F.~Pretorius.
\newblock Evolution of binary black-hole spacetimes.
\newblock {\em Phys. Rev. Lett.}, 95:121101, 2005, {arXiv:gr-qc/0507014}.

\bibitem{Pretorius:2006tp}
F.~Pretorius.
\newblock Simulation of binary black hole spacetimes with a harmonic evolution
  scheme.
\newblock {\em Class. Quantum Grav.}, 23:S529--S552, 2006,
  {arXiv:gr-qc/0602115}.

\bibitem{Pretorius:2007nq}
F.~Pretorius.
\newblock Binary black hole coalescence.
\newblock In M.~Colpi, P.~Casella, V.~Gorini, U.~Moschella, and A.~Possenti,
  editors, {\em Physics of Relativistic Objects in Compact Binaries: from Birth
  to Coalescence}, pages 305--369. Springer, Heidelberg, Germany, 2009,
  {arXiv:0710.1338 [gr-qc]}.

\bibitem{Pretorius:2007jn}
F.~Pretorius and D.~Khurana.
\newblock Black hole mergers and unstable circular orbits.
\newblock {\em Class. Quantum Grav.}, 24:S83--S108, 2007,
  {arXiv:gr-qc/0702084}.

\bibitem{2006Sci...312..719P}
D.~J. {Price} and S.~{Rosswog}.
\newblock {Producing Ultrastrong Magnetic Fields in Neutron Star Mergers}.
\newblock {\em Science}, 312:719--722, May 2006, astro-ph/0603845.

\bibitem{Read:2013zra}
J.~S. Read et~al.
\newblock {Matter effects on binary neutron star waveforms}.
\newblock {\em Phys.Rev.}, D88:044042, 2013, 1306.4065.

\bibitem{1988Natur.333..523R}
M.~J. {Rees}.
\newblock {Tidal disruption of stars by black holes of 10 to the 6th-10 to the
  8th solar masses in nearby galaxies}.
\newblock {\em \nat}, 333:523--528, June 1988.

\bibitem{Reynolds:2013rva}
C.~S. Reynolds.
\newblock {The Spin of Supermassive Black Holes}.
\newblock {\em Class.Quant.Grav.}, 30:244004, 2013, 1307.3246.

\bibitem{Rezzolla:2010fd}
L.~Rezzolla et~al.
\newblock {Accurate evolutions of unequal-mass neutron-star binaries:
  properties of the torus and short GRB engines}.
\newblock {\em Class. Quantum Grav.}, 27:114105, 2010, {arXiv:1001.3074
  [gr-qc]}.

\bibitem{Roedig:2011rn}
C.~Roedig and A.~Sesana.
\newblock {Origin and Implications of high eccentricities in massive black hole
  binaries at sub-pc scales}.
\newblock {\em J.Phys.Conf.Ser.}, 363:012035, 2012, 1111.3742.

\bibitem{Ruffert:1996by}
M.~Ruffert, H.-T. Janka, K.~Takahashi, and G.~Sch\"{a}fer.
\newblock Coalescing neutron stars -- a step towards physical models, ii.
  neutrino emission, neutron tori, and gamma-ray bursts.
\newblock {\em Astron. Astrophys.}, 319:122--153, 1997,
  {arXiv:astro-ph/9606181}.

\bibitem{Samsing:2013kua}
J.~Samsing, M.~MacLeod, and E.~Ramirez-Ruiz.
\newblock {The Formation of Eccentric Compact Binary Inspirals and the Role of
  Gravitational Wave Emission in Binary-Single Stellar Encounters}.
\newblock 2013, 1308.2964.

\bibitem{Sarbach:2012pr}
O.~Sarbach and M.~Tiglio.
\newblock {Continuum and Discrete Initial-Boundary-Value Problems and
  Einstein's Field Equations}.
\newblock {\em Living Rev.Rel.}, 15:9, 2012, 1203.6443.

\bibitem{AGN_downsizing1}
E.~{Scannapieco}, J.~{Silk}, and R.~{Bouwens}.
\newblock {AGN Feedback Causes Downsizing}.
\newblock {\em Astrophys. J. Lett.}, 635:L13--L16, Dec. 2005, astro-ph/0511116.

\bibitem{Scheel:2008rj}
M.~A. Scheel, M.~Boyle, T.~Chu, L.~E. Kidder, K.~D. Matthews, and H.~P.
  Pfeiffer.
\newblock High-accuracy waveforms for binary black hole inspiral, merger, and
  ringdown.
\newblock {\em Phys. Rev. D}, 79:024003, 2009, {arXiv:0810.1767 [gr-qc]}.

\bibitem{Schmidt:2010it}
P.~Schmidt et~al.
\newblock {Tracking the precession of compact binaries from their
  gravitational-wave signal}.
\newblock {\em Phys. Rev. D}, 84:024046, 2011, {arXiv:1012.2879 [gr-qc]}.

\bibitem{Schnetter:2003rb}
E.~Schnetter, S.~H. Hawley, and I.~Hawke.
\newblock {Evolutions in 3D numerical relativity using fixed mesh refinement}.
\newblock {\em Class. Quantum Grav.}, 21:1465--1488, 2004,
  {arXiv:gr-qc/0310042}.

\bibitem{2013arXiv1307.3542S}
J.~D. Schnittman.
\newblock {Astrophysics of Super-massive Black Hole Mergers}.
\newblock {\em Class.Quant.Grav.}, 30:244007, 2013, 1307.3542.

\bibitem{Sekiguchi:2011zd}
Y.~Sekiguchi et~al.
\newblock {Gravitational waves and neutrino emission from the merger of binary
  neutron stars}.
\newblock {\em Phys.Rev.Lett.}, 107:051102, 2011, 1105.2125.

\bibitem{Seoane:2013qna}
P.~A. Seoane et~al.
\newblock {The Gravitational Universe}.
\newblock 2013, 1305.5720.

\bibitem{Sesana:2010wy}
A.~Sesana, J.~Gair, E.~Berti, and M.~Volonteri.
\newblock Reconstructing the massive black hole cosmic history through
  gravitational waves.
\newblock {\em Phys. Rev. D}, 83:044036, 2011, {arXiv:1011.5893 [astro-ph]}.

\bibitem{Seto:2013wwa}
N.~Seto.
\newblock {Highly Eccentric Kozai Mechanism and Gravitational-Wave Observation
  for Neutron Star Binaries}.
\newblock {\em Phys.Rev.Lett.}, 111:061106, 2013, 1304.5151.

\bibitem{Shankar:2009ub}
F.~Shankar.
\newblock {The Demography of Super-Massive Black Holes: Growing Monsters at the
  Heart of Galaxies}.
\newblock {\em New Astron.Rev.}, 53:57--77, 2009, 0907.5213.

\bibitem{Shapiro:2009uy}
S.~L. Shapiro.
\newblock {Filling the disk hollow following binary black hole merger: The
  transient accretion afterglow}.
\newblock {\em Phys.Rev.}, D81:024019, 2010, 0912.2345.

\bibitem{Shapiro:1991zza}
S.~L. Shapiro and S.~A. Teukolsky.
\newblock {Formation of naked singularities: The violation of cosmic
  censorship}.
\newblock {\em Phys. Rev. Lett.}, 66:994--997, 1991.

\bibitem{Shibata:2013pra}
M.~Shibata, K.~Taniguchi, H.~Okawa, and A.~Buonanno.
\newblock {Coalescence of binary neutron stars in a scalar-tensor theory of
  gravity}.
\newblock 2013, 1310.0627.

\bibitem{soltan82}
A.~{Soltan}.
\newblock {Masses of quasars}.
\newblock {\em Mon. Not. R. Astron. Soc.}, 200:115--122, July 1982.

\bibitem{Somiya:2011np}
K.~Somiya.
\newblock {Detector configuration of KAGRA: The Japanese cryogenic
  gravitational-wave detector}.
\newblock {\em Class.Quant.Grav.}, 29:124007, 2012, 1111.7185.

\bibitem{bhns_astro_letter}
B.~C. Stephens, W.~E. East, and F.~Pretorius.
\newblock {Eccentric Black Hole-Neutron Star Mergers}.
\newblock {\em Astrophys. J. Lett.}, 737(1):L5, 2011, 1105.3175.

\bibitem{Stephens:2008hu}
B.~C. Stephens, S.~L. Shapiro, and Y.~T. Liu.
\newblock {Collapse of magnetized hypermassive neutron stars in general
  relativity: Disk evolution and outflows}.
\newblock {\em Phys.Rev.}, D77:044001, 2008, 0802.0200.

\bibitem{2011MNRAS.412...75S}
N.~{Stone} and A.~{Loeb}.
\newblock {Prompt tidal disruption of stars as an electromagnetic signature of
  supermassive black hole coalescence}.
\newblock {\em \mnras}, 412:75--80, 2011, 1004.4833.

\bibitem{Tai:2014bfa}
K.~S. Tai, S.~T. McWilliams, and F.~Pretorius.
\newblock {Detecting gravitational waves from highly eccentric compact
  binaries}.
\newblock 2014, 1403.7754.

\bibitem{2013Natur.500..547T}
N.~R. {Tanvir} et~al.
\newblock {A `kilonova' associated with the short-duration {$\gamma$}-ray burst
  GRB130603B}.
\newblock {\em \nat}, 500:547--549, 2013, 1306.4971.

\bibitem{Taracchini:2013rva}
A.~Taracchini, A.~Buonanno, Y.~Pan, T.~Hinderer, M.~Boyle, et~al.
\newblock {Effective-one-body model for black-hole binaries with generic mass
  ratios and spins}.
\newblock 2013, 1311.2544.

\bibitem{2010ApJ...711...50T}
A.~{Tchekhovskoy}, R.~{Narayan}, and J.~C. {McKinney}.
\newblock {Black Hole Spin and The Radio Loud/Quiet Dichotomy of Active
  Galactic Nuclei}.
\newblock {\em \apj}, 711:50--63, Mar. 2010, 0911.2228.

\bibitem{2008PhRvD..78h1501T}
W.~{Tichy} and P.~{Marronetti}.
\newblock {Final mass and spin of black-hole mergers}.
\newblock {\em \prd}, 78(8):081501, Oct. 2008, 0807.2985.

\bibitem{Tsang:2013mca}
D.~Tsang.
\newblock {Shattering Flares During Close Encounters of Neutron Stars}.
\newblock {\em Astrophys.J.}, 777:103, 2013, 1307.3554.

\bibitem{Tsang:2011ad}
D.~Tsang et~al.
\newblock {Resonant Shattering of Neutron Star Crusts}.
\newblock {\em Phys.Rev.Lett.}, 108:011102, 2012, 1110.0467.

\bibitem{volonteri_sikora}
M.~{Volonteri}, M.~{Sikora}, and J.-P. {Lasota}.
\newblock {Black Hole Spin and Galactic Morphology}.
\newblock {\em Astrophys. J.}, 667:704--713, Oct. 2007, 0706.3900.

\bibitem{volonteri13}
M.~{Volonteri}, M.~{Sikora}, J.-P. {Lasota}, and A.~{Merloni}.
\newblock {The Evolution of Active Galactic Nuclei and their Spins}.
\newblock {\em Astrophys. J.}, 775:94, Oct. 2013, 1210.1025.

\bibitem{2013arXiv1312.1357W}
C.~L. {Wainwright} et~al.
\newblock {Simulating the universe(s): from cosmic bubble collisions to
  cosmological observables with numerical relativity}.
\newblock 2013, 1312.1357.
\newblock [arXiv:1312.1357].

\bibitem{1997gr.qc....10068W}
R.~M. {Wald}.
\newblock {Gravitational Collapse and Cosmic Censorship}.
\newblock {\em ArXiv General Relativity and Quantum Cosmology e-prints}, Oct.
  1997, gr-qc/9710068.

\bibitem{Wen:2002km}
L.~Wen.
\newblock {On the Eccentricity Distribution of Coalescing Black Hole Binaries
  Driven by the Kozai Mechanism in Globular Clusters}.
\newblock {\em Astrophys. J.}, 598:419--430, 2003, astro-ph/0211492.

\bibitem{1993tegp.book.....W}
C.~M. {Will}.
\newblock {\em {Theory and Experiment in Gravitational Physics}}.
\newblock Mar. 1993.

\bibitem{Will:2006LR}
C.~M. Will.
\newblock The confrontation between general relativity and experiment.
\newblock {\em Living Rev. Relativity}, 9(3), 2006, {arXiv:gr-qc/0510072}.
\newblock \url{http://www.livingreviews.org/lrr-2006-3}.

\bibitem{Wongwathanarat:2012zp}
A.~Wongwathanarat, H.-T. Janka, and E.~Mueller.
\newblock {Three-dimensional neutrino-driven supernovae: Neutron star kicks,
  spins, and asymmetric ejection of nucleosynthesis products}.
\newblock {\em AA 552,}, A126, 2013, 1210.8148.

\bibitem{2013PhRvD..88h3509X}
B.~{Xue}, D.~{Garfinkle}, F.~{Pretorius}, and P.~J. {Steinhardt}.
\newblock {Nonperturbative analysis of the evolution of cosmological
  perturbations through a nonsingular bounce}.
\newblock {\em \prd}, 88(8):083509, Oct. 2013, 1308.3044.

\bibitem{2014arXiv1404.1435Y}
C.-M. {Yoo} and H.~{Okawa}.
\newblock {Black Hole Universe with \$$\backslash$Lambda\$}.
\newblock {\em ArXiv e-prints}, Apr. 2014, 1404.1435.

\bibitem{2013PhRvL.111p1102Y}
C.-M. {Yoo}, H.~{Okawa}, and K.-i. {Nakao}.
\newblock {Black-Hole Universe: Time Evolution}.
\newblock {\em Physical Review Letters}, 111(16):161102, Oct. 2013, 1306.1389.

\bibitem{Yunes:2009ke}
N.~Yunes and F.~Pretorius.
\newblock {Fundamental theoretical bias in gravitational wave astrophysics and
  the parametrized post-Einsteinian framework}.
\newblock {\em Phys. Rev. D}, 80:122003, 2009, {arXiv:0909.3328 [gr-qc]}.

\bibitem{Zhao:2009yp}
X.~Zhao and G.~J. Mathews.
\newblock {Effects of structure formation on the expansion rate of the
  universe: an estimate from numerical simulations}.
\newblock {\em Phys.Rev.}, D83:023524, 2011, 0912.4750.

\end{thebibliography}
\bibliographystyle{habbrv}

\end{document}